\documentclass[aps,
prb,
reprint,
superscriptaddress
]{revtex4-2}

\usepackage[T1]{fontenc}
\usepackage[utf8]{inputenc}

\usepackage{amsmath,amssymb,graphicx,bm,xspace,array,longtable}
\usepackage[linktocpage=true]{hyperref}

\graphicspath{{./figures/}}

\newcommand{\sxc}{_{\mathrm{xc}}}
\newcommand{\rrscan}{r$^2$SCAN\xspace}
\newcommand{\lcwpbe}{LC-$\omega$PBE\xspace}
\newcommand{\nup}{n_\uparrow}
\newcommand{\ndn}{n_\downarrow}
\newcommand{\br}{\bm{r}}

\newcolumntype{R}[1]{>{\raggedleft\let\newline\\\arraybackslash\hspace{0pt}}m{#1}}

\begin{document}

\title[Understanding density errors]{Understanding density driven errors for reaction barrier heights}

\author{Aaron D. Kaplan}
\email{kaplan@temple.edu}
\affiliation{Department of Physics, Temple University, Philadelphia, PA 19122}

\author{Chandra Shahi}
\affiliation{Department of Physics, Temple University, Philadelphia, PA 19122}

\author{Pradeep Bhetwal}
\affiliation{Department of Physics, Temple University, Philadelphia, PA 19122}

\author{Raj K. Sah}
\affiliation{Department of Physics, Temple University, Philadelphia, PA 19122}

\author{John P. Perdew}
\email{perdew@temple.edu}
\affiliation{Department of Physics, Temple University, Philadelphia, PA 19122}
%\alsoaffiliation{Department of Chemistry, Temple University, Philadelphia, PA 19122}
\affiliation{Department of Chemistry, Temple University, Philadelphia, PA 19122}

\date{\today}

\begin{abstract}
    Delocalization errors, such as charge-transfer and some self-interaction errors, plague computationally-efficient and otherwise-accurate density functional approximations (DFAs).
    Evaluating a semi-local DFA non-self-consistently on the Hartree-Fock (HF) density is often recommended as a computationally inexpensive remedy for delocalization errors.
    For sophisticated meta-GGAs like SCAN, this approach can achieve remarkable accuracy.
    This HF-DFT (also known as DFA@HF) is often presumed to work, when it significantly improves over the DFA, because the HF density is more accurate than the self-consistent DFA density, in those cases.
    By applying the metrics of density-corrected density functional theory (DFT), we show that HF-DFT works for barrier heights by making a \textit{localizing} charge transfer error or density over-correction, thereby producing a somewhat-reliable cancellation of density- and functional-driven errors for the energy.
    A quantitative analysis of the charge transfer errors in a few randomly-selected transition states confirms this trend.
    We do not have the exact functional and exact electron densities that would be needed to evaluate the exact density- and functional-driven errors for the large BH76 database of barrier heights.
    Instead, we have identified and employed three fully-non-local proxy functionals (the SCAN 50\% global hybrid, the range-separated hybrid \lcwpbe{}, and SCAN-FLOSIC) and their self-consistent proxy densities.
    These functionals are chosen because they yield reasonably accurate self-consistent barrier heights, and because their self-consistent total energies are nearly piecewise linear in fractional electron number - two important points of similarity to the exact functional.
    We argue that density-driven errors of the energy in a self-consistent density functional calculation are second-order in the density error, and that large density-driven errors arise primarily from incorrect electron transfers over length scales larger than the diameter of an atom.
\end{abstract}

\maketitle

\section{Introduction}

Stretched radical bonds (such as those arising in the transition states of chemical reactions) present an unusual challenge for semi-local density functional approximations (DFAs).
Within Kohn-Sham density functional theory \cite{kohn1965}, semi-local (SL) DFAs are single integrals over an exchange-correlation energy density $e\sxc$
\begin{equation}
    E\sxc^\text{SL}[\nup,\ndn] = \int e\sxc^\text{SL}(\nup(\br),\ndn(\br),...) \, d^3 r,
\end{equation}
where $\nup(\br)$ and $\ndn(\br)$ are the up- and down-spin densities, respectively.
The local spin-density approximation (LSDA) \cite{kohn1965,perdew1992} employs only the spin densities, whereas the generalized gradient approximation (GGA) also employs their gradients, and the meta-GGA further employs the positive orbital kinetic energy densities $\tau_\uparrow(\br), \, \tau_\downarrow(\br)$ (and/or the Laplacians of the spin densities).

The semi-local exchange-correlation hole \cite{perdew1996a} is too localized to describe stretched radical bonds well, as demonstrated for the simple case of stretched H$_2^+$ \cite{becke2003}.
In the limit of large bond length, this molecule has half an electron, and half of the electron's exact exchange hole, on each nuclear center.
The semi-local functionals then make the energy too negative, an error that cannot be fixed by spin-symmetry breaking.
But the transition states we will study here dissociate to product molecules of integer charge.

For a molecular complex A--B, imagine constructing a fictive dividing plane orthogonal to  the bond axis such that the exact electron density on either side integrates to the number of electrons for the isolated fragments \cite{dasgupta2022}.
In the dissociation limit, the electron density on atom A (B) should integrate to integer-valued $N_\text{A}$ ($N_\text{B}$) electrons \cite{perdew1982}.
For finite separations, a semi-local DFA can transfer electrons across this dividing plane \cite{dasgupta2022}, as it underestimates the Perdew-Parr-Levy-Balduz (PPLB) \cite{perdew1982} straight-line condition.
Such delocalizing charge-transfer errors have been found \cite{dasgupta2022} to correlate with errors in the binding energy of a water dimer.
Conversely, Hartree-Fock theory \cite{szabo1982} overestimates \cite{cohen2008a} the PPLB condition, and can produce an erroneous charge transfer across the dividing plane in the opposite direction.
Both phenomena will be demonstrated here.

It has been shown, e.g., in ref \citenum{vydrov2007}, that a self-interaction correction (SIC) to a semi-local DFA can decrease its violation of the PPLB condition.
A SIC removes all one-electron self-interaction errors (the erroneous interaction of each electron with itself contained in the Hartree energy) from a semi-local DFA, however a general-purpose SIC has been elusive.
The commonly-used Perdew-Zunger (PZ) SIC \cite{perdew1981} is not invariant under a unitary transformation of the occupied Kohn-Sham orbitals used to construct it \cite{pederson2014}.
Thus, a different representation of the orbitals that yields the same spin densities (a unitary transformation of the canonical Kohn-Sham orbitals) can lead to a lower SIC total energy.
The Fermi-L\"owdin orbital (FLO) SIC method \cite{pederson2014} was proposed to minimize the SIC energy over a subset of localizing unitary transformations.

However, the existence of noded real-valued FLOs in stretched molecules has proved challenging for FLOSIC \cite{shahi2019}, suggesting the use of complex-valued Fermi-L\"owdin orbitals \cite{withanage2022}.
The situation is similar in PZSIC with a more general unitary transformation, where the use of complex-valued orbitals has been shown \cite{lehtola2016} to inhibit spontaneous symmetry breaking and further lower the total energy over PZSIC with real-valued orbitals.
Alternatively, the SIC energy density for real orbitals can be scaled down locally in many-electron regions \cite{zope2019}, but the ``gauge'' of the self-exchange-correlation energy density can be incompatible with that of the self-Hartree energy density \cite{bhattarai2020}.
This leads to errors in the scaled-down SIC.
We use ``gauge'' in the classical electromagnetism sense: The divergence of any function $\bm{G}\sxc$ that vanishes sufficiently rapidly at the boundary surface, $\text{bdy}\, \Omega$, of the integration volume $\Omega$, can be added to the exchange-correlation energy density $e\sxc$ and yield the same integrated exchange-correlation energy,
\begin{align}
  \int_\Omega \left[ e\sxc + \nabla \cdot \bm{G}\sxc \right] d^3r
  &= \int_\Omega e\sxc \, d^3r + \int_{\text{bdy}\, \Omega} \bm{G}\sxc \cdot d\bm{S} \nonumber \\
  & = \int_\Omega e\sxc \, d^3r.
\end{align}
Besides LSDA, which is in the same gauge as the Hartree energy density, the use of a compliance function \cite{bhattarai2020}, like those used for local hybrids \cite{tao2008}, is needed to put both energy densities in the same gauge.

Other methods, such as the localized orbital scaling correction (LOSC), have been developed to restore the correct fractional charge \cite{perdew1982} and spin \cite{cohen2008} constraints to semi-local DFAs.
A local hybrid, which locally mixes short-range exact (single-determinant) exchange with semi-local exchange, was recently trained to approximately satisfy these constraints \cite{kirkpatrick2021}.
However, both methods are computationally intensive, and neither can be generalized to all systems (particularly solids).

A conceptually simpler method to reduce the violation of the PPLB condition treats the short- and long-range parts of the Coulomb interaction separately \cite{savin1995,leininger1997}.
The short-range part is given by a semi-local hole model that correctly finds a cusp in the zero-separation or on-top exchange-correlation hole \cite{burke1998}.
The long-range part is given by a wavefunction method; in the work of ref \citenum{leininger1997}, the configuration interaction (CI) method was used.
As CI cannot be used to treat solids, more generally-applicable models use exact exchange \cite{hisayoshi2001} for the long range part.
The partitioning is typically done using the error function \cite{adamson1999} with an effective screening or range-separation parameter $\omega$.

These long-range-corrected (range-separated) hybrid DFAs better satisfy the PPLB fractional charge condition \cite{vydrov2007}, leading to a reduction in ``delocalization'' errors \cite{henderson2009}.
For systems with one electron (generally, systems with fractional electron number $0 < N \leq 1$), self-interaction errors and delocalization errors are identical \cite{li2017}.
Reducing delocalization errors can also reduce self-interaction errors \cite{ruzsinszky2007}, and vice versa.
Even single-empirical-parameter range-separated hybrids like LC-$\omega$PBE \cite{vydrov2006} predict barrier heights (BHs) with near-chemical accuracy (about 1 kcal/mol absolute error).
Full exact exchange at long range is correct for molecules and wide-gap insulators, but effectively predicts zero screening.
This yields poor predictions for solid-state metals and narrow-gap insulators, as well as metallic clusters \cite{gerber2007}.

The framework of density-corrected (DC) DFT \cite{kim2013,wasserman2017,sim2018,vuckovic2019} permits a quantitative analysis of errors made by DFAs.
DC-DFT expresses the energy error as the sum of two in-principle evaluable metrics: the error made by an approximate energy functional $E_\text{approx}$ in predicting the exact energy $E_\text{exact}$,
\begin{equation}
    \Delta E_\text{F} = E_\text{approx}[n_\text{exact}] - E_\text{exact}[n_\text{exact}], \label{eq:fde}
\end{equation}
also called the functional-driven error; and the error made by a DFA in predicting the exact density $n_\text{exact}$
\begin{equation}
    \Delta E_\text{D} =  E_\text{approx}[n_\text{approx}] - E_\text{approx}[n_\text{exact}], \label{eq:dde}
\end{equation}
or the density-driven error.
These definitions are easily generalized from total energies to total energy differences.
Given a density computed with a correlated wavefunction method such as coupled cluster (CC) or configuration interaction (CI), it would appear straightforward to evaluate eq \ref{eq:dde}.
However, the quantum chemical correlation energy, defined as the error made by the Hartree-Fock approximation, differs from the definition of the Kohn-Sham correlation energy.
Moreover, the Kohn-Sham kinetic energy is, like the Hartree-Fock kinetic energy, found for non-interacting electrons.
Correlated wavefunction methods produce an ``interacting'' kinetic energy without an easily-separable non-interacting component.
Thus, to evaluate eq \ref{eq:dde} using a correlated wavefunction density, one must invert \cite{nam2020} the density to yield a local Kohn-Sham potential and its associated orbitals and density.

Such Kohn-Sham inversion procedures have been tested primarily for closed-shell systems \cite{shi2021}.
Reference \cite{nam2020} studied the OH...Cl$^-$ complex, which is similar to a transition state used in the BH76 set, O...H...Cl, and estimated the uncertainty in the inversion to be less than 0.5 kcal/mol (less than the minimum absolute barrier height in BH76), from the discrepancy between different inversion methods.
However, this estimate of the  uncertainty accounts only for the difference in methods and not the uncertainty in energies computed from those densities.
Here, we have chosen to instead find a computationally feasible method for estimating density- and functional-driven errors without making recourse to a Kohn-Sham inversion scheme.

Some densities that are multi-reference in character, such as un-stretched singlet C$_2$, have been shown not to be non-interacting pure-state $v$-representable \cite{schipper1998}.
The Kohn-Sham inversion process depends crucially upon the $v$-representability of the density.
This issue can be circumvented by permitting fractional occupancies of the Kohn-Sham orbitals in the inversion process, but an accurate and reliable inversion could also require a multi-configurational correlated wavefunction approach as input.
Transition states of chemical reactions, which may have multi-reference character, are thus likely to challenge Kohn-Sham inversion methods.

A system is considered ``normal'' in DC-DFT when the errors made by the energy functional are larger than those in its self-consistent density, $|\Delta E_\text{F}| > |\Delta E_\text{D}|$.
An abnormal system has density-driven errors of comparable or larger magnitude to its functional-driven errors, $|\Delta E_\text{F}| \lesssim |\Delta E_\text{D}|$ \cite{wasserman2017}.
Systems with a large sensitivity to perturbations in the Kohn-Sham potential, including those with small HOMO-LUMO gaps, are said to exhibit density sensitivity \cite{kim2013}.

Applying PZSIC to a semi-local DFA can reduce the density- and functional-driven errors in abnormal systems and thus improve, e.g., reaction barrier heights \cite{patchkovskii2002,mishra2022}.
However, it has long been known \cite{scuseria1992,janesko2008,verma2012} that applying semi-local DFAs to the Hartree-Fock density tends to produce highly accurate reaction barrier heights, and other quantum chemical quantities \cite{oliphant1994}.
This methodology, variously called DFA@HF to indicate a DFA evaluated at the HF density, or HF-DFT, is relatively inexpensive but loses the benefits of self-consistency, like computation of forces from the Hellman-Feynman theorem \cite{levy1985}.
However, it is possible to optimize geometries using a non-variational implementation \cite{verma2012} of HF-DFT.

HF-DFT has achieved good accuracy in broad chemical tests \cite{santra2021}, and even chemical accuracy for water clusters, liquid water, and similar systems \cite{dasgupta2022,dasgupta2021,song2022a}.
In this work, we explore when and why DFA@HF works well.
We use the quantitative metrics of DC-DFT to explain the remarkable accuracy of SCAN@HF for barrier heights.
We had expected to find a significant reduction in the density-driven error of eq \ref{eq:dde}, but we found instead a significant cancellation of that error with the functional-driven error of eq \ref{eq:fde}.
This is achieved by changing the sign of the density-driven error from negative to positive.
The possibility of an error cancellation in HF-DFT was mentioned recently in refs \citenum{janesko2017} and \citenum{crisostomo2022}.
The following section details the computational methods used in this work.
Barrier height and charge transfer errors are analyzed in the ``Results and Discussion'' section; approximate density- and functional-driven errors and density sensitivities are analyzed in the ``Analysis of density- and functional-driven errors'' section.

\section{Computational Methods \label{sec:comp_metd}}

The calculations of the BH76 \cite{zhao2005,goerigk2017} set of 76 forward and backward reaction barrier heights (BHs) were done in PySCF \cite{qiming2015,qiming2018,qiming2020}.
For calculations with a PZSIC, we used the NRLMOL-FLOSIC code \cite{NRLMOL}.
For the Hartree-Fock and density functional calculations run with PySCF, all systems with zero total spin were run as spin restricted; this includes six transition states.
All FLOSIC calculations were run spin-unrestricted, which allows for spin-contamination.

In PySCF, total energies were converged to $10^{-7}$ Hartree, and the densest numerical integration grid (size 9) was used.
In a few cases (such as Cl with the LSDA or the N$_2$H$_3$ transition state with HF), a level shift was needed to stabilize convergence; the DIIS (direct inversion in the iterative subspace) method \cite{pulay1980} was used throughout.
We a used spherical representation of the basis sets for aug-cc-pVQZ and def2-QZVP, and Cartesian representation for the NRLMOL basis set, consistent with the NRLMOL-FLOSIC code \cite{NRLMOL}.
Charge, multiplicity, and structural data were taken from ref \citenum{goerigk2017}.
All requisite input files can be found at the public code repository \cite{code_repo}, and all data can be found at the public data repository \cite{data_repo}.
The data repository includes relaxed \rrscan{} Fermi orbital descriptors (described below).

Fermi-L\"owdin orbital (FLO) SIC \cite{pederson2014,pederson2015,yang2017,yamamoto2019} calculations were performed using the NRLMOL-FLOSIC code \cite{NRLMOL}.
Although ref \citenum{withanage2022} recently described a modification to the FLOSIC code utilizing complex-valued orbitals, we have restricted our work to real-valued orbitals.
All total energies were converged to $10^{-6}$ Hartree.
These calculations employed the NRLMOL or density-functional optimized \cite{porezag1999} basis set in the Cartesian representation.
As before, charge, multiplicity, and structural data were taken from ref \citenum{goerigk2017}.

To perform FLOSIC calculations, an initial set of Fermi orbital descriptors (FODs) must be supplied by the user.
Two different methods \cite{schwalbe2019} were used to generate initial FODs, both of which are packages within the PyFLOSIC \cite{schwalbe2020} code: PyCOM and fodMC.
PyCOM computes the FOD as the center of mass of each orbital density; it is DFA-dependent but otherwise conceptually simple.
fodMC non-deterministically optimizes the FOD positions according to user-supplied chemical bonding information and an inter-FOD Coulomb-like repulsion.
fodMC is thus DFA-independent, but requires chemical intuition and must be modified to handle charged atoms.
The forces on the FODs were optimized to less than $5\times 10^{-4}$ Hartree/bohr.

All rungs of the Jacob's ladder hierarchy \cite{perdew2001} of density functional approximations (DFAs) up to the single hybrid level were considered.
As a stand-in for the Kohn-Sham exact exchange only approximation, we used the Hartree-Fock approximation (HF) \cite{szabo1982}.
For standard Kohn-Sham DFAs, we employed the LSDA with the Perdew-Wang parameterization  \cite{perdew1992} of the uniform electron gas correlation energy, the Perdew-Burke-Ernzerhof (PBE) generalized gradient approximation (GGA) \cite{perdew1996}, and the Becke exchange \cite{becke1988} with Lee-Yang-Parr correlation \cite{lee1988,miehlich1989} (BLYP) GGA.

At the meta-GGA level, we used the non-empirical strongly-constrained and appropriately normed (SCAN) \cite{sun2015} and revised-regularized SCAN (\rrscan{}) \cite{furness2020} functionals.
We also consider two empirical meta-GGAs that are known \cite{goerigk2017} to perform with the accuracy of hybrids for BH76: M06-L \cite{zhao2006} and MN15-L \cite{yu2016}, both of which are fitted to the BH76 set, in addition to other chemical reaction data.
Our motivation is to understand how HF-DFT on the one hand, and fitting to barrier heights on the other hand, affect the density- and functional-driven errors of eqs \ref{eq:fde} and \ref{eq:dde}.

At the hybrid level, we considered the long-range corrected, range-separated hybrid \lcwpbe{} \cite{vydrov2006}, which is a one-parameter generalization of PBE, and thus minimally-empirical, as well as the empirical B3LYP \cite{becke1993,stephens1994}.

Three sets of calculations have been performed: with the aug-cc-pVQZ \cite{dunning1989}, def2-QZVP \cite{weigend2003,weigend2005}, and NRLMOL or density-functional optimized \cite{porezag1999,pritchard2019} basis sets.
The aug-cc-pVQZ set is of comparable size to the def2-QZVP set, and the latter was recommended for general use (especially BH76) in ref \citenum{goerigk2017}.
The aug-cc-pVQZ set gives results comparable to def2-QZVP, which we show in Table \ref{tab:ak_pyscf_aug-cc-pvqz} and Supporting Information (SI) Table \ref{tab:ak_pyscf_def2-qzvp}, respectively.
Whereas the aug-cc-pVQZ and def2-QZVP sets include Gaussian-type orbitals with orbital angular momentum 3 and higher, the NRLMOL basis set includes only $s$, $p$, and $d$ shells.
The BH76 error statistics computed with the much smaller NRLMOL basis set, shown in SI Table \ref{tab:ak_pyscf_nrlmol}, are surprisingly close to those computed with either quadruple-zeta set.

As the data computed with the aug-cc-pVQZ and def2-QZVP bases agree to within the uncertainty of the reference energies, we will perform analysis on data computed with the (slightly) larger aug-cc-pVQZ set.
Nearly identical conclusions could be drawn from the def2-QZVP data.

\section{Results and Discussion \label{sec:rd}}

In all cases, applying a semi-local DFA, except MN15-L, to the HF density yields significantly lower BH errors, seen in Table \ref{tab:ak_pyscf_aug-cc-pvqz}.
However, applying a minimally empirical non-local DFA like \lcwpbe{} to the HF density yields \textit{worse} reaction BHs than those found self-consistently.
This latter tendency was noted in ref \citenum{janesko2008}.
B3LYP evaluated on the HF density makes a roughly 40\% lower mean absolute deviation (MAD) than self-consistent B3LYP.
Applying semi-local DFAs and B3LYP to the \lcwpbe{} density yields a minor $10-15\%$ decrease in the MADs, compared to the 50--60\% reduction in the MADs using the HF density, also noted previously in ref \citenum{janesko2008}.
All this suggests that the B3LYP densities in the transition states are closer to those of a semi-local DFA than to the HF density.

\begin{ruledtabular}
    \begin{table*}[ht]
        \centering
        \begin{tabular}{l|rrr|rrr|rrr}
BH76 & \multicolumn{3}{c|}{MD} & \multicolumn{3}{c|}{MAD} & \multicolumn{3}{c}{RMSD}\\ 
DFA & @DFA & @HF & @LC-$\omega$PBE & @DFA & @HF & @LC-$\omega$PBE & @DFA & @HF & @LC-$\omega$PBE \\ \hline 
HF & 10.62 &  &  & 11.27 &  &  & 13.18 &  & \\  
LSDA & -15.30 & -5.09 & -13.16 & 15.39 & 7.82 & 13.38 & 17.84 & 9.98 & 15.61\\  
PBE & -8.88 & -0.92 & -7.70 & 8.93 & 3.85 & 7.76 & 10.26 & 5.34 & 9.10\\  
BLYP & -8.05 & -0.17 & -6.80 & 8.06 & 2.84 & 6.85 & 9.28 & 4.63 & 8.05\\  
SCAN & -7.44 & -1.70 & -6.85 & 7.50 & 3.05 & 6.94 & 8.22 & 4.03 & 7.65\\  
r$^2$SCAN & -6.91 & -1.10 & -6.35 & 6.96 & 2.84 & 6.42 & 7.75 & 4.08 & 7.18\\  
M06-L & -3.58 & 2.64 & -2.77 & 3.84 & 3.17 & 3.25 & 4.86 & 4.73 & 4.19\\  
MN15-L & -0.87 & 4.94 & -0.32 & 1.80 & 5.37 & 1.85 & 2.65 & 6.67 & 2.43\\  
B3LYP & -4.35 & 1.04 & -3.90 & 4.41 & 2.71 & 3.99 & 5.24 & 3.99 & 4.82\\  
LC-$\omega$PBE & 0.60 & 4.11 &  & 1.87 & 4.18 &  & 2.49 & 5.69 & \\ \hline 
BH76RC & \multicolumn{3}{c|}{MD} & \multicolumn{3}{c|}{MAD} & \multicolumn{3}{c}{RMSD}\\ 
DFA & @DFA & @HF & @LC-$\omega$PBE & @DFA & @HF & @LC-$\omega$PBE & @DFA & @HF & @LC-$\omega$PBE \\ \hline 
HF & -0.27 &  &  & 8.54 &  &  & 11.31 &  & \\  
LSDA & 0.53 & -0.54 & -0.41 & 8.72 & 6.59 & 7.79 & 11.24 & 8.64 & 10.35\\  
PBE & 1.04 & 0.70 & 0.97 & 4.09 & 2.86 & 3.81 & 6.00 & 4.10 & 5.65\\  
BLYP & 0.77 & 0.60 & 0.74 & 3.26 & 2.36 & 3.10 & 4.35 & 3.06 & 4.11\\  
SCAN & -0.06 & -0.50 & -0.11 & 3.12 & 2.70 & 3.04 & 4.18 & 3.40 & 3.99\\  
r$^2$SCAN & 0.06 & -0.41 & -0.00 & 2.98 & 2.59 & 2.89 & 4.03 & 3.20 & 3.86\\  
M06-L & 1.58 & 1.11 & 1.54 & 2.77 & 2.39 & 2.71 & 4.16 & 3.13 & 3.85\\  
MN15-L & 1.19 & 0.94 & 1.28 & 2.34 & 2.41 & 2.40 & 3.14 & 3.10 & 3.08\\  
B3LYP & -0.20 & -0.22 & -0.20 & 2.07 & 1.87 & 2.04 & 2.66 & 2.23 & 2.62\\  
LC-$\omega$PBE & -0.54 & -0.63 &  & 2.19 & 1.81 &  & 2.74 & 2.35 & \\  
        \end{tabular}
        \caption{BH76 error statistics (in kcal/mol) using PySCF and the aug-cc-pVQZ \cite{dunning1989} basis set.
        Mean deviations (MDs), mean absolute deviations (MADs), and root-mean-squared deviations (RMSDs) are reported.
        Reference reaction energies are taken from ref \cite{goerigk2017}.
        Self-consistent results are reported in the ``@DFA'' columns.
        The ``@HF'' columns report non-selfconsistent results for the DFA evaluated at the Hartree-Fock densities. 
        Analogously, the ``@LC-$\omega$PBE'' columns report non-selfconsistent results using the LC-$\omega$PBE densities.
        The BH76 MADs for DFA@\lcwpbe{} in this table are rather similar to those of DFA@DM21m in Table S2 of ref \cite{kirkpatrick2021}, although DM21m is a version of DM21 that was not trained to satisfy the PPLB straight-line condition.
        Using the reference data from ref \cite{goerigk2017}, the mean (mean absolute) barrier height in BH76 is 18.19 kcal/mol (18.61 kcal/mol).
        Four reactions have negative barrier heights.
        The rates of chemical reactions are sensitive to barrier heights.
        }
        \label{tab:ak_pyscf_aug-cc-pvqz}
    \end{table*}
\end{ruledtabular}

The reaction energies of the BH76 set, also called BH76RC \cite{goerigk2017}, show that the density driven errors primarily derive from the transition states.
BH76 and BH76RC error statistics are presented in Table \ref{tab:ak_pyscf_aug-cc-pvqz} using the aug-cc-pVQZ basis set, and SI Tables \ref{tab:ak_pyscf_def2-qzvp} and \ref{tab:ak_pyscf_nrlmol} using the def2-QZVP and NRLMOL basis sets, respectively.
Applying a semi-local DFA to the Hartree-Fock or the \lcwpbe{} density still yields a decrease in the MADs, but to a much smaller degree than for the reaction BHs.
As the BH76RC set does not use the BH76 transition states to compute reaction energies, this demonstrates that errors made by semi-local DFAs computed in barrier heights are density-driven for the transition states.
Of course, there are smaller density-driven errors that impact the reaction energies, and the accuracy of the DFA for reaction energies is limited by its functional error.

As expected by the variational principle, evaluating a DFA non-self-consistently on a different density will give a higher total energy than when evaluated on its true self-consistent density.
To correct the too-low barrier heights of semi-local DFAs, the positive shift in the transition state total energies has to be larger than the corresponding shift in the reactant and product state energies.
This is confirmed in SI Table \ref{tab:dE_HF_DFA}, which shows that the transition state energies increase on average by about 6--10 kcal/mol more than the reactant and product states for the LSDA, B3LYP, and the GGAs and meta-GGAs considered here when evaluated on the HF density.
The \lcwpbe{} energy shifts differ by 5 kcal/mol on average.
Thus the poorer accuracy of MN15-L@HF and \lcwpbe{}@HF cannot be explained solely by the relative energy shifts.

Our tests indicate that all semi-local DFAs considered here show strong density-sensitivity for the transition states, which feature stretched radical bonds.
To see how the density-driven errors decrease as single-determinant exchange is admixed into a semi-local DFA, we define the family of hybrid meta-GGAs SX-$x$
\begin{align}
    E\sxc^{\text{SX}-x}[n_\uparrow,n_\downarrow] =& x E_\mathrm{x}^\text{EXX} + (1 - x) E_\mathrm{x}^\text{SCAN}[n_\uparrow,n_\downarrow] \label{eq:SX-x}\\
    & + E_\mathrm{c}^\text{SCAN}[n_\uparrow,n_\downarrow]. \nonumber
\end{align}
SX-0 is equivalent to SCAN; SX-1 uses the Kohn-Sham exact exchange only approximation (EXOA) with SCAN correlation.

Table \ref{tab:sx_hyb_mads} presents the MAD in the BH76 set for LSDA, PBE, SCAN, \rrscan{}, and \lcwpbe{} evaluated non-self-consistently on the SX-$x$ density as a function of $x$.
This data is also plotted in SI fig \ref{fig:sx_hyb_mads}.
SI Table \ref{tab:R2X_hyb_mads} and fig \ref{fig:R2X_hyb_mads} present a similar analysis using an \rrscan{} global hybrid, called R2X-$x$.
These show that $0.25 \leq x \leq 0.5$ yields an \rrscan{} global hybrid of comparable accuracy as \lcwpbe{} for BH76.

For all semi-local DFAs, increasing the admixture of single-determinant exchange improves the BH76 MAD.
Evaluating \rrscan{} on the SCAN (SX-0) density yields no marked improvement in the barrier heights, while evaluating PBE and LSDA on the SCAN density yields minor improvements in the barrier heights.
However, the \lcwpbe{}@SX-$x$ MADs are virtually identical to the self-consistent \lcwpbe{} MADs for $x=0.25, \, 0.5$, implying that the SX-0.25 and SX-0.5 densities are of roughly the same quality as those of LC-$\omega$PBE.
As $x=0.5$ provides the lowest BH76 MAD and the closest \lcwpbe{} MAD, we will proceed assuming that SX-0.5 is the best SCAN global hybrid for BH76.

\begin{ruledtabular}

\begin{table}
    \centering
    \begin{tabular}{rrrrrrr}
$x$ & LSDA & PBE & SCAN & r$^2$SCAN & \lcwpbe{} & SX \\
SCF & 15.39 & 8.93 & 7.50 & 6.96 & 1.87 &  \\ \hline
0.00 & 14.43 & 8.64 & 7.50 & 6.92 & 2.08 & 7.50 \\
0.10 & 14.13 & 8.42 & 7.45 & 6.85 & 1.97 & 5.88 \\
0.25 & 13.51 & 7.92 & 7.21 & 6.59 & 1.87 & 3.77 \\
0.50 & 12.05 & 6.65 & 6.36 & 5.71 & 1.86 & 2.69 \\
0.75 & 10.41 & 5.17 & 5.01 & 4.35 & 2.31 & 4.45 \\
1.00 & 8.62 & 3.95 & 3.63 & 3.14 & 3.20 & 6.37 \\ \hline
@HF & 7.82 & 3.85 & 3.05 & 2.84 & 4.18 &  \\
    \end{tabular}
    \caption{Self-consistent (SCF), and non-self-consistent (DFA@SX-$x$ and @HF) BH76 MADs for the DFAs considered here, using the aug-cc-pVQZ basis set.
    $x$ represents the fraction of exact single-determinant exchange mixed with the SCAN exchange energy in eq \ref{eq:SX-x}.
    When a numeric value of $x$ is given, the MAD is DFA@SX-$x$.
    The ``@HF'' row evaluates DFA@HF.
    For a visual representation of this figure, see Appendix Figure \ref{fig:sx_hyb_mads}.
    }
    \label{tab:sx_hyb_mads}
\end{table}

\end{ruledtabular}

\begin{ruledtabular}

\begin{table}
    \centering
    \begin{tabular}{rR{2cm}rrr}
    & \multicolumn{2}{c}{CH$_3$...F..Cl} & F...H...H & Cl...H...H \\
$x$ & BH Error (kcal/mol) & $\Delta N$ & $\Delta N$ & $\Delta N$ \\ \hline
0.00 & 12.55 & -0.0702 & -0.0606 & -0.0209 \\
0.10 & 12.48 & -0.0645 & -0.0501 & -0.0195 \\
0.25 & 12.09 & -0.0539 & -0.0332 & -0.0174 \\
0.50 & 10.33 & -0.0302 & -0.0048 & -0.0135 \\
0.75 & 6.85 & -0.0013 & 0.0188 & -0.0090 \\
0.85 & 5.03 & 0.0105 & 0.0264 & -0.0071 \\
1.00 & 1.93 & 0.0270 & 0.0359 & -0.0039 \\ \hline
@HF & 0.94 & 0.0357 & 0.0479 & 0.0107 \\
    \end{tabular}
    \caption{Barrier height (BH) errors, from SX-$x$ and SCAN@HF, for the reaction CH$_3$ + FCl $\rightarrow$ CH$_3$F + Cl, and their corresponding charge transfer errors $\Delta N$ for the transition state, CH$_3$...F...Cl.
    The barrier height errors shown here are averages of the absolute errors in the forwards and reverse reaction barrier heights.
    Charge transfer errors are also shown for the F...H...H transition state, used in the reaction F + H$_2$ $\rightarrow$ H + HF, and the Cl...H...H transition state, used for the reaction H + HCl $\rightarrow$ H$_2$ + Cl.
    The charge transfer or density errors are evaluated on the CH$_3$ side of the transition state CH$_3$...F...Cl, and on the H...H side of the other two transition states.
    Self-consistent densities are taken from the SCAN global hybrid SX-$x$, eq \ref{eq:SX-x}, and BH errors are SCAN@SX-$x$.}
    \label{tab:SCAN_at_SX_DE}
\end{table}

\end{ruledtabular}

Table \ref{tab:SCAN_at_SX_DE} quantifies the density errors of the SX-$x$ densities found using eq \ref{eq:SX-x} for the reactions CH$_3$ + FCl $\rightarrow$ CH$_3$F + Cl, F + H$_2$ $\rightarrow$ H + HF, and H + HCl $\rightarrow$ H$_2$ + Cl.
For these reactions, the transition states are CH$_3$...F...Cl, F...H...H, and Cl...H...H, respectively.
As in the introduction, a fictive dividing plane \cite{dasgupta2022} is placed between the CH$_3$ and the F...Cl complex such that the computed orbital optimized CCD density has 9 electrons on the CH$_3$ side, and 26 electrons on the F...Cl side.
Analogously, a plane is defined such that there are 9 (17) electrons on the F (Cl) side, and 2 electrons on the H$_2$ side, of the F...H...H (Cl...H...H) complex.
Assuming the orbital-optimized CCD density to be nearly exact, we can then estimate the charge transfer error $\Delta N$ across the dividing plane for any given approximate density by integrating the charge density on both sides of the plane.
This method was introduced in ref \citenum{dasgupta2022} to analyze charge transfer errors in water clusters.

For $x \lesssim 0.8$ ($x\gtrsim 0.8$), electrons are transferred from (to) the CH$_3$ side of the CH$_3$...F..Cl transition state.
However, the SCAN@SX-$x$ BH errors minimize when there is a charge transfer in the opposite direction as for the SCAN@SX-1 and SCAN@HF BH errors.
Thus, while the SX-0.8 or correlated-wavefunction densities transfer no charge across the neutral plane, SCAN's description of the transition state benefits from a localizing charge transfer error.
As is well-known \cite{perdew1996b}, there is no single optimal mixing parameter $x$ for a global hybrid - the optimal $x$ is system-  and property-dependent.
We modify the value of $x$ to study the nature of charge-transfer errors, and do not advocate for optimizing the value of $x$ in practical applications.
A linear fit to the data would yield zero charge transfer at $x=0.76$ for CH$_3$...F...Cl and $x=0.59$ for F...H...H.
For Cl...H...H, even the 100\% SCAN global hybrid makes a delocalizing charge transfer error; only the HF density is sufficiently localizing to produce a charge transfer of the opposite sign.
In all cases, the magnitude of the charge transferred by the HF density is less than what is transferred by the SCAN density.
Our findings are consistent across different transition states.
The over-localization of the HF density in SCAN@HF cancels much of the functional-driven error of SCAN.

\begin{figure}
    \centering
    \includegraphics[width=\columnwidth]{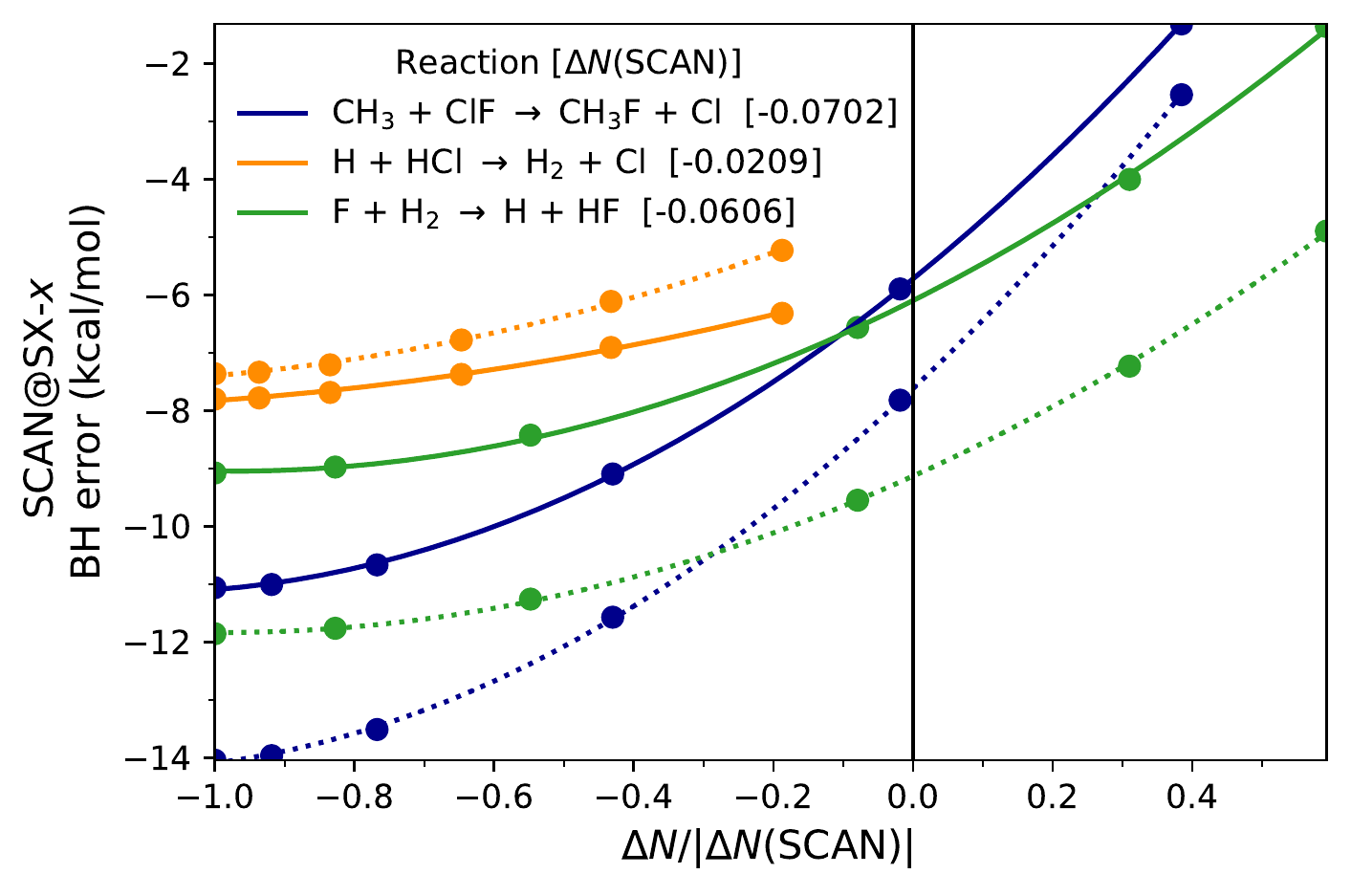}
    \caption{Barrier height errors of SCAN@SX-$x$ as a function of the integrated charge transfer errors $\Delta N$ made by the SX-$x$ global hybrid.
    The values of $\Delta N$ are taken from Table \ref{tab:SCAN_at_SX_DE}.
    Forward reactions are shown in solid lines, and reverse reactions in dotted lines.
    To present all three reactions in the same figure, the charge transfer errors are normalized to the SCAN (SX-0) values of $\Delta N$, presented in square brackets in the legend.
    The SCAN errors are quadratic, consistent with eq \ref{eq:dde_taylor}.
    }
    \label{fig:scan_pplb}
\end{figure}

The density-driven error of a self-consistent density functional calculation, eq \ref{eq:dde}, can be found from a functional Taylor expansion \cite{vuckovic2019} of $E_\text{approx}[n]$ about its minimum, in powers of $n_\text{exact} - n_\text{approx}$:
\begin{align}
    \Delta E_\text{D} &= -\frac{1}{2} \int d^3 r \int d^3 r'
    \frac{\delta^2 E_\text{approx}}{\delta n(\br) \delta n(\br')}\bigg|_{n_\text{approx}}
    \label{eq:dde_taylor} \\
    & \times \left[ n_\text{exact}(\br) - n_\text{approx}(\br) \right]
    \left[ n_\text{exact}(\br') - n_\text{approx}(\br') \right]
    \nonumber
    %& + \mathcal{O}\left[ n_\text{exact}(\br) - n_\text{approx}(\br) \right]^3.
    %\nonumber
\end{align}
See sec \ref{sec:ctan} for a derivation of this expansion.
The series has been truncated at second order, since we found for the reactions in Table \ref{tab:SCAN_at_SX_DE} that the SCAN errors of the barrier heights were accurately quadratic in the charge transfer errors $\Delta N$, see fig \ref{fig:scan_pplb}.

Approximate density functionals like SCAN make small density errors for atoms or compact molecules with integer electron numbers \cite{dasgupta2022,medvedev2017}, but much larger density errors associated with electron charge transfers over distances much greater than the size of an atom \cite{perdew1982,perdew1985a}.
For example \cite{perdew1985a}, for highly-stretched NaCl, 0.4 electrons are incorrectly transfered from neutral Na to neutral Cl by the LSDA, which lacks the derivative discontinuity \cite{perdew1982} of the exact functional that guarantees an integer number of electrons on each atom.
Of course, the electron charge transfer errors in the stretched bonds of transition states are not as large as those in highly-stretched NaCl, but they are still larger than more localized density errors, and their energetic effect is magnified by the second-order dependence of the density-driven error on the density error.

The density-driven error of a self-consistent density functional calculation, as defined in eq \ref{eq:dde} and discussed above, is always negative or zero by the variational principle.
For a non-self-consistent calculation, like SCAN@HF, we define the density-driven error as in eq \ref{eq:de_dfa_at_hf} (below), which can be positive if the SCAN energy evaluated on the Hartree-Fock density is higher than the SCAN energy evaluated on the proxy-exact density.

\begin{ruledtabular}
\begin{table}
  \centering
  \caption{BH76 error statistics (in kcal/mol) using the FLOSIC code and its default NRLMOL basis set \cite{porezag1999}. 
  The DFA rows present self-consistent DFA results.
  ``DFA-FLOSIC'' indicates a fully self-consistent calculation using the DFA with a Fermi-L\"owdin orbital self-interaction correction.
  ``DFA@FLOSIC'' indicates a DFA evaluated non-selfconsistently on its self-consistent DFA-FLOSIC density.}
    \begin{tabular}{lrrrr}
          & LSDA & PBE & SCAN  & r$^2$SCAN \\ \hline
    \multicolumn{5}{c}{DFA} \\
    MD & -15.36 & -9.88 & -7.74 & -7.24 \\
    MAD & 15.46 & 10.49 & 7.81  & 7.30 \\ \hline
    \multicolumn{5}{c}{DFA-FLOSIC} \\
    MD & 0.26  & 3.04  & 0.61  & 1.34 \\
    MAD & 5.70 & 5.26  & 3.79  & 3.54 \\ \hline
    \multicolumn{5}{c}{ DFA@FLOSIC} \\
    MD & -10.93 & -5.02 & -5.76 & -5.23 \\
    MAD & 11.17 & 5.34  & 5.85  & 5.31 \\
    \end{tabular}
  \label{tab:bh76_sic}
\end{table}
\end{ruledtabular}

Table \ref{tab:bh76_sic} demonstrates the effect of an SIC on barrier heights, both self-consistently (DFA-FLOSIC) and by non-self-consistently applying a DFA to its self-consistent FLOSIC density (DFA@FLOSIC).
By considering both cases, we can partially separate the role of density-driven and functional-driven errors.

While hybrid functionals also reduce self-interaction errors, they cannot fully remove them in a systematic way.
Moreover, global hybrids, like SX-0.5, will have an incorrect decay of the exchange-correlation potential in a finite system.
If $a$ is the mixing parameter of a global hybrid, its exchange-correlation potential $v\sxc$ will decay as $-a/r$ at large-$r$, instead of the correct \cite{sham1985,almbladh1985} behavior, $\lim_{r\to\infty} v\sxc = -1/r$.
By construction, long-range corrected range-separated hybrids, like \lcwpbe{}, will exhibit the correct decay of $v\sxc$.
A non-empirical SIC, such as FLOSIC, fully eliminates one-electron self-interaction errors, and restores the $-1/r$ decay of $v\sxc$ in a finite system regardless of the underlying approximation used.
Thus the SIC analysis is relevant for understanding the degree to which hybrids eliminate one-electron self-interaction errors.

Using the LSDA-FLOSIC density but retaining the LSDA energy functional, LSDA@FLOSIC's BH76 MAD is 72\% that of LSDA, suggesting a decrease in density-driven error.
Further changing the energy functional using self-consistent LSDA-FLOSIC further reduces the MAD to 37\% that of LSDA.
Thus LSDA makes significant density-driven and functional-driven errors that are of comparable magnitude.

Conversely, the PBE-FLOSIC (PBE@FLOSIC) BH76 MAD is 50\% (51\%) that of self-consistent PBE.
Thus PBE's errors for BH76 are primarily density-driven.
The SCAN and \rrscan{} MADs decrease by roughly 25\% when evaluated on their FLOSIC densities, and a further 25\% when evaluated self-consistently with FLOSIC.

The HF density is not the most correct density in the sense that it best mimics the exact density.
As seen in Table \ref{tab:SCAN_at_SX_DE}, HF makes charge transfer errors of comparable magnitude to SCAN but of opposite sign.
This is despite the well-known fact, also shown here in Tables \ref{tab:ak_pyscf_aug-cc-pvqz} and \ref{tab:sx_hyb_mads}, that non-self-consistently evaluating a semi-local DFA on the HF density produces very accurate BHs.

\subsection{Analysis of density- and functional-driven errors \label{sec:dde_fde}}

We can now give an interpretation to the observations of the previous paragraph: all BH computations involve an energy difference of the form
\begin{equation}
    \Delta E_\text{BH} = E_\text{TS} - \sum_{i=\text{R,P}} E_i,
\end{equation}
where $E_\text{TS}$ is the total energy of the transition state (TS), and the summation is over the total energies of the (R)eactants or (P)roducts.
Suppose, as argued here, that $E_\text{TS}$ is much more sensitive to perturbations in the Kohn-Sham potential or density.
If a DFA makes $\Delta E_\text{BH}$ too low, as in virtually all the BHs considered here, then evaluating it non-self-consistently on DFA2 (DFA@DFA2) will necessarily raise the energy by the variational principle.
The reaction energies (RE) of BH76RC are the non-positive quantities
\begin{equation}
    \Delta E_\text{RE} = E_\text{P} - E_\text{R}.
\end{equation}

\begin{ruledtabular}

\begin{table*}
    \centering
    \begin{tabular}{lrrrrrr}
 & \multicolumn{6}{c}{Proxy reference or proxy exact} \\
 & \multicolumn{2}{c}{\lcwpbe{}} & \multicolumn{2}{c}{SX-0.5} & \multicolumn{2}{c}{SCAN-FLOSIC} \\ 
\textit{BH76} & MFE & MDE & MFE & MDE & MFE & MDE \\ \hline 
LSDA & -13.76 & -2.14 & -11.61 & -3.52 & -11.53 & -4.45 \\  
PBE & -8.30 & -1.18 & -6.42 & -2.29 & -6.23 & -3.66 \\  
BLYP & -7.40 & -1.25 & -5.60 & -2.28 &  &  \\  
SCAN & -7.46 & -0.58 & -6.11 & -1.16 & -6.38 & -1.97 \\  
\rrscan{} & -6.95 & -0.57 & -5.47 & -1.27 & -7.85 & 0.00 \\  
M06-L & -3.37 & -0.81 & -1.67 & -1.73 &  &  \\  
MN15-L & -0.92 & -0.54 & 0.70 & -1.40 &  &  \\  
B3LYP & -4.50 & -0.46 & -3.13 & -1.06 &  &  \\  
\lcwpbe{} &  &  & 1.16 & -0.39 &  &  \\  
LSDA@HF & -13.76 & 8.06 & -11.61 & 6.69 & -11.53 & 5.54 \\  
PBE@HF & -8.30 & 6.78 & -6.42 & 5.68 & -6.23 & 4.41 \\  
SCAN@HF & -7.46 & 5.15 & -6.11 & 4.58 & -6.38 & 3.81 \\  
\rrscan{}@HF & -6.95 & 5.25 & -5.47 & 4.54 & -7.85 & 5.88 \\ \hline 
 & \multicolumn{2}{c}{\lcwpbe{}} & \multicolumn{2}{c}{SX-0.5} & \multicolumn{2}{c}{SCAN-FLOSIC} \\ 
\textit{BH76RC} & MFE & MDE & MFE & MDE & MFE & MDE \\ \hline 
LSDA & 0.13 & 0.94 & 1.30 & 0.93 & -0.45 & 0.26 \\  
PBE & 1.51 & 0.08 & 2.64 & 0.10 & 0.35 & 0.73 \\  
BLYP & 1.27 & 0.04 & 2.47 & 0.00 &  &  \\  
SCAN & 0.43 & 0.05 & 1.48 & 0.15 & -0.12 & 0.19 \\  
\rrscan{} & 0.53 & 0.07 & 1.60 & 0.16 & 0.19 & 0.00 \\  
M06-L & 2.08 & 0.04 & 3.14 & 0.14 &  &  \\  
MN15-L & 1.82 & -0.09 & 2.85 & 0.04 &  &  \\  
B3LYP & 0.34 & 0.00 & 1.55 & -0.05 &  &  \\  
\lcwpbe{} &  &  & 1.17 & -0.01 &  &  \\  
LSDA@HF & 0.13 & -0.13 & 1.30 & -0.14 & -0.45 & -0.08 \\  
PBE@HF & 1.51 & -0.26 & 2.64 & -0.24 & 0.35 & 0.44 \\  
SCAN@HF & 0.43 & -0.39 & 1.48 & -0.29 & -0.12 & -0.23 \\  
\rrscan{}@HF & 0.53 & -0.41 & 1.60 & -0.32 & 0.19 & -0.45 \\
    \end{tabular}
    \caption{Mean functional-driven errors (MFEs) and mean density-driven errors (MDEs) for the reaction energy differences in the BH76 and BH76RC sets.
    The vertical columns indicate which DFA is used as a proxy for the exact functional and density in eqs \ref{eq:fde} and \ref{eq:dde}.
    For the same metrics applied to the single-point total energies for the individual molecules in both sets, see Table \ref{tab:bh76_fde_dde_SP} in the Appendix.
    In the SCAN@HF row, the functional- and density-driven errors are computed according to eqs \ref{eq:fe_scan_at_hf} and \ref{eq:de_scan_at_hf}.
    Because the NRLMOL code cannot (at the time of writing) perform hybrid DFA or HF calculations, the SCAN-FLOSIC-proxy MFE and MDE for SCAN@HF were found from reaction energy differences computed with the NRLMOL basis set in PySCF (see Table \ref{tab:ak_pyscf_nrlmol} especially).
    }
    \label{tab:bh76_fde_dde_dE}
\end{table*}

\end{ruledtabular}

At this point, we may only assert that DFA2 produces a more accurate density than the original DFA if the functional-driven error of the original DFA is negligible.
Thus one may easily obtain ``the right answer for the wrong reason.''
For the reasons outlined in the Introduction, we cannot evaluate eqs \ref{eq:fde} and \ref{eq:dde} using a correlated-wavefunction method.
We can, however, replace $E_\text{exact}$ with $E_\text{ref}$ and its self-consistent density $n_\text{ref}$, and use a reference (ref) method we believe produces a highly accurate density.
We select SCAN-FLOSIC, \lcwpbe{}, and SX-0.5 as proxies for the unavailable exact functional.
These three functionals are chosen because they produce total energies of open separated subsystems that vary almost linearly with electron number between adjacent integers \cite{vydrov2007} (while the exact functional is exactly linear), and because their self-consistent barrier heights are nearly correct (while those of the exact functional are fully correct).

We demand that the proxies adhere closely to the PPLB condition as we cannot construct a point-wise error metric that measures closeness to the exact density.
If an approximate functional gives an almost straight-line interpolation between total energies of adjacent integer electron number, then it is unlikely to make energetically-relevant charge-transfer errors.
As argued in ref \cite{sim2018}, such point-wise metrics, including the $L^2$ norm used in Kohn-Sham inversion techniques
\begin{equation}
    L^2(n_1,n_2) = \int d^3 r \, |n_1(\br) - n_2(\br)|^2,
\end{equation}
are highly-sensitive to the grid and basis set used.
Moreover, the $L^2$ norm in this sense has dimensions of inverse cubic length.
Because the exact functional is variational, one can determine the relative correctness of different densities by ordering them by lowest total energy.
One might attempt something similar using a variational correlated wavefunction method, such as configuration interaction.
However, we are unaware of software that is capable of using orbitals from our proxies as input to a configuration interaction calculation, and development of such software is well-beyond the scope of this work.

While SCAN is expected to be accurate when evaluated on the exact densities of the reactants and products, the SCAN energy evaluated on the exact density of the transition state is expected to be too negative, because the SCAN energy is too low for stretched radical bonds.
Thus SCAN is expected to make a negative functional-driven error for the barrier heights, and SCAN@HF is expected to make a partly-canceling positive density-driven error.
We will now quantify this interpretation.

As functions of the electron number between adjacent integers on an open separated subsystem, the total energy varies sub-linearly for semi-local functionals and super-linearly for HF.
In a collection of open isolated subsystems, semi-local functionals can incorrectly assign non-integer electron number to each subsystem, while the HF and exact functionals will not.
The HF energy, however, cannot be regarded as a proxy for the exact energy, since the HF barrier heights are much too high.
Thus we use the SCAN energy in lieu of the HF energy.

SCAN@HF and refinements like \rrscan{}-DC4@HF \cite{song2022a} work remarkably well for water clusters and liquid water.
Our earlier and simpler interpretation \cite{dasgupta2021,dasgupta2022} was that SCAN makes negligible functional-driven error for this problem, and that the HF density is almost exact for the cluster.
Our current interpretation, based on our detailed analysis of BH76, suggests that, at least for barrier heights, a more nuanced interpretation is likely to be more correct: For the normal covalent bonds in the reactants and products, self-consistent SCAN yields both an accurate total energy and an accurate electron density \cite{dasgupta2022}.
Earlier work \cite{dasgupta2022} on the binding energies of water clusters found that SCAN@HF was more accurate than SCAN@SCAN-FLOSIC, and concluded that the HF density was closer to the exact density than the SCAN-FLOSIC density, but that conclusion may have to be revisited.

Table \ref{tab:bh76_fde_dde_dE} presents the mean functional-driven errors (MFEs) and mean density-driven errors (MDEs) for the reaction energy differences in the BH76 and BH76RC sets.
By considering \textit{mean} errors, we see exactly why the density-correction works.
When we use DFA@HF, such as SCAN@HF, as a ``target'' DFA to evaluate errors for, the functional-driven and ``generalized density-driven'' errors are computed as
\begin{align}
    \Delta E_\text{F}(\text{DFA@HF}) &=E_\text{DFA}[n_\text{ref}] - E_\text{ref}[n_\text{ref}] \label{eq:fe_dfa_at_hf} \\
    \Delta E_\text{D}(\text{DFA@HF}) &= E_\text{DFA}[n_\text{HF}] - E_\text{DFA}[n_\text{ref}] \label{eq:de_dfa_at_hf},
\end{align}
where ``ref'' is one of the proxies.
In the formalism of DC-DFT, eq \ref{eq:de_dfa_at_hf} is not a density-driven error, but an energetically-measured ``density-driven difference'' \cite{vuckovic2019}, because the self-consistent density does not appear explicitly there.
However, eq \ref{eq:de_dfa_at_hf} can be written as the difference of two density-driven errors,
\begin{align}
    \Delta E_\text{D}(\text{DFA@HF}) =& \left(E_\text{DFA}[n_\text{DFA}] - E_\text{DFA}[n_\text{ref}] \right) \\
    & -
    \left(E_\text{DFA}[n_\text{DFA}] - E_\text{DFA}[n_\text{HF}] \right).
    \nonumber
\end{align}
Therefore, eq \ref{eq:de_dfa_at_hf} energetically measures the difference between the reference density (here, a proxy for the exact density) and the Hartree-Fock density using the variational minimum of a self-consistent DFA calculation as a common reference point.
Then we find that the negative functional-driven error of SCAN is mostly canceled by a positive density-driven error from the too-localized HF density.
Note that the sum of eqs \ref{eq:fe_dfa_at_hf} and \ref{eq:de_dfa_at_hf} is exactly the total error of SCAN@HF only when $E_\text{ref}[n_\text{ref}]=E_\text{exact}[n_\text{exact}]$.
The BH76RC MFEs and MDEs do not show similar cancellation, except SCAN@HF applied to the \lcwpbe{} proxy.

SI Table \ref{tab:bh76_fde_dde_SP} performs a similar analysis for the individual molecules in the sets, with transition state and reactant/product subsets considered separately.
While the density-driven errors for the transition states are generally twice those of the reactant and product states, these metrics have clearer interpretations when applied to energy differences.
However, Table \ref{tab:bh76_fde_dde_SP} provides a check of our methods: the variational principle demands that the density-driven error of a DFA total energy be negative, as the self-consistent density will always yield a lower total energy than the proxy density.
By considering only total energies and not energy differences, Table \ref{tab:bh76_fde_dde_SP} confirms our methodology: every self-consistent DFA there (excluding SCAN@HF) has negative average density-driven errors.
As SCAN@HF is not a variational method, its density-driven errors can be positive.

As Table \ref{tab:bh76_fde_dde_dE} presents average density-driven errors in energy \textit{differences}, the sign of the density-driven error will not necessarily be negative.
The self-consistent barrier heights of (most) semi-local DFAs are systematically too low, and by the variational principle, the barrier heights found from evaluation on a different density will be higher.
Thus the BH76 mean density errors are negative.
Semi-local DFAs make less systematic errors for the BH76RC set, and thus the sign of the density-driven error will depend on how closely the self-consistent density resembles the reference density.
While the chosen three nonlocal functionals are suitable proxies for the exact functional for the barrier heights, it is very possible that self-consistent SCAN or \rrscan{} would be better proxies for the reaction energies, which involve only molecules in equilibrium \cite{dasgupta2022}.

We note two important observations regarding empirical vs. non- and semi-empirical DFAs.
BLYP performs largely the same as PBE for the BH76 set, demonstrating the consistency of density-driven errors.
The average B3LYP functional- and density-driven errors applied to any self-consistent reference (thus excluding SCAN@HF) lie nearly halfway between those of \rrscan{} and \lcwpbe{}.
The Minnesota functionals M06-L and MN15-L make slightly larger density-driven errors for the barrier heights than do SCAN or \rrscan{}, but their fitting to barrier heights reduces their functional-driven errors for this property.

\begin{ruledtabular}
    \begin{table}[]
        \centering
        \begin{tabular}{lrrr}
DFA & ADS & AMIN & AMAX \\ \hline 
\multicolumn{4}{c}{BH76} \\ 
LSDA & $10.33 \pm 8.07$ & 1.48 & 49.25 \\  
PBE & $8.37 \pm 6.62$ & 0.41 & 39.16 \\  
BLYP & $8.11 \pm 6.66$ & 0.46 & 38.50 \\  
SCAN & $5.62 \pm 4.54$ & 0.14 & 25.78 \\  
\rrscan{} & $5.89 \pm 4.87$ & 0.29 & 27.76 \\  
M06-L & $6.16 \pm 4.71$ & 0.02 & 28.51 \\  
MN15-L & $5.63 \pm 5.21$ & 0.07 & 29.12 \\  
SX-0.25 & $2.48 \pm 2.58$ & 0.00 & 13.29 \\  
B3LYP & $5.13 \pm 4.58$ & 0.23 & 25.70 \\  
\lcwpbe{} & $2.42 \pm 3.40$ & 0.01 & 20.55 \\ \hline 
\multicolumn{4}{c}{BH76RC} \\ 
LSDA & $2.59 \pm 2.78$ & 0.08 & 11.87 \\  
PBE & $2.38 \pm 2.76$ & 0.04 & 9.78 \\  
BLYP & $2.30 \pm 2.67$ & 0.04 & 10.01 \\  
SCAN & $2.16 \pm 2.60$ & 0.03 & 7.65 \\  
\rrscan{} & $2.35 \pm 2.67$ & 0.02 & 8.27 \\
M06-L & $2.26 \pm 2.03$ & 0.02 & 7.73 \\  
MN15-L & $2.02 \pm 2.16$ & 0.06 & 7.78 \\  
SX-0.25 & $1.64 \pm 2.33$ & 0.00 & 6.90 \\  
B3LYP & $1.75 \pm 2.21$ & 0.00 & 7.38 \\  
\lcwpbe{} & $1.68 \pm 1.96$ & 0.05 & 5.85 \\
        \end{tabular}
        \caption{Mean absolute density sensitivity (ADS) and its standard deviation, evaluated using eq \ref{eq:ds} on all reactions in the BH76 and BH76RC sets.
        All units are kcal/mol.
        Absolute minimum (AMIN) and maximum (AMAX) density sensitivities $S$ are reported.
        All calculations used the aug-cc-pVQZ basis set \cite{dunning1989}.
        SCAN0 is the more common name of the SX-0.25 global hybrid; its self-consistent BH76 MAD is 3.77 kcal/mol, and its @HF MAD is 2.76 kcal/mol.}
     \label{tab:dens_sens}
    \end{table}
\end{ruledtabular}

An alternative metric of density sensitivity is provided by ref \citenum{sim2018}:
\begin{equation}
    S = |E_\text{approx}[n_\text{LSDA}] - E_\text{approx}[n_\text{HF}] |, \label{eq:ds}
\end{equation}
a simple total energy difference between DFA@LSDA and DFA@HF.
When $S$ is above a threshold, 2 kcal/mol in ref \citenum{sim2018}, a system is expected to exhibit strong density sensitivities that could be cured by evaluation on the HF density.
Table \ref{tab:dens_sens} reports the mean absolute density sensitivities for the DFAs considered here, as well as their standard deviations and extreme errors.
Note that the sensitivities are evaluated for the reactions in the BH76 and BH76RC sets, and not individual molecules.
The density sensitivities of SX-0.25 and \lcwpbe{} are both about 2.5 kcal/mol for BH76, reasonably close to the 2 kcal/mol threshold.
Note that ref \citenum{song2022} advocates for using a higher or lower cutoff when physically motivated.

However, the semi-local DFAs -- LSDA, PBE, SCAN, and \rrscan{} -- show much larger density sensitivities for BH76, ranging from about 6 kcal/mol for SCAN and \rrscan{} to 10 kcal/mol for LSDA.
Moreover, B3LYP shows density sensitivity comparable to SCAN and \rrscan{}.
Despite their hybrid-level performance for BH76, M06-L and MN15-L show comparable density sensitivity to the less-accurate semi-local DFAs, suggesting that their self-consistent densities will not be accurate.
That all semi-local DFAs and B3LYP exhibit a strong density sensitivity for BH76 is consistent with Table \ref{tab:ak_pyscf_aug-cc-pvqz}, where all semi-local MADs decrease when they are applied to the HF density.

For the BH76RC set, no DFA shows a strong density sensitivity, as the maximum average sensitivity is 2.5 kcal/mol from LSDA.
The density sensitivities of  the hybrids lie within the 2 kcal/mol threshold.
This observation is consistent with the BH76RC error statistics in Table \ref{tab:ak_pyscf_aug-cc-pvqz}, at least for the meta-GGAs and hybrids.
Applying a more sophisticated DFA to the HF density yields no marked decrease in the BH76RC MAD, suggesting that the errors are primarily functional-driven for this set.

In Table \ref{tab:bh76_fde_dde_dE}, the functional- and density-driven errors have been calculated from the following equations
\begin{align}
   \Delta E_\text{F} &= E_\text{approx}[n_\text{proxy}] - E_\text{proxy}[n_\text{proxy}], \\
   \Delta E_\text{D} &= E_\text{approx}[n_\text{approx}] - E_\text{approx}[n_\text{proxy}].
\end{align}
Section \ref{sec:exact_v_prox} shows that a plausible alternative, eq \ref{eq:fde_no_prox_en}, which employs the exact barrier height energies, leads to nearly the same mean errors and conclusions.
Section \ref{sec:fde_dde_full} presents the individual functional- and density-driven errors from which Table \ref{tab:bh76_fde_dde_dE} was constructed.
There are some clear outliers, which is normal in density-functional theory, but overall, the individual numbers tell the same story that the mean errors tell: The self-consistent LSDA, GGA, and meta-GGA functionals tend to make large negative functional- and smaller negative density-driven errors, while SCAN@HF over-corrects the electron density, leading to positive density-driven errors that tend to cancel the negative functional-driven errors.

\section{Conclusions}

This work applies the metrics of density-corrected (DC) DFT \cite{kim2013,wasserman2017,sim2018} and quantitative measures of charge-transfer errors \cite{dasgupta2022} to establish why DFA@HF often produces accurate energies for systems where self-consistent DFAs fail badly.
A standard set of reaction barrier heights (BHs), BH76, with their corresponding reaction energies, BH76RC, was selected to compute these metrics \cite{goerigk2017}.
Accurate geometries and reference energies are publicly available for these sets.
Moreover, the BH76 BHs consider transition states with stretched radical bonds, whereas the BH76RC reactions require only the more-compact reactant and product states.

For the BH76 barrier heights, self-consistent density functional approximations improve in mean absolute error (Table \ref{tab:ak_pyscf_aug-cc-pvqz}) from LSDA to PBE to SCAN or \rrscan{}.
The mean functional-driven and density-driven errors (Table \ref{tab:bh76_fde_dde_dE}) improve in the same sequence.

While it is often presumed that HF yields more accurate densities for these abnormal systems, we demonstrated that DFA@HF works for barrier heights because HF makes a charge transfer error of the opposite sign as most semi-local DFAs.
The self-consistent DFA density makes a delocalizing charge transfer error, and HF produces a localizing charge transfer error.
Thus DFA@HF benefits from a cancellation of functional- and density-driven errors via evaluation on a too-local density.

The metrics of DC-DFT require non-self-consistent evaluation of a DFA on the exact density.
As we generally cannot efficiently and reliably invert a high-level wavefunction density to yield a local Kohn-Sham potential, we established three proxies for the exact functional.
A good proxy should yield low errors for the test set in consideration, and should either (a), very nearly  satisfy the Perdew-Parr-Levy-Balduz theorem \cite{perdew1982} on the exchange correlation energy, such as SCAN-FLOSIC and the range-separated hybrid \lcwpbe{} \cite{vydrov2006}; or (b), yield low charge transfer errors for the test set, such as the SCAN global hybrid with 50\% exact exchange admixture (called SX-0.5 here, sometimes called SCAN50).
Note that the optimal fraction of exact exchange admixture will vary with the test set and property.
Accurate prediction of the energies at hand (here, reaction barrier heights) is a necessary, but insufficient condition for selecting a proxy for the exact functional.
A good proxy must also yield small density-driven errors, quantified by its adherence to the PPLB condition or its demonstrated lack of charge transfer errors for the systems at hand.

Careful inspection of SI Tables \ref{tab:BH76_errors_LCwPBE}, \ref{tab:BH76_errors_S50X}, and \ref{tab:BH76_errors_SCAN-FLOSIC} will reveal that both SX-0.5 and SCAN-FLOSIC make substantially larger errors for the first twelve reaction barriers than the last sixty-four.
For both the first twelve and last sixty-four reactions, \lcwpbe{} makes a 1.87 kcal/mol MAD.
SX-0.5 (SCAN-FLOSIC) makes a 4.08 (4.99) kcal/mol MAD for the first twelve, and 2.43 (3.56) kcal/mol MAD for the last sixty-four.
We therefore view \lcwpbe{} as the most reliable proxy for this set, SX-0.5 as the second-most reliable, and SCAN-FLOSIC as the third-most reliable proxy.

The energy metrics used here, motivated by the formalism of DC-DFT, suggest, by the weight of evidence, that SCAN@HF works by a strong cancellation of functional- and generalized density-driven errors.
Most self-consistent semi-local DFAs do not show similar cancellation of these errors.
We note no strong differences between PBE and BLYP for the BH76 set in terms of reaction BHs, density-, and functional-driven errors.
These conclusions hold using any of the proxies selected here.

We also evaluated the density sensitivity \cite{sim2018}, a metric used to determine when a density correction is needed.
While the 25\% SCAN global hybrid (SX-0.25, or more commonly, SCAN0) and \lcwpbe{} show very low sensitivity for the BH76 BHs, all semi-local DFAs (including M06-L and MN15-L) and B3LYP show high density sensitivity.
The relatively high density sensitivity indicates that a self-consistent density is unlikely to be reliable.
The BH76RC reaction energies show low density sensitivity for all considered DFAs, indicating that the errors for this set are primarily functional-driven.

Our error-cancellation explanation for the success of HF-DFT for barrier heights is not intended to apply to all the situations in which HF-DFT works.
For the binding energy curves of heteronuclear molecules \cite{nam2020}, HF-DFT works because it eliminates a large density-driven error.
In the dissociation limit, HF correctly yields separate atoms of integer charge, while semi-local functionals do not \cite{perdew1982}.
HF-DFT also works for the electron affinities of atoms (in the complete-basis-set limit) by eliminating a density-driven error.

We know that the SCAN-like functionals make significantly negative functional-driven errors for stretched radical bonds (as shown for stretched H$_2^+$ in ref \citenum{shahi2019} and for transition states here), so the error cancellation we have found here in SCAN@HF for barrier heights seems reasonable in retrospect.
We do not yet know if the SCAN-like functionals make large functional-driven errors in other realistic situations.

Density-corrected DFT (also known as DC-DFT, HF-DFT or DFA@HF) provides a useful correction to an energy difference predicted by an approximate density functional.
The non-local functionals that we have used as proxies for the exact functional are not intended to replace or compete with DC-DFT, even in the restricted realm of barrier heights, and probably could not because of the outliers shown in the SI.
These proxies serve a different purpose, suggesting by the weight of the evidence (but not proving) that DC-DFT for barrier heights works, not by correcting the density but by over-correcting it, leading to a cancellation of errors.
The reliability of this error cancellation remains to be explained.
Table \ref{tab:bh76_fde_dde_dE} shows that, for SCAN@HF (a success story for DC-DFT), the mean functional-driven error of the BH76 barrier heights is of order -6.1 to -7.5 kcal/mol, and the mean density driven error is of order +3.8 to +5.2 kcal/mol.

We are now working \cite{shahi2022} on accurate but demanding Kohn-Sham inversions of the CCSD(T) densities to better resolve functional- and density-driven errors for a few typical systems.

\begin{acknowledgements}
CS and JPP were supported by the U.S. Department of Energy, Office of Science, Office of Basic Energy Sciences, as part of the Computational Chemical Sciences Program, under Award No. DE-SC0018331.
JPP, RKS, and PB were supported by National Science Foundation grant DMR-1939528.
ADK thanks Temple University for a presidential fellowship.
We thank Aron Cohen and Kieron Burke for brief discussions of density corrections.
We thank Kieron Burke for reading our manuscript and for suggesting consistency tests for our proxy functionals like those in Sec. S1 of the Supporting Information.
Kieron Burke and the referees made other
helpful suggestions.
\end{acknowledgements}

\bibliographystyle{apsrev4-2}
\bibliography{bh_dc}

\begin{thebibliography}{85}%
\makeatletter
\providecommand \@ifxundefined [1]{%
 \@ifx{#1\undefined}
}%
\providecommand \@ifnum [1]{%
 \ifnum #1\expandafter \@firstoftwo
 \else \expandafter \@secondoftwo
 \fi
}%
\providecommand \@ifx [1]{%
 \ifx #1\expandafter \@firstoftwo
 \else \expandafter \@secondoftwo
 \fi
}%
\providecommand \natexlab [1]{#1}%
\providecommand \enquote  [1]{``#1''}%
\providecommand \bibnamefont  [1]{#1}%
\providecommand \bibfnamefont [1]{#1}%
\providecommand \citenamefont [1]{#1}%
\providecommand \href@noop [0]{\@secondoftwo}%
\providecommand \href [0]{\begingroup \@sanitize@url \@href}%
\providecommand \@href[1]{\@@startlink{#1}\@@href}%
\providecommand \@@href[1]{\endgroup#1\@@endlink}%
\providecommand \@sanitize@url [0]{\catcode `\\12\catcode `\$12\catcode
  `\&12\catcode `\#12\catcode `\^12\catcode `\_12\catcode `\%12\relax}%
\providecommand \@@startlink[1]{}%
\providecommand \@@endlink[0]{}%
\providecommand \url  [0]{\begingroup\@sanitize@url \@url }%
\providecommand \@url [1]{\endgroup\@href {#1}{\urlprefix }}%
\providecommand \urlprefix  [0]{URL }%
\providecommand \Eprint [0]{\href }%
\providecommand \doibase [0]{https://doi.org/}%
\providecommand \selectlanguage [0]{\@gobble}%
\providecommand \bibinfo  [0]{\@secondoftwo}%
\providecommand \bibfield  [0]{\@secondoftwo}%
\providecommand \translation [1]{[#1]}%
\providecommand \BibitemOpen [0]{}%
\providecommand \bibitemStop [0]{}%
\providecommand \bibitemNoStop [0]{.\EOS\space}%
\providecommand \EOS [0]{\spacefactor3000\relax}%
\providecommand \BibitemShut  [1]{\csname bibitem#1\endcsname}%
\let\auto@bib@innerbib\@empty
%</preamble>
\bibitem [{\citenamefont {Kohn}\ and\ \citenamefont {Sham}(1965)}]{kohn1965}%
  \BibitemOpen
  \bibfield  {author} {\bibinfo {author} {\bibfnamefont {W.}~\bibnamefont
  {Kohn}}\ and\ \bibinfo {author} {\bibfnamefont {L.~J.}\ \bibnamefont
  {Sham}},\ }\href {https://doi.org/10.1103/PhysRev.140.A1133} {\bibfield
  {journal} {\bibinfo  {journal} {Phys. Rev.}\ }\textbf {\bibinfo {volume}
  {140}},\ \bibinfo {pages} {A1133} (\bibinfo {year} {1965})}\BibitemShut
  {NoStop}%
\bibitem [{\citenamefont {Perdew}\ and\ \citenamefont
  {Wang}(1992)}]{perdew1992}%
  \BibitemOpen
  \bibfield  {author} {\bibinfo {author} {\bibfnamefont {J.~P.}\ \bibnamefont
  {Perdew}}\ and\ \bibinfo {author} {\bibfnamefont {Y.}~\bibnamefont {Wang}},\
  }\href {https://doi.org/10.1103/PhysRevB.45.13244} {\bibfield  {journal}
  {\bibinfo  {journal} {Phys. Rev. B}\ }\textbf {\bibinfo {volume} {45}},\
  \bibinfo {pages} {13244} (\bibinfo {year} {1992})}\BibitemShut {NoStop}%
\bibitem [{\citenamefont {Perdew}\ \emph
  {et~al.}(1996{\natexlab{a}})\citenamefont {Perdew}, \citenamefont {Burke},\
  and\ \citenamefont {Wang}}]{perdew1996a}%
  \BibitemOpen
  \bibfield  {author} {\bibinfo {author} {\bibfnamefont {J.~P.}\ \bibnamefont
  {Perdew}}, \bibinfo {author} {\bibfnamefont {K.}~\bibnamefont {Burke}},\ and\
  \bibinfo {author} {\bibfnamefont {Y.}~\bibnamefont {Wang}},\ }\href
  {https://doi.org/10.1103/PhysRevB.54.16533} {\bibfield  {journal} {\bibinfo
  {journal} {Phys. Rev. B}\ }\textbf {\bibinfo {volume} {54}},\ \bibinfo
  {pages} {16533} (\bibinfo {year} {1996}{\natexlab{a}})}\BibitemShut {NoStop}%
\bibitem [{\citenamefont {Becke}(2003)}]{becke2003}%
  \BibitemOpen
  \bibfield  {author} {\bibinfo {author} {\bibfnamefont {A.~D.}\ \bibnamefont
  {Becke}},\ }\href {https://doi.org/10.1063/1.1589733} {\bibfield  {journal}
  {\bibinfo  {journal} {J. Chem. Phys.}\ }\textbf {\bibinfo {volume} {119}},\
  \bibinfo {pages} {2972} (\bibinfo {year} {2003})}\BibitemShut {NoStop}%
\bibitem [{\citenamefont {Dasgupta}\ \emph {et~al.}(2022)\citenamefont
  {Dasgupta}, \citenamefont {Shahi}, \citenamefont {Bhetwal}, \citenamefont
  {Perdew},\ and\ \citenamefont {Paesani}}]{dasgupta2022}%
  \BibitemOpen
  \bibfield  {author} {\bibinfo {author} {\bibfnamefont {S.}~\bibnamefont
  {Dasgupta}}, \bibinfo {author} {\bibfnamefont {C.}~\bibnamefont {Shahi}},
  \bibinfo {author} {\bibfnamefont {P.}~\bibnamefont {Bhetwal}}, \bibinfo
  {author} {\bibfnamefont {J.~P.}\ \bibnamefont {Perdew}},\ and\ \bibinfo
  {author} {\bibfnamefont {F.}~\bibnamefont {Paesani}},\ }\href
  {https://doi.org/10.1021/acs.jctc.2c00313} {\bibfield  {journal} {\bibinfo
  {journal} {J. Chem. Theory Comput.}\ }\textbf {\bibinfo {volume} {18}},\
  \bibinfo {pages} {4745} (\bibinfo {year} {2022})}\BibitemShut {NoStop}%
\bibitem [{\citenamefont {Perdew}\ \emph {et~al.}(1982)\citenamefont {Perdew},
  \citenamefont {Parr}, \citenamefont {Levy},\ and\ \citenamefont
  {Balduz~Jr.}}]{perdew1982}%
  \BibitemOpen
  \bibfield  {author} {\bibinfo {author} {\bibfnamefont {J.~P.}\ \bibnamefont
  {Perdew}}, \bibinfo {author} {\bibfnamefont {R.~G.}\ \bibnamefont {Parr}},
  \bibinfo {author} {\bibfnamefont {M.}~\bibnamefont {Levy}},\ and\ \bibinfo
  {author} {\bibfnamefont {J.~L.}\ \bibnamefont {Balduz~Jr.}},\ }\href
  {https://doi.org/10.1103/PhysRevLett.49.1691} {\bibfield  {journal} {\bibinfo
   {journal} {Phys. Rev. Lett.}\ }\textbf {\bibinfo {volume} {49}},\ \bibinfo
  {pages} {1691} (\bibinfo {year} {1982})}\BibitemShut {NoStop}%
\bibitem [{\citenamefont {Szabo}\ and\ \citenamefont
  {Ostlund}(1982)}]{szabo1982}%
  \BibitemOpen
  \bibfield  {author} {\bibinfo {author} {\bibfnamefont {A.}~\bibnamefont
  {Szabo}}\ and\ \bibinfo {author} {\bibfnamefont {N.~S.}\ \bibnamefont
  {Ostlund}},\ }\href@noop {} {\emph {\bibinfo {title} {Modern Quantum
  Chemistry}}}\ (\bibinfo  {publisher} {Macmillan},\ \bibinfo {year}
  {1982})\BibitemShut {NoStop}%
\bibitem [{\citenamefont {Cohen}\ \emph
  {et~al.}(2008{\natexlab{a}})\citenamefont {Cohen}, \citenamefont
  {Mori-S\'anchez},\ and\ \citenamefont {Yang}}]{cohen2008a}%
  \BibitemOpen
  \bibfield  {author} {\bibinfo {author} {\bibfnamefont {A.~J.}\ \bibnamefont
  {Cohen}}, \bibinfo {author} {\bibfnamefont {P.}~\bibnamefont
  {Mori-S\'anchez}},\ and\ \bibinfo {author} {\bibfnamefont {W.}~\bibnamefont
  {Yang}},\ }\href {https://doi.org/10.1103/PhysRevB.77.115123} {\bibfield
  {journal} {\bibinfo  {journal} {Phys. Rev. B}\ }\textbf {\bibinfo {volume}
  {77}},\ \bibinfo {pages} {115123} (\bibinfo {year}
  {2008}{\natexlab{a}})}\BibitemShut {NoStop}%
\bibitem [{\citenamefont {Vydrov}\ \emph {et~al.}(2007)\citenamefont {Vydrov},
  \citenamefont {Scuseria},\ and\ \citenamefont {Perdew}}]{vydrov2007}%
  \BibitemOpen
  \bibfield  {author} {\bibinfo {author} {\bibfnamefont {O.~A.}\ \bibnamefont
  {Vydrov}}, \bibinfo {author} {\bibfnamefont {G.~E.}\ \bibnamefont
  {Scuseria}},\ and\ \bibinfo {author} {\bibfnamefont {J.~P.}\ \bibnamefont
  {Perdew}},\ }\href {https://doi.org/10.1063/1.2723119} {\bibfield  {journal}
  {\bibinfo  {journal} {J. Chem. Phys.}\ }\textbf {\bibinfo {volume} {126}},\
  \bibinfo {pages} {154109} (\bibinfo {year} {2007})}\BibitemShut {NoStop}%
\bibitem [{\citenamefont {Perdew}\ and\ \citenamefont
  {Zunger}(1981)}]{perdew1981}%
  \BibitemOpen
  \bibfield  {author} {\bibinfo {author} {\bibfnamefont {J.~P.}\ \bibnamefont
  {Perdew}}\ and\ \bibinfo {author} {\bibfnamefont {A.}~\bibnamefont
  {Zunger}},\ }\href {https://doi.org/https://doi.org/10.1103/PhysRevB.23.5048}
  {\bibfield  {journal} {\bibinfo  {journal} {Phys. Rev. B}\ }\textbf {\bibinfo
  {volume} {23}},\ \bibinfo {pages} {5048} (\bibinfo {year}
  {1981})}\BibitemShut {NoStop}%
\bibitem [{\citenamefont {Pederson}\ \emph {et~al.}(2014)\citenamefont
  {Pederson}, \citenamefont {Ruzsinszky},\ and\ \citenamefont
  {Perdew}}]{pederson2014}%
  \BibitemOpen
  \bibfield  {author} {\bibinfo {author} {\bibfnamefont {M.~R.}\ \bibnamefont
  {Pederson}}, \bibinfo {author} {\bibfnamefont {A.}~\bibnamefont
  {Ruzsinszky}},\ and\ \bibinfo {author} {\bibfnamefont {J.~P.}\ \bibnamefont
  {Perdew}},\ }\href {https://doi.org/10.1063/1.4869581} {\bibfield  {journal}
  {\bibinfo  {journal} {J. Chem. Phys.}\ }\textbf {\bibinfo {volume} {140}},\
  \bibinfo {pages} {121103} (\bibinfo {year} {2014})}\BibitemShut {NoStop}%
\bibitem [{\citenamefont {Shahi}\ \emph {et~al.}(2019)\citenamefont {Shahi},
  \citenamefont {Bhattarai}, \citenamefont {Wagle}, \citenamefont {Santra},
  \citenamefont {Schwalbe}, \citenamefont {Hahn}, \citenamefont {Kortus},
  \citenamefont {Jackson}, \citenamefont {Peralta}, \citenamefont {Trepte},
  \citenamefont {Lehtola}, \citenamefont {Nepal}, \citenamefont {Myneni},
  \citenamefont {Neupane}, \citenamefont {Adhikari}, \citenamefont
  {Ruzsinszky}, \citenamefont {Yamamoto}, \citenamefont {Baruah}, \citenamefont
  {Zope},\ and\ \citenamefont {Perdew}}]{shahi2019}%
  \BibitemOpen
  \bibfield  {author} {\bibinfo {author} {\bibfnamefont {C.}~\bibnamefont
  {Shahi}}, \bibinfo {author} {\bibfnamefont {P.}~\bibnamefont {Bhattarai}},
  \bibinfo {author} {\bibfnamefont {K.}~\bibnamefont {Wagle}}, \bibinfo
  {author} {\bibfnamefont {B.}~\bibnamefont {Santra}}, \bibinfo {author}
  {\bibfnamefont {S.}~\bibnamefont {Schwalbe}}, \bibinfo {author}
  {\bibfnamefont {T.}~\bibnamefont {Hahn}}, \bibinfo {author} {\bibfnamefont
  {J.}~\bibnamefont {Kortus}}, \bibinfo {author} {\bibfnamefont {K.~A.}\
  \bibnamefont {Jackson}}, \bibinfo {author} {\bibfnamefont {J.~E.}\
  \bibnamefont {Peralta}}, \bibinfo {author} {\bibfnamefont {K.}~\bibnamefont
  {Trepte}}, \bibinfo {author} {\bibfnamefont {S.}~\bibnamefont {Lehtola}},
  \bibinfo {author} {\bibfnamefont {N.~K.}\ \bibnamefont {Nepal}}, \bibinfo
  {author} {\bibfnamefont {H.}~\bibnamefont {Myneni}}, \bibinfo {author}
  {\bibfnamefont {B.}~\bibnamefont {Neupane}}, \bibinfo {author} {\bibfnamefont
  {S.}~\bibnamefont {Adhikari}}, \bibinfo {author} {\bibfnamefont
  {A.}~\bibnamefont {Ruzsinszky}}, \bibinfo {author} {\bibfnamefont
  {Y.}~\bibnamefont {Yamamoto}}, \bibinfo {author} {\bibfnamefont
  {T.}~\bibnamefont {Baruah}}, \bibinfo {author} {\bibfnamefont {R.~R.}\
  \bibnamefont {Zope}},\ and\ \bibinfo {author} {\bibfnamefont {J.~P.}\
  \bibnamefont {Perdew}},\ }\href {https://doi.org/10.1063/1.5087065}
  {\bibfield  {journal} {\bibinfo  {journal} {J. Chem. Phys.}\ }\textbf
  {\bibinfo {volume} {150}},\ \bibinfo {pages} {174102} (\bibinfo {year}
  {2019})}\BibitemShut {NoStop}%
\bibitem [{\citenamefont {Withanage}\ \emph {et~al.}(2022)\citenamefont
  {Withanage}, \citenamefont {Jackson},\ and\ \citenamefont
  {Pederson}}]{withanage2022}%
  \BibitemOpen
  \bibfield  {author} {\bibinfo {author} {\bibfnamefont {K.~P.~K.}\
  \bibnamefont {Withanage}}, \bibinfo {author} {\bibfnamefont {K.~A.}\
  \bibnamefont {Jackson}},\ and\ \bibinfo {author} {\bibfnamefont {M.~R.}\
  \bibnamefont {Pederson}},\ }\href {https://doi.org/10.1063/5.0091212}
  {\bibfield  {journal} {\bibinfo  {journal} {J. Chem. Phys.}\ }\textbf
  {\bibinfo {volume} {156}},\ \bibinfo {pages} {231103} (\bibinfo {year}
  {2022})}\BibitemShut {NoStop}%
\bibitem [{\citenamefont {Lehtola}\ \emph {et~al.}(2016)\citenamefont
  {Lehtola}, \citenamefont {Head-Gordon},\ and\ \citenamefont
  {Jónsson}}]{lehtola2016}%
  \BibitemOpen
  \bibfield  {author} {\bibinfo {author} {\bibfnamefont {S.}~\bibnamefont
  {Lehtola}}, \bibinfo {author} {\bibfnamefont {M.}~\bibnamefont
  {Head-Gordon}},\ and\ \bibinfo {author} {\bibfnamefont {H.}~\bibnamefont
  {Jónsson}},\ }\href {https://doi.org/10.1021/acs.jctc.6b00347} {\bibfield
  {journal} {\bibinfo  {journal} {J. Chem. Theory Comput.}\ }\textbf {\bibinfo
  {volume} {12}},\ \bibinfo {pages} {3195} (\bibinfo {year}
  {2016})}\BibitemShut {NoStop}%
\bibitem [{\citenamefont {Zope}\ \emph {et~al.}(2019)\citenamefont {Zope},
  \citenamefont {Yamamoto}, \citenamefont {Diaz}, \citenamefont {Baruah},
  \citenamefont {Peralta}, \citenamefont {Jackson}, \citenamefont {Santra},\
  and\ \citenamefont {Perdew}}]{zope2019}%
  \BibitemOpen
  \bibfield  {author} {\bibinfo {author} {\bibfnamefont {R.~R.}\ \bibnamefont
  {Zope}}, \bibinfo {author} {\bibfnamefont {Y.}~\bibnamefont {Yamamoto}},
  \bibinfo {author} {\bibfnamefont {C.~M.}\ \bibnamefont {Diaz}}, \bibinfo
  {author} {\bibfnamefont {T.}~\bibnamefont {Baruah}}, \bibinfo {author}
  {\bibfnamefont {J.~E.}\ \bibnamefont {Peralta}}, \bibinfo {author}
  {\bibfnamefont {K.~A.}\ \bibnamefont {Jackson}}, \bibinfo {author}
  {\bibfnamefont {B.}~\bibnamefont {Santra}},\ and\ \bibinfo {author}
  {\bibfnamefont {J.~P.}\ \bibnamefont {Perdew}},\ }\href
  {https://doi.org/10.1063/1.5129533} {\bibfield  {journal} {\bibinfo
  {journal} {J. Chem. Phys.}\ }\textbf {\bibinfo {volume} {151}},\ \bibinfo
  {pages} {214108} (\bibinfo {year} {2019})}\BibitemShut {NoStop}%
\bibitem [{\citenamefont {Bhattarai}\ \emph {et~al.}(2020)\citenamefont
  {Bhattarai}, \citenamefont {Wagle}, \citenamefont {Shahi}, \citenamefont
  {Yamamoto}, \citenamefont {Romero}, \citenamefont {Santra}, \citenamefont
  {Zope}, \citenamefont {Jackson}, \citenamefont {Peralta},\ and\ \citenamefont
  {Perdew}}]{bhattarai2020}%
  \BibitemOpen
  \bibfield  {author} {\bibinfo {author} {\bibfnamefont {P.}~\bibnamefont
  {Bhattarai}}, \bibinfo {author} {\bibfnamefont {K.}~\bibnamefont {Wagle}},
  \bibinfo {author} {\bibfnamefont {C.}~\bibnamefont {Shahi}}, \bibinfo
  {author} {\bibfnamefont {Y.}~\bibnamefont {Yamamoto}}, \bibinfo {author}
  {\bibfnamefont {S.}~\bibnamefont {Romero}}, \bibinfo {author} {\bibfnamefont
  {B.}~\bibnamefont {Santra}}, \bibinfo {author} {\bibfnamefont {R.~R.}\
  \bibnamefont {Zope}}, \bibinfo {author} {\bibfnamefont {K.~A.}\ \bibnamefont
  {Jackson}}, \bibinfo {author} {\bibfnamefont {J.~E.}\ \bibnamefont
  {Peralta}},\ and\ \bibinfo {author} {\bibfnamefont {J.~P.}\ \bibnamefont
  {Perdew}},\ }\href {https://doi.org/10.1063/5.0010375} {\bibfield  {journal}
  {\bibinfo  {journal} {J. Chem. Phys.}\ }\textbf {\bibinfo {volume} {152}},\
  \bibinfo {pages} {214109} (\bibinfo {year} {2020})}\BibitemShut {NoStop}%
\bibitem [{\citenamefont {Tao}\ \emph {et~al.}(2008)\citenamefont {Tao},
  \citenamefont {Staroverov}, \citenamefont {Scuseria},\ and\ \citenamefont
  {Perdew}}]{tao2008}%
  \BibitemOpen
  \bibfield  {author} {\bibinfo {author} {\bibfnamefont {J.}~\bibnamefont
  {Tao}}, \bibinfo {author} {\bibfnamefont {V.~N.}\ \bibnamefont {Staroverov}},
  \bibinfo {author} {\bibfnamefont {G.~E.}\ \bibnamefont {Scuseria}},\ and\
  \bibinfo {author} {\bibfnamefont {J.~P.}\ \bibnamefont {Perdew}},\ }\href
  {https://doi.org/10.1103/PhysRevA.77.012509} {\bibfield  {journal} {\bibinfo
  {journal} {Phys. Rev. A}\ }\textbf {\bibinfo {volume} {77}},\ \bibinfo
  {pages} {012509} (\bibinfo {year} {2008})}\BibitemShut {NoStop}%
\bibitem [{\citenamefont {Cohen}\ \emph
  {et~al.}(2008{\natexlab{b}})\citenamefont {Cohen}, \citenamefont
  {Mori-Sánchez},\ and\ \citenamefont {Yang}}]{cohen2008}%
  \BibitemOpen
  \bibfield  {author} {\bibinfo {author} {\bibfnamefont {A.~J.}\ \bibnamefont
  {Cohen}}, \bibinfo {author} {\bibfnamefont {P.}~\bibnamefont
  {Mori-Sánchez}},\ and\ \bibinfo {author} {\bibfnamefont {W.}~\bibnamefont
  {Yang}},\ }\href {https://doi.org/10.1063/1.2987202} {\bibfield  {journal}
  {\bibinfo  {journal} {J. Chem. Phys.}\ }\textbf {\bibinfo {volume} {129}},\
  \bibinfo {pages} {121104} (\bibinfo {year} {2008}{\natexlab{b}})}\BibitemShut
  {NoStop}%
\bibitem [{\citenamefont {Kirkpatrick}\ \emph {et~al.}(2021)\citenamefont
  {Kirkpatrick}, \citenamefont {McMorrow}, \citenamefont {Turban},
  \citenamefont {Gaunt}, \citenamefont {Spencer}, \citenamefont {Matthews},
  \citenamefont {Obika}, \citenamefont {Thiry}, \citenamefont {Fortunato},
  \citenamefont {Pfau}, \citenamefont {Castellanos}, \citenamefont {Petersen},
  \citenamefont {Nelson}, \citenamefont {Kohli}, \citenamefont {Mori-Sánchez},
  \citenamefont {Hassabis},\ and\ \citenamefont {Cohen}}]{kirkpatrick2021}%
  \BibitemOpen
  \bibfield  {author} {\bibinfo {author} {\bibfnamefont {J.}~\bibnamefont
  {Kirkpatrick}}, \bibinfo {author} {\bibfnamefont {B.}~\bibnamefont
  {McMorrow}}, \bibinfo {author} {\bibfnamefont {D.~H.~P.}\ \bibnamefont
  {Turban}}, \bibinfo {author} {\bibfnamefont {A.~L.}\ \bibnamefont {Gaunt}},
  \bibinfo {author} {\bibfnamefont {J.~S.}\ \bibnamefont {Spencer}}, \bibinfo
  {author} {\bibfnamefont {A.~G. D.~G.}\ \bibnamefont {Matthews}}, \bibinfo
  {author} {\bibfnamefont {A.}~\bibnamefont {Obika}}, \bibinfo {author}
  {\bibfnamefont {L.}~\bibnamefont {Thiry}}, \bibinfo {author} {\bibfnamefont
  {M.}~\bibnamefont {Fortunato}}, \bibinfo {author} {\bibfnamefont
  {D.}~\bibnamefont {Pfau}}, \bibinfo {author} {\bibfnamefont {L.~R.}\
  \bibnamefont {Castellanos}}, \bibinfo {author} {\bibfnamefont
  {S.}~\bibnamefont {Petersen}}, \bibinfo {author} {\bibfnamefont {A.~W.~R.}\
  \bibnamefont {Nelson}}, \bibinfo {author} {\bibfnamefont {P.}~\bibnamefont
  {Kohli}}, \bibinfo {author} {\bibfnamefont {P.}~\bibnamefont
  {Mori-Sánchez}}, \bibinfo {author} {\bibfnamefont {D.}~\bibnamefont
  {Hassabis}},\ and\ \bibinfo {author} {\bibfnamefont {A.~J.}\ \bibnamefont
  {Cohen}},\ }\href {https://doi.org/10.1126/science.abj6511} {\bibfield
  {journal} {\bibinfo  {journal} {Science}\ }\textbf {\bibinfo {volume}
  {374}},\ \bibinfo {pages} {1385} (\bibinfo {year} {2021})}\BibitemShut
  {NoStop}%
\bibitem [{\citenamefont {Savin}(1995)}]{savin1995}%
  \BibitemOpen
  \bibfield  {author} {\bibinfo {author} {\bibfnamefont {A.}~\bibnamefont
  {Savin}},\ }\href
  {http://libproxy.temple.edu/login?url=https://search.ebscohost.com/login.aspx?direct=true&db=e000xna&AN=564471&site=ehost-live&scope=site}
  {\emph {\bibinfo {title} {Recent Advances In Density Functional Methods, Part
  I.}}},\ edited by\ \bibinfo {editor} {\bibfnamefont {C.}~\bibnamefont
  {Delano~Pun}},\ \bibinfo {series} {Recent Advances in Computational
  Chemistry}\ No.\ \bibinfo {number} {vol. 1}\ (\bibinfo  {publisher} {World
  Scientific},\ \bibinfo {year} {1995})\ Chap.~\bibinfo {chapter}
  {4}\BibitemShut {NoStop}%
\bibitem [{\citenamefont {Leininger}\ \emph {et~al.}(1997)\citenamefont
  {Leininger}, \citenamefont {Stoll}, \citenamefont {Werner},\ and\
  \citenamefont {Savin}}]{leininger1997}%
  \BibitemOpen
  \bibfield  {author} {\bibinfo {author} {\bibfnamefont {T.}~\bibnamefont
  {Leininger}}, \bibinfo {author} {\bibfnamefont {H.}~\bibnamefont {Stoll}},
  \bibinfo {author} {\bibfnamefont {H.-J.}\ \bibnamefont {Werner}},\ and\
  \bibinfo {author} {\bibfnamefont {A.}~\bibnamefont {Savin}},\ }\href
  {https://doi.org/https://doi.org/10.1016/S0009-2614(97)00758-6} {\bibfield
  {journal} {\bibinfo  {journal} {Chem. Phys. Lett.}\ }\textbf {\bibinfo
  {volume} {275}},\ \bibinfo {pages} {151} (\bibinfo {year}
  {1997})}\BibitemShut {NoStop}%
\bibitem [{\citenamefont {Burke}\ \emph {et~al.}(1998)\citenamefont {Burke},
  \citenamefont {Perdew},\ and\ \citenamefont {Ernzerhof}}]{burke1998}%
  \BibitemOpen
  \bibfield  {author} {\bibinfo {author} {\bibfnamefont {K.}~\bibnamefont
  {Burke}}, \bibinfo {author} {\bibfnamefont {J.~P.}\ \bibnamefont {Perdew}},\
  and\ \bibinfo {author} {\bibfnamefont {M.}~\bibnamefont {Ernzerhof}},\ }\href
  {https://doi.org/10.1063/1.476976} {\bibfield  {journal} {\bibinfo  {journal}
  {J. Chem. Phys.}\ }\textbf {\bibinfo {volume} {109}},\ \bibinfo {pages}
  {3760} (\bibinfo {year} {1998})}\BibitemShut {NoStop}%
\bibitem [{\citenamefont {Iikura}\ \emph {et~al.}(2001)\citenamefont {Iikura},
  \citenamefont {Tsuneda}, \citenamefont {Yanai},\ and\ \citenamefont
  {Hirao}}]{hisayoshi2001}%
  \BibitemOpen
  \bibfield  {author} {\bibinfo {author} {\bibfnamefont {H.}~\bibnamefont
  {Iikura}}, \bibinfo {author} {\bibfnamefont {T.}~\bibnamefont {Tsuneda}},
  \bibinfo {author} {\bibfnamefont {T.}~\bibnamefont {Yanai}},\ and\ \bibinfo
  {author} {\bibfnamefont {K.}~\bibnamefont {Hirao}},\ }\href
  {https://doi.org/10.1063/1.1383587} {\bibfield  {journal} {\bibinfo
  {journal} {J. Chem. Phys.}\ }\textbf {\bibinfo {volume} {115}},\ \bibinfo
  {pages} {3540} (\bibinfo {year} {2001})}\BibitemShut {NoStop}%
\bibitem [{\citenamefont {Adamson}\ \emph {et~al.}(1999)\citenamefont
  {Adamson}, \citenamefont {Dombroski},\ and\ \citenamefont
  {Gill}}]{adamson1999}%
  \BibitemOpen
  \bibfield  {author} {\bibinfo {author} {\bibfnamefont {R.~D.}\ \bibnamefont
  {Adamson}}, \bibinfo {author} {\bibfnamefont {J.~P.}\ \bibnamefont
  {Dombroski}},\ and\ \bibinfo {author} {\bibfnamefont {P.~M.~W.}\ \bibnamefont
  {Gill}},\ }\href
  {https://doi.org/https://doi-org.libproxy.temple.edu/10.1002/(SICI)1096-987X(19990715)20:9<921::AID-JCC3>3.0.CO;2-K}
  {\bibfield  {journal} {\bibinfo  {journal} {J. Comput. Chem.}\ }\textbf
  {\bibinfo {volume} {20}},\ \bibinfo {pages} {921} (\bibinfo {year}
  {1999})}\BibitemShut {NoStop}%
\bibitem [{\citenamefont {Henderson}\ \emph {et~al.}(2009)\citenamefont
  {Henderson}, \citenamefont {Izmaylov}, \citenamefont {Scalmani},\ and\
  \citenamefont {Scuseria}}]{henderson2009}%
  \BibitemOpen
  \bibfield  {author} {\bibinfo {author} {\bibfnamefont {T.~M.}\ \bibnamefont
  {Henderson}}, \bibinfo {author} {\bibfnamefont {A.~F.}\ \bibnamefont
  {Izmaylov}}, \bibinfo {author} {\bibfnamefont {G.}~\bibnamefont {Scalmani}},\
  and\ \bibinfo {author} {\bibfnamefont {G.~E.}\ \bibnamefont {Scuseria}},\
  }\href {https://doi.org/10.1063/1.3185673} {\bibfield  {journal} {\bibinfo
  {journal} {J. Chem. Phys.}\ }\textbf {\bibinfo {volume} {131}},\ \bibinfo
  {pages} {044108} (\bibinfo {year} {2009})}\BibitemShut {NoStop}%
\bibitem [{\citenamefont {Li}\ \emph {et~al.}(2017)\citenamefont {Li},
  \citenamefont {Zheng}, \citenamefont {Su},\ and\ \citenamefont
  {Yang}}]{li2017}%
  \BibitemOpen
  \bibfield  {author} {\bibinfo {author} {\bibfnamefont {C.}~\bibnamefont
  {Li}}, \bibinfo {author} {\bibfnamefont {X.}~\bibnamefont {Zheng}}, \bibinfo
  {author} {\bibfnamefont {N.~Q.}\ \bibnamefont {Su}},\ and\ \bibinfo {author}
  {\bibfnamefont {W.}~\bibnamefont {Yang}},\ }\href
  {https://doi.org/10.1093/nsr/nwx111} {\bibfield  {journal} {\bibinfo
  {journal} {National Science Review}\ }\textbf {\bibinfo {volume} {5}},\
  \bibinfo {pages} {203} (\bibinfo {year} {2017})}\BibitemShut {NoStop}%
\bibitem [{\citenamefont {Ruzsinszky}\ \emph {et~al.}(2007)\citenamefont
  {Ruzsinszky}, \citenamefont {Perdew}, \citenamefont {Csonka}, \citenamefont
  {Vydrov},\ and\ \citenamefont {Scuseria}}]{ruzsinszky2007}%
  \BibitemOpen
  \bibfield  {author} {\bibinfo {author} {\bibfnamefont {A.}~\bibnamefont
  {Ruzsinszky}}, \bibinfo {author} {\bibfnamefont {J.~P.}\ \bibnamefont
  {Perdew}}, \bibinfo {author} {\bibfnamefont {G.~I.}\ \bibnamefont {Csonka}},
  \bibinfo {author} {\bibfnamefont {O.~A.}\ \bibnamefont {Vydrov}},\ and\
  \bibinfo {author} {\bibfnamefont {G.~E.}\ \bibnamefont {Scuseria}},\ }\href
  {https://doi.org/10.1063/1.2566637} {\bibfield  {journal} {\bibinfo
  {journal} {J. Chem. Phys.}\ }\textbf {\bibinfo {volume} {126}},\ \bibinfo
  {pages} {104102} (\bibinfo {year} {2007})}\BibitemShut {NoStop}%
\bibitem [{\citenamefont {Vydrov}\ and\ \citenamefont
  {Scuseria}(2006)}]{vydrov2006}%
  \BibitemOpen
  \bibfield  {author} {\bibinfo {author} {\bibfnamefont {O.~A.}\ \bibnamefont
  {Vydrov}}\ and\ \bibinfo {author} {\bibfnamefont {G.~E.}\ \bibnamefont
  {Scuseria}},\ }\href {https://doi.org/10.1063/1.2409292} {\bibfield
  {journal} {\bibinfo  {journal} {J. Chem. Phys.}\ }\textbf {\bibinfo {volume}
  {125}},\ \bibinfo {pages} {234109} (\bibinfo {year} {2006})}\BibitemShut
  {NoStop}%
\bibitem [{\citenamefont {Gerber}\ \emph {et~al.}(2007)\citenamefont {Gerber},
  \citenamefont {Ángyán}, \citenamefont {Marsman},\ and\ \citenamefont
  {Kresse}}]{gerber2007}%
  \BibitemOpen
  \bibfield  {author} {\bibinfo {author} {\bibfnamefont {I.~C.}\ \bibnamefont
  {Gerber}}, \bibinfo {author} {\bibfnamefont {J.~G.}\ \bibnamefont
  {Ángyán}}, \bibinfo {author} {\bibfnamefont {M.}~\bibnamefont {Marsman}},\
  and\ \bibinfo {author} {\bibfnamefont {G.}~\bibnamefont {Kresse}},\ }\href
  {https://doi.org/10.1063/1.2759209} {\bibfield  {journal} {\bibinfo
  {journal} {J. Chem. Phys.}\ }\textbf {\bibinfo {volume} {127}},\ \bibinfo
  {pages} {054101} (\bibinfo {year} {2007})}\BibitemShut {NoStop}%
\bibitem [{\citenamefont {Kim}\ \emph {et~al.}(2013)\citenamefont {Kim},
  \citenamefont {Sim},\ and\ \citenamefont {Burke}}]{kim2013}%
  \BibitemOpen
  \bibfield  {author} {\bibinfo {author} {\bibfnamefont {M.-C.}\ \bibnamefont
  {Kim}}, \bibinfo {author} {\bibfnamefont {E.}~\bibnamefont {Sim}},\ and\
  \bibinfo {author} {\bibfnamefont {K.}~\bibnamefont {Burke}},\ }\href
  {https://doi.org/10.1103/PhysRevLett.111.073003} {\bibfield  {journal}
  {\bibinfo  {journal} {Phys. Rev. Lett.}\ }\textbf {\bibinfo {volume} {111}},\
  \bibinfo {pages} {073003} (\bibinfo {year} {2013})}\BibitemShut {NoStop}%
\bibitem [{\citenamefont {Wasserman}\ \emph {et~al.}(2017)\citenamefont
  {Wasserman}, \citenamefont {Nafziger}, \citenamefont {Jiang}, \citenamefont
  {Kim}, \citenamefont {Sim},\ and\ \citenamefont {Burke}}]{wasserman2017}%
  \BibitemOpen
  \bibfield  {author} {\bibinfo {author} {\bibfnamefont {A.}~\bibnamefont
  {Wasserman}}, \bibinfo {author} {\bibfnamefont {J.}~\bibnamefont {Nafziger}},
  \bibinfo {author} {\bibfnamefont {K.}~\bibnamefont {Jiang}}, \bibinfo
  {author} {\bibfnamefont {M.-C.}\ \bibnamefont {Kim}}, \bibinfo {author}
  {\bibfnamefont {E.}~\bibnamefont {Sim}},\ and\ \bibinfo {author}
  {\bibfnamefont {K.}~\bibnamefont {Burke}},\ }\href
  {https://doi.org/10.1146/annurev-physchem-052516-044957} {\bibfield
  {journal} {\bibinfo  {journal} {Annu. Rev. Phys. Chem.}\ }\textbf {\bibinfo
  {volume} {68}},\ \bibinfo {pages} {555} (\bibinfo {year} {2017})},\ \bibinfo
  {note} {pMID: 28463652}\BibitemShut {NoStop}%
\bibitem [{\citenamefont {Sim}\ \emph {et~al.}(2018)\citenamefont {Sim},
  \citenamefont {Song},\ and\ \citenamefont {Burke}}]{sim2018}%
  \BibitemOpen
  \bibfield  {author} {\bibinfo {author} {\bibfnamefont {E.}~\bibnamefont
  {Sim}}, \bibinfo {author} {\bibfnamefont {S.}~\bibnamefont {Song}},\ and\
  \bibinfo {author} {\bibfnamefont {K.}~\bibnamefont {Burke}},\ }\href
  {https://doi.org/10.1021/acs.jpclett.8b02855} {\bibfield  {journal} {\bibinfo
   {journal} {J. Phys. Chem. Lett.}\ }\textbf {\bibinfo {volume} {9}},\
  \bibinfo {pages} {6385} (\bibinfo {year} {2018})}\BibitemShut {NoStop}%
\bibitem [{\citenamefont {Vuckovic}\ \emph {et~al.}(2019)\citenamefont
  {Vuckovic}, \citenamefont {Song}, \citenamefont {Kozlowski}, \citenamefont
  {Sim},\ and\ \citenamefont {Burke}}]{vuckovic2019}%
  \BibitemOpen
  \bibfield  {author} {\bibinfo {author} {\bibfnamefont {S.}~\bibnamefont
  {Vuckovic}}, \bibinfo {author} {\bibfnamefont {S.}~\bibnamefont {Song}},
  \bibinfo {author} {\bibfnamefont {J.}~\bibnamefont {Kozlowski}}, \bibinfo
  {author} {\bibfnamefont {E.}~\bibnamefont {Sim}},\ and\ \bibinfo {author}
  {\bibfnamefont {K.}~\bibnamefont {Burke}},\ }\href
  {https://doi.org/10.1021/acs.jctc.9b00826} {\bibfield  {journal} {\bibinfo
  {journal} {J. Chem. Theory Computat.}\ }\textbf {\bibinfo {volume} {15}},\
  \bibinfo {pages} {6636} (\bibinfo {year} {2019})}\BibitemShut {NoStop}%
\bibitem [{\citenamefont {Nam}\ \emph {et~al.}(2020)\citenamefont {Nam},
  \citenamefont {Song}, \citenamefont {Sim},\ and\ \citenamefont
  {Burke}}]{nam2020}%
  \BibitemOpen
  \bibfield  {author} {\bibinfo {author} {\bibfnamefont {S.}~\bibnamefont
  {Nam}}, \bibinfo {author} {\bibfnamefont {S.}~\bibnamefont {Song}}, \bibinfo
  {author} {\bibfnamefont {E.}~\bibnamefont {Sim}},\ and\ \bibinfo {author}
  {\bibfnamefont {K.}~\bibnamefont {Burke}},\ }\href
  {https://doi.org/10.1021/acs.jctc.0c00391} {\bibfield  {journal} {\bibinfo
  {journal} {J. Chem. Theory Comput.}\ }\textbf {\bibinfo {volume} {16}},\
  \bibinfo {pages} {5014} (\bibinfo {year} {2020})},\ \bibinfo {note} {pMID:
  32667787}\BibitemShut {NoStop}%
\bibitem [{\citenamefont {Shi}\ and\ \citenamefont
  {Wasserman}(2021)}]{shi2021}%
  \BibitemOpen
  \bibfield  {author} {\bibinfo {author} {\bibfnamefont {Y.}~\bibnamefont
  {Shi}}\ and\ \bibinfo {author} {\bibfnamefont {A.}~\bibnamefont
  {Wasserman}},\ }\href {https://doi.org/10.1021/acs.jpclett.1c00752}
  {\bibfield  {journal} {\bibinfo  {journal} {J. Phys. Chem. Lett.}\ }\textbf
  {\bibinfo {volume} {12}},\ \bibinfo {pages} {5308} (\bibinfo {year}
  {2021})},\ \bibinfo {note} {pMID: 34061541}\BibitemShut {NoStop}%
\bibitem [{\citenamefont {Schipper}\ \emph {et~al.}(1998)\citenamefont
  {Schipper}, \citenamefont {Gritsenko},\ and\ \citenamefont
  {Baerends}}]{schipper1998}%
  \BibitemOpen
  \bibfield  {author} {\bibinfo {author} {\bibfnamefont {P.~R.~T.}\
  \bibnamefont {Schipper}}, \bibinfo {author} {\bibfnamefont {O.~V.}\
  \bibnamefont {Gritsenko}},\ and\ \bibinfo {author} {\bibfnamefont {E.~J.}\
  \bibnamefont {Baerends}},\ }\href {https://doi.org/10.1007/s002140050343}
  {\bibfield  {journal} {\bibinfo  {journal} {Theor. Chem. Acc.}\ }\textbf
  {\bibinfo {volume} {99}},\ \bibinfo {pages} {329} (\bibinfo {year}
  {1998})}\BibitemShut {NoStop}%
\bibitem [{\citenamefont {Patchkovskii}\ and\ \citenamefont
  {Ziegler}(2002)}]{patchkovskii2002}%
  \BibitemOpen
  \bibfield  {author} {\bibinfo {author} {\bibfnamefont {S.}~\bibnamefont
  {Patchkovskii}}\ and\ \bibinfo {author} {\bibfnamefont {T.}~\bibnamefont
  {Ziegler}},\ }\href {https://doi.org/10.1063/1.1468640} {\bibfield  {journal}
  {\bibinfo  {journal} {J. Chem. Phys.}\ }\textbf {\bibinfo {volume} {116}},\
  \bibinfo {pages} {7806} (\bibinfo {year} {2002})}\BibitemShut {NoStop}%
\bibitem [{\citenamefont {Mishra}\ \emph {et~al.}(2022)\citenamefont {Mishra},
  \citenamefont {Yamamoto}, \citenamefont {Johnson}, \citenamefont {Jackson},
  \citenamefont {Zope},\ and\ \citenamefont {Baruah}}]{mishra2022}%
  \BibitemOpen
  \bibfield  {author} {\bibinfo {author} {\bibfnamefont {P.}~\bibnamefont
  {Mishra}}, \bibinfo {author} {\bibfnamefont {Y.}~\bibnamefont {Yamamoto}},
  \bibinfo {author} {\bibfnamefont {J.~K.}\ \bibnamefont {Johnson}}, \bibinfo
  {author} {\bibfnamefont {K.~A.}\ \bibnamefont {Jackson}}, \bibinfo {author}
  {\bibfnamefont {R.~R.}\ \bibnamefont {Zope}},\ and\ \bibinfo {author}
  {\bibfnamefont {T.}~\bibnamefont {Baruah}},\ }\href
  {https://doi.org/10.1063/5.0070893} {\bibfield  {journal} {\bibinfo
  {journal} {J. Chem. Phys.}\ }\textbf {\bibinfo {volume} {156}},\ \bibinfo
  {pages} {014306} (\bibinfo {year} {2022})}\BibitemShut {NoStop}%
\bibitem [{\citenamefont {Scuseria}(1992)}]{scuseria1992}%
  \BibitemOpen
  \bibfield  {author} {\bibinfo {author} {\bibfnamefont {G.~E.}\ \bibnamefont
  {Scuseria}},\ }\href {https://doi.org/10.1063/1.463977} {\bibfield  {journal}
  {\bibinfo  {journal} {J. Chem. Phys.}\ }\textbf {\bibinfo {volume} {97}},\
  \bibinfo {pages} {7528} (\bibinfo {year} {1992})}\BibitemShut {NoStop}%
\bibitem [{\citenamefont {Janesko}\ and\ \citenamefont
  {Scuseria}(2008)}]{janesko2008}%
  \BibitemOpen
  \bibfield  {author} {\bibinfo {author} {\bibfnamefont {B.~G.}\ \bibnamefont
  {Janesko}}\ and\ \bibinfo {author} {\bibfnamefont {G.~E.}\ \bibnamefont
  {Scuseria}},\ }\href {https://doi.org/10.1063/1.2940738} {\bibfield
  {journal} {\bibinfo  {journal} {J. Chem. Phys.}\ }\textbf {\bibinfo {volume}
  {128}},\ \bibinfo {pages} {244112} (\bibinfo {year} {2008})}\BibitemShut
  {NoStop}%
\bibitem [{\citenamefont {Verma}\ \emph {et~al.}(2012)\citenamefont {Verma},
  \citenamefont {Perera},\ and\ \citenamefont {Bartlett}}]{verma2012}%
  \BibitemOpen
  \bibfield  {author} {\bibinfo {author} {\bibfnamefont {P.}~\bibnamefont
  {Verma}}, \bibinfo {author} {\bibfnamefont {A.}~\bibnamefont {Perera}},\ and\
  \bibinfo {author} {\bibfnamefont {R.~J.}\ \bibnamefont {Bartlett}},\ }\href
  {https://doi.org/https://doi.org/10.1016/j.cplett.2011.12.017} {\bibfield
  {journal} {\bibinfo  {journal} {Chem. Phys. Lett.}\ }\textbf {\bibinfo
  {volume} {524}},\ \bibinfo {pages} {10} (\bibinfo {year} {2012})}\BibitemShut
  {NoStop}%
\bibitem [{\citenamefont {Oliphant}\ and\ \citenamefont
  {Bartlett}(1994)}]{oliphant1994}%
  \BibitemOpen
  \bibfield  {author} {\bibinfo {author} {\bibfnamefont {N.}~\bibnamefont
  {Oliphant}}\ and\ \bibinfo {author} {\bibfnamefont {R.~J.}\ \bibnamefont
  {Bartlett}},\ }\href {https://doi.org/10.1063/1.467064} {\bibfield  {journal}
  {\bibinfo  {journal} {J. Chem. Phys.}\ }\textbf {\bibinfo {volume} {100}},\
  \bibinfo {pages} {6550} (\bibinfo {year} {1994})}\BibitemShut {NoStop}%
\bibitem [{\citenamefont {Levy}\ and\ \citenamefont {Perdew}(1985)}]{levy1985}%
  \BibitemOpen
  \bibfield  {author} {\bibinfo {author} {\bibfnamefont {M.}~\bibnamefont
  {Levy}}\ and\ \bibinfo {author} {\bibfnamefont {J.~P.}\ \bibnamefont
  {Perdew}},\ }\href {https://doi.org/10.1103/PhysRevA.32.2010} {\bibfield
  {journal} {\bibinfo  {journal} {Phys. Rev. A}\ }\textbf {\bibinfo {volume}
  {32}},\ \bibinfo {pages} {2010} (\bibinfo {year} {1985})}\BibitemShut
  {NoStop}%
\bibitem [{\citenamefont {Santra}\ and\ \citenamefont
  {Martin}(2021)}]{santra2021}%
  \BibitemOpen
  \bibfield  {author} {\bibinfo {author} {\bibfnamefont {G.}~\bibnamefont
  {Santra}}\ and\ \bibinfo {author} {\bibfnamefont {J.~M.~L.}\ \bibnamefont
  {Martin}},\ }\href {https://doi.org/10.1021/acs.jctc.0c01055} {\bibfield
  {journal} {\bibinfo  {journal} {J. Chem. Theory Comput.}\ }\textbf {\bibinfo
  {volume} {17}},\ \bibinfo {pages} {1368} (\bibinfo {year} {2021})},\ \bibinfo
  {note} {pMID: 33625863}\BibitemShut {NoStop}%
\bibitem [{\citenamefont {Dasgupta}\ \emph {et~al.}(2021)\citenamefont
  {Dasgupta}, \citenamefont {Lambros}, \citenamefont {Perdew},\ and\
  \citenamefont {Paesani}}]{dasgupta2021}%
  \BibitemOpen
  \bibfield  {author} {\bibinfo {author} {\bibfnamefont {S.}~\bibnamefont
  {Dasgupta}}, \bibinfo {author} {\bibfnamefont {E.}~\bibnamefont {Lambros}},
  \bibinfo {author} {\bibfnamefont {J.}~\bibnamefont {Perdew}},\ and\ \bibinfo
  {author} {\bibfnamefont {F.}~\bibnamefont {Paesani}},\ }\href
  {https://doi.org/10.1038/s41467-021-26618-9} {\bibfield  {journal} {\bibinfo
  {journal} {Nature Commun.}\ }\textbf {\bibinfo {volume} {12}},\ \bibinfo
  {pages} {6359} (\bibinfo {year} {2021})}\BibitemShut {NoStop}%
\bibitem [{\citenamefont {Song}\ \emph
  {et~al.}(2022{\natexlab{a}})\citenamefont {Song}, \citenamefont {Vuckovic},
  \citenamefont {Kim}, \citenamefont {Sim},\ and\ \citenamefont
  {Burke}}]{song2022a}%
  \BibitemOpen
  \bibfield  {author} {\bibinfo {author} {\bibfnamefont {S.}~\bibnamefont
  {Song}}, \bibinfo {author} {\bibfnamefont {S.}~\bibnamefont {Vuckovic}},
  \bibinfo {author} {\bibfnamefont {Y.}~\bibnamefont {Kim}}, \bibinfo {author}
  {\bibfnamefont {E.}~\bibnamefont {Sim}},\ and\ \bibinfo {author}
  {\bibfnamefont {K.}~\bibnamefont {Burke}}} (\bibinfo {year}
  {2022}{\natexlab{a}}),\ \bibinfo {note} {arXiv:2207.04169}\BibitemShut
  {NoStop}%
\bibitem [{\citenamefont {Janesko}(2017)}]{janesko2017}%
  \BibitemOpen
  \bibfield  {author} {\bibinfo {author} {\bibfnamefont {B.~G.}\ \bibnamefont
  {Janesko}},\ }\href {https://doi.org/10.1039/C6CP08108H} {\bibfield
  {journal} {\bibinfo  {journal} {Phys. Chem. Chem. Phys.}\ }\textbf {\bibinfo
  {volume} {19}},\ \bibinfo {pages} {4793} (\bibinfo {year}
  {2017})}\BibitemShut {NoStop}%
\bibitem [{\citenamefont {Crisostomo}\ \emph {et~al.}(2022)\citenamefont
  {Crisostomo}, \citenamefont {Pederson}, \citenamefont {Kozlowski},
  \citenamefont {Kalita}, \citenamefont {Cancio}, \citenamefont {Datchev},
  \citenamefont {Wasserman}, \citenamefont {Song},\ and\ \citenamefont
  {Burke}}]{crisostomo2022}%
  \BibitemOpen
  \bibfield  {author} {\bibinfo {author} {\bibfnamefont {S.}~\bibnamefont
  {Crisostomo}}, \bibinfo {author} {\bibfnamefont {R.}~\bibnamefont
  {Pederson}}, \bibinfo {author} {\bibfnamefont {J.}~\bibnamefont {Kozlowski}},
  \bibinfo {author} {\bibfnamefont {B.}~\bibnamefont {Kalita}}, \bibinfo
  {author} {\bibfnamefont {A.~C.}\ \bibnamefont {Cancio}}, \bibinfo {author}
  {\bibfnamefont {K.}~\bibnamefont {Datchev}}, \bibinfo {author} {\bibfnamefont
  {A.}~\bibnamefont {Wasserman}}, \bibinfo {author} {\bibfnamefont
  {S.}~\bibnamefont {Song}},\ and\ \bibinfo {author} {\bibfnamefont
  {K.}~\bibnamefont {Burke}}} (\bibinfo {year} {2022}),\ \bibinfo {note}
  {arXiv:2207.05794}\BibitemShut {NoStop}%
\bibitem [{\citenamefont {Zhao}\ \emph {et~al.}(2005)\citenamefont {Zhao},
  \citenamefont {Gonz{\'{a}}lez-Garc{\'i}a},\ and\ \citenamefont
  {Truhlar}}]{zhao2005}%
  \BibitemOpen
  \bibfield  {author} {\bibinfo {author} {\bibfnamefont {Y.}~\bibnamefont
  {Zhao}}, \bibinfo {author} {\bibfnamefont {N.}~\bibnamefont
  {Gonz{\'{a}}lez-Garc{\'i}a}},\ and\ \bibinfo {author} {\bibfnamefont {D.~G.}\
  \bibnamefont {Truhlar}},\ }\href {https://doi.org/10.1021/jp045141s}
  {\bibfield  {journal} {\bibinfo  {journal} {J. Phys. Chem. A}\ }\textbf
  {\bibinfo {volume} {109}},\ \bibinfo {pages} {2012} (\bibinfo {year}
  {2005})}\BibitemShut {NoStop}%
\bibitem [{\citenamefont {Goerigk}\ \emph {et~al.}(2017)\citenamefont
  {Goerigk}, \citenamefont {Hansen}, \citenamefont {Bauer}, \citenamefont
  {Ehrlich}, \citenamefont {Najibi},\ and\ \citenamefont
  {Grimme}}]{goerigk2017}%
  \BibitemOpen
  \bibfield  {author} {\bibinfo {author} {\bibfnamefont {L.}~\bibnamefont
  {Goerigk}}, \bibinfo {author} {\bibfnamefont {A.}~\bibnamefont {Hansen}},
  \bibinfo {author} {\bibfnamefont {C.}~\bibnamefont {Bauer}}, \bibinfo
  {author} {\bibfnamefont {S.}~\bibnamefont {Ehrlich}}, \bibinfo {author}
  {\bibfnamefont {A.}~\bibnamefont {Najibi}},\ and\ \bibinfo {author}
  {\bibfnamefont {S.}~\bibnamefont {Grimme}},\ }\href
  {https://doi.org/10.1039/C7CP04913G} {\bibfield  {journal} {\bibinfo
  {journal} {Phys. Chem. Chem. Phys.}\ }\textbf {\bibinfo {volume} {19}},\
  \bibinfo {pages} {32184} (\bibinfo {year} {2017})},\ \bibinfo {note} {and
  website
  \url{http://www.thch.uni-bonn.de/tc.old/downloads/GMTKN/GMTKN55/BH76.html}}\BibitemShut
  {NoStop}%
\bibitem [{\citenamefont {Sun}(2015)}]{qiming2015}%
  \BibitemOpen
  \bibfield  {author} {\bibinfo {author} {\bibfnamefont {Q.}~\bibnamefont
  {Sun}},\ }\href {https://doi.org/https://doi.org/10.1002/jcc.23981}
  {\bibfield  {journal} {\bibinfo  {journal} {J. Comput. Chem.}\ }\textbf
  {\bibinfo {volume} {36}},\ \bibinfo {pages} {1664} (\bibinfo {year}
  {2015})}\BibitemShut {NoStop}%
\bibitem [{\citenamefont {Sun}\ \emph {et~al.}(2018)\citenamefont {Sun},
  \citenamefont {Berkelbach}, \citenamefont {Blunt}, \citenamefont {Booth},
  \citenamefont {Guo}, \citenamefont {Li}, \citenamefont {Liu}, \citenamefont
  {McClain}, \citenamefont {Sayfutyarova}, \citenamefont {Sharma},
  \citenamefont {Wouters},\ and\ \citenamefont {Chan}}]{qiming2018}%
  \BibitemOpen
  \bibfield  {author} {\bibinfo {author} {\bibfnamefont {Q.}~\bibnamefont
  {Sun}}, \bibinfo {author} {\bibfnamefont {T.~C.}\ \bibnamefont {Berkelbach}},
  \bibinfo {author} {\bibfnamefont {N.~S.}\ \bibnamefont {Blunt}}, \bibinfo
  {author} {\bibfnamefont {G.~H.}\ \bibnamefont {Booth}}, \bibinfo {author}
  {\bibfnamefont {S.}~\bibnamefont {Guo}}, \bibinfo {author} {\bibfnamefont
  {Z.}~\bibnamefont {Li}}, \bibinfo {author} {\bibfnamefont {J.}~\bibnamefont
  {Liu}}, \bibinfo {author} {\bibfnamefont {J.~D.}\ \bibnamefont {McClain}},
  \bibinfo {author} {\bibfnamefont {E.~R.}\ \bibnamefont {Sayfutyarova}},
  \bibinfo {author} {\bibfnamefont {S.}~\bibnamefont {Sharma}}, \bibinfo
  {author} {\bibfnamefont {S.}~\bibnamefont {Wouters}},\ and\ \bibinfo {author}
  {\bibfnamefont {G.~K.-L.}\ \bibnamefont {Chan}},\ }\href
  {https://doi.org/https://doi.org/10.1002/wcms.1340} {\bibfield  {journal}
  {\bibinfo  {journal} {WIREs Comput. Mol. Sci.}\ }\textbf {\bibinfo {volume}
  {8}},\ \bibinfo {pages} {e1340} (\bibinfo {year} {2018})}\BibitemShut
  {NoStop}%
\bibitem [{\citenamefont {Sun}\ \emph {et~al.}(2020)\citenamefont {Sun},
  \citenamefont {Zhang}, \citenamefont {Banerjee}, \citenamefont {Bao},
  \citenamefont {Barbry}, \citenamefont {Blunt}, \citenamefont {Bogdanov},
  \citenamefont {Booth}, \citenamefont {Chen}, \citenamefont {Cui},
  \citenamefont {Eriksen}, \citenamefont {Gao}, \citenamefont {Guo},
  \citenamefont {Hermann}, \citenamefont {Hermes}, \citenamefont {Koh},
  \citenamefont {Koval}, \citenamefont {Lehtola}, \citenamefont {Li},
  \citenamefont {Liu}, \citenamefont {Mardirossian}, \citenamefont {McClain},
  \citenamefont {Motta}, \citenamefont {Mussard}, \citenamefont {Pham},
  \citenamefont {Pulkin}, \citenamefont {Purwanto}, \citenamefont {Robinson},
  \citenamefont {Ronca}, \citenamefont {Sayfutyarova}, \citenamefont
  {Scheurer}, \citenamefont {Schurkus}, \citenamefont {Smith}, \citenamefont
  {Sun}, \citenamefont {Sun}, \citenamefont {Upadhyay}, \citenamefont {Wagner},
  \citenamefont {Wang}, \citenamefont {White}, \citenamefont {Whitfield},
  \citenamefont {Williamson}, \citenamefont {Wouters}, \citenamefont {Yang},
  \citenamefont {Yu}, \citenamefont {Zhu}, \citenamefont {Berkelbach},
  \citenamefont {Sharma}, \citenamefont {Sokolov},\ and\ \citenamefont
  {Chan}}]{qiming2020}%
  \BibitemOpen
  \bibfield  {author} {\bibinfo {author} {\bibfnamefont {Q.}~\bibnamefont
  {Sun}}, \bibinfo {author} {\bibfnamefont {X.}~\bibnamefont {Zhang}}, \bibinfo
  {author} {\bibfnamefont {S.}~\bibnamefont {Banerjee}}, \bibinfo {author}
  {\bibfnamefont {P.}~\bibnamefont {Bao}}, \bibinfo {author} {\bibfnamefont
  {M.}~\bibnamefont {Barbry}}, \bibinfo {author} {\bibfnamefont {N.~S.}\
  \bibnamefont {Blunt}}, \bibinfo {author} {\bibfnamefont {N.~A.}\ \bibnamefont
  {Bogdanov}}, \bibinfo {author} {\bibfnamefont {G.~H.}\ \bibnamefont {Booth}},
  \bibinfo {author} {\bibfnamefont {J.}~\bibnamefont {Chen}}, \bibinfo {author}
  {\bibfnamefont {Z.-H.}\ \bibnamefont {Cui}}, \bibinfo {author} {\bibfnamefont
  {J.~J.}\ \bibnamefont {Eriksen}}, \bibinfo {author} {\bibfnamefont
  {Y.}~\bibnamefont {Gao}}, \bibinfo {author} {\bibfnamefont {S.}~\bibnamefont
  {Guo}}, \bibinfo {author} {\bibfnamefont {J.}~\bibnamefont {Hermann}},
  \bibinfo {author} {\bibfnamefont {M.~R.}\ \bibnamefont {Hermes}}, \bibinfo
  {author} {\bibfnamefont {K.}~\bibnamefont {Koh}}, \bibinfo {author}
  {\bibfnamefont {P.}~\bibnamefont {Koval}}, \bibinfo {author} {\bibfnamefont
  {S.}~\bibnamefont {Lehtola}}, \bibinfo {author} {\bibfnamefont
  {Z.}~\bibnamefont {Li}}, \bibinfo {author} {\bibfnamefont {J.}~\bibnamefont
  {Liu}}, \bibinfo {author} {\bibfnamefont {N.}~\bibnamefont {Mardirossian}},
  \bibinfo {author} {\bibfnamefont {J.~D.}\ \bibnamefont {McClain}}, \bibinfo
  {author} {\bibfnamefont {M.}~\bibnamefont {Motta}}, \bibinfo {author}
  {\bibfnamefont {B.}~\bibnamefont {Mussard}}, \bibinfo {author} {\bibfnamefont
  {H.~Q.}\ \bibnamefont {Pham}}, \bibinfo {author} {\bibfnamefont
  {A.}~\bibnamefont {Pulkin}}, \bibinfo {author} {\bibfnamefont
  {W.}~\bibnamefont {Purwanto}}, \bibinfo {author} {\bibfnamefont {P.~J.}\
  \bibnamefont {Robinson}}, \bibinfo {author} {\bibfnamefont {E.}~\bibnamefont
  {Ronca}}, \bibinfo {author} {\bibfnamefont {E.~R.}\ \bibnamefont
  {Sayfutyarova}}, \bibinfo {author} {\bibfnamefont {M.}~\bibnamefont
  {Scheurer}}, \bibinfo {author} {\bibfnamefont {H.~F.}\ \bibnamefont
  {Schurkus}}, \bibinfo {author} {\bibfnamefont {J.~E.~T.}\ \bibnamefont
  {Smith}}, \bibinfo {author} {\bibfnamefont {C.}~\bibnamefont {Sun}}, \bibinfo
  {author} {\bibfnamefont {S.-N.}\ \bibnamefont {Sun}}, \bibinfo {author}
  {\bibfnamefont {S.}~\bibnamefont {Upadhyay}}, \bibinfo {author}
  {\bibfnamefont {L.~K.}\ \bibnamefont {Wagner}}, \bibinfo {author}
  {\bibfnamefont {X.}~\bibnamefont {Wang}}, \bibinfo {author} {\bibfnamefont
  {A.}~\bibnamefont {White}}, \bibinfo {author} {\bibfnamefont {J.~D.}\
  \bibnamefont {Whitfield}}, \bibinfo {author} {\bibfnamefont {M.~J.}\
  \bibnamefont {Williamson}}, \bibinfo {author} {\bibfnamefont
  {S.}~\bibnamefont {Wouters}}, \bibinfo {author} {\bibfnamefont
  {J.}~\bibnamefont {Yang}}, \bibinfo {author} {\bibfnamefont {J.~M.}\
  \bibnamefont {Yu}}, \bibinfo {author} {\bibfnamefont {T.}~\bibnamefont
  {Zhu}}, \bibinfo {author} {\bibfnamefont {T.~C.}\ \bibnamefont {Berkelbach}},
  \bibinfo {author} {\bibfnamefont {S.}~\bibnamefont {Sharma}}, \bibinfo
  {author} {\bibfnamefont {A.~Y.}\ \bibnamefont {Sokolov}},\ and\ \bibinfo
  {author} {\bibfnamefont {G.~K.-L.}\ \bibnamefont {Chan}},\ }\href
  {https://doi.org/10.1063/5.0006074} {\bibfield  {journal} {\bibinfo
  {journal} {J. Chem. Phys.}\ }\textbf {\bibinfo {volume} {153}},\ \bibinfo
  {pages} {024109} (\bibinfo {year} {2020})}\BibitemShut {NoStop}%
\bibitem [{NRL(2022)}]{NRLMOL}%
  \BibitemOpen
  \href@noop {} {} (\bibinfo {year} {2022}),\ \bibinfo {note} {{R. R. Zope, T.
  Baruah, Y. Yamamoto, L. Basurto, C. Diaz, J. Peralta, and K. A. Jackson,
  FLOSIC 0.1.2, based on the NRLMOL Code of M. R. Pederson.}}\BibitemShut
  {Stop}%
\bibitem [{\citenamefont {Pulay}(1980)}]{pulay1980}%
  \BibitemOpen
  \bibfield  {author} {\bibinfo {author} {\bibfnamefont {P.}~\bibnamefont
  {Pulay}},\ }\href
  {https://doi.org/https://doi.org/10.1016/0009-2614(80)80396-4} {\bibfield
  {journal} {\bibinfo  {journal} {Chem. Phys. Lett.}\ }\textbf {\bibinfo
  {volume} {73}},\ \bibinfo {pages} {393} (\bibinfo {year} {1980})}\BibitemShut
  {NoStop}%
\bibitem [{\citenamefont {Kaplan}(2022)}]{code_repo}%
  \BibitemOpen
  \bibfield  {author} {\bibinfo {author} {\bibfnamefont {A.~D.}\ \bibnamefont
  {Kaplan}},\ }\href@noop {} {\bibinfo {title} {{BH76-PySCF-PyFLOSIC}}}
  (\bibinfo {year} {2022}),\ \bibinfo {note} {public code repository:
  \url{https://github.com/esoteric-ephemera/BH76-PySCF-PyFLOSIC}}\BibitemShut
  {NoStop}%
\bibitem [{\citenamefont {Kaplan}\ \emph {et~al.}(2022)\citenamefont {Kaplan},
  \citenamefont {Shahi}, \citenamefont {Sah}, \citenamefont {Bhetwal},\ and\
  \citenamefont {Perdew}}]{data_repo}%
  \BibitemOpen
  \bibfield  {author} {\bibinfo {author} {\bibfnamefont {A.~D.}\ \bibnamefont
  {Kaplan}}, \bibinfo {author} {\bibfnamefont {C.}~\bibnamefont {Shahi}},
  \bibinfo {author} {\bibfnamefont {R.~K.}\ \bibnamefont {Sah}}, \bibinfo
  {author} {\bibfnamefont {P.}~\bibnamefont {Bhetwal}},\ and\ \bibinfo {author}
  {\bibfnamefont {J.~P.}\ \bibnamefont {Perdew}},\ }\href@noop {} {\bibinfo
  {title} {{Data for ``Understanding density driven errors via reaction barrier
  heights''}}} (\bibinfo {year} {2022}),\ \bibinfo {note} {zenodo data
  repository, DOI:10.5281/zenodo.7075664}\BibitemShut {NoStop}%
\bibitem [{\citenamefont {Pederson}(2015)}]{pederson2015}%
  \BibitemOpen
  \bibfield  {author} {\bibinfo {author} {\bibfnamefont {M.~R.}\ \bibnamefont
  {Pederson}},\ }\href {https://doi.org/10.1063/1.4907592} {\bibfield
  {journal} {\bibinfo  {journal} {J. Chem. Phys.}\ }\textbf {\bibinfo {volume}
  {142}},\ \bibinfo {pages} {064112} (\bibinfo {year} {2015})}\BibitemShut
  {NoStop}%
\bibitem [{\citenamefont {Yang}\ \emph {et~al.}(2017)\citenamefont {Yang},
  \citenamefont {Pederson},\ and\ \citenamefont {Perdew}}]{yang2017}%
  \BibitemOpen
  \bibfield  {author} {\bibinfo {author} {\bibfnamefont {Z.-h.}\ \bibnamefont
  {Yang}}, \bibinfo {author} {\bibfnamefont {M.~R.}\ \bibnamefont {Pederson}},\
  and\ \bibinfo {author} {\bibfnamefont {J.~P.}\ \bibnamefont {Perdew}},\
  }\href {https://doi.org/10.1103/PhysRevA.95.052505} {\bibfield  {journal}
  {\bibinfo  {journal} {Phys. Rev. A}\ }\textbf {\bibinfo {volume} {95}},\
  \bibinfo {pages} {052505} (\bibinfo {year} {2017})}\BibitemShut {NoStop}%
\bibitem [{\citenamefont {Yamamoto}\ \emph {et~al.}(2019)\citenamefont
  {Yamamoto}, \citenamefont {Diaz}, \citenamefont {Basurto}, \citenamefont
  {Jackson}, \citenamefont {Baruah},\ and\ \citenamefont
  {Zope}}]{yamamoto2019}%
  \BibitemOpen
  \bibfield  {author} {\bibinfo {author} {\bibfnamefont {Y.}~\bibnamefont
  {Yamamoto}}, \bibinfo {author} {\bibfnamefont {C.~M.}\ \bibnamefont {Diaz}},
  \bibinfo {author} {\bibfnamefont {L.}~\bibnamefont {Basurto}}, \bibinfo
  {author} {\bibfnamefont {K.~A.}\ \bibnamefont {Jackson}}, \bibinfo {author}
  {\bibfnamefont {T.}~\bibnamefont {Baruah}},\ and\ \bibinfo {author}
  {\bibfnamefont {R.~R.}\ \bibnamefont {Zope}},\ }\href
  {https://doi.org/10.1063/1.5120532} {\bibfield  {journal} {\bibinfo
  {journal} {J. Chem. Phys.}\ }\textbf {\bibinfo {volume} {151}},\ \bibinfo
  {pages} {154105} (\bibinfo {year} {2019})}\BibitemShut {NoStop}%
\bibitem [{\citenamefont {Porezag}\ and\ \citenamefont
  {Pederson}(1999)}]{porezag1999}%
  \BibitemOpen
  \bibfield  {author} {\bibinfo {author} {\bibfnamefont {D.}~\bibnamefont
  {Porezag}}\ and\ \bibinfo {author} {\bibfnamefont {M.~R.}\ \bibnamefont
  {Pederson}},\ }\href {https://doi.org/10.1103/PhysRevA.60.2840} {\bibfield
  {journal} {\bibinfo  {journal} {Phys. Rev. A}\ }\textbf {\bibinfo {volume}
  {60}},\ \bibinfo {pages} {2840} (\bibinfo {year} {1999})}\BibitemShut
  {NoStop}%
\bibitem [{\citenamefont {Schwalbe}\ \emph {et~al.}(2019)\citenamefont
  {Schwalbe}, \citenamefont {Trepte}, \citenamefont {Fiedler}, \citenamefont
  {Johnson}, \citenamefont {Kraus}, \citenamefont {Hahn}, \citenamefont
  {Peralta}, \citenamefont {Jackson},\ and\ \citenamefont
  {Kortus}}]{schwalbe2019}%
  \BibitemOpen
  \bibfield  {author} {\bibinfo {author} {\bibfnamefont {S.}~\bibnamefont
  {Schwalbe}}, \bibinfo {author} {\bibfnamefont {K.}~\bibnamefont {Trepte}},
  \bibinfo {author} {\bibfnamefont {L.}~\bibnamefont {Fiedler}}, \bibinfo
  {author} {\bibfnamefont {A.~I.}\ \bibnamefont {Johnson}}, \bibinfo {author}
  {\bibfnamefont {J.}~\bibnamefont {Kraus}}, \bibinfo {author} {\bibfnamefont
  {T.}~\bibnamefont {Hahn}}, \bibinfo {author} {\bibfnamefont {J.~E.}\
  \bibnamefont {Peralta}}, \bibinfo {author} {\bibfnamefont {K.~A.}\
  \bibnamefont {Jackson}},\ and\ \bibinfo {author} {\bibfnamefont
  {J.}~\bibnamefont {Kortus}},\ }\href
  {https://doi.org/https://doi.org/10.1002/jcc.26062} {\bibfield  {journal}
  {\bibinfo  {journal} {J. Comput. Chem.}\ }\textbf {\bibinfo {volume} {40}},\
  \bibinfo {pages} {2843} (\bibinfo {year} {2019})}\BibitemShut {NoStop}%
\bibitem [{\citenamefont {Schwalbe}\ \emph {et~al.}(2020)\citenamefont
  {Schwalbe}, \citenamefont {Fiedler}, \citenamefont {Kraus}, \citenamefont
  {Kortus}, \citenamefont {Trepte},\ and\ \citenamefont
  {Lehtola}}]{schwalbe2020}%
  \BibitemOpen
  \bibfield  {author} {\bibinfo {author} {\bibfnamefont {S.}~\bibnamefont
  {Schwalbe}}, \bibinfo {author} {\bibfnamefont {L.}~\bibnamefont {Fiedler}},
  \bibinfo {author} {\bibfnamefont {J.}~\bibnamefont {Kraus}}, \bibinfo
  {author} {\bibfnamefont {J.}~\bibnamefont {Kortus}}, \bibinfo {author}
  {\bibfnamefont {K.}~\bibnamefont {Trepte}},\ and\ \bibinfo {author}
  {\bibfnamefont {S.}~\bibnamefont {Lehtola}},\ }\href
  {https://doi.org/10.1063/5.0012519} {\bibfield  {journal} {\bibinfo
  {journal} {J. Chem. Phys.}\ }\textbf {\bibinfo {volume} {153}},\ \bibinfo
  {pages} {084104} (\bibinfo {year} {2020})},\ \bibinfo {note} {see especially
  the code repository, \url{https://github.com/pyflosic}}\BibitemShut {NoStop}%
\bibitem [{\citenamefont {Perdew}\ and\ \citenamefont
  {Schmidt}(2001)}]{perdew2001}%
  \BibitemOpen
  \bibfield  {author} {\bibinfo {author} {\bibfnamefont {J.~P.}\ \bibnamefont
  {Perdew}}\ and\ \bibinfo {author} {\bibfnamefont {K.}~\bibnamefont
  {Schmidt}},\ }in\ \href {https://doi.org/10.1063/1.1390175} {\emph {\bibinfo
  {booktitle} {Density Functional Theory and Its Applications to Materials}}},\
  Vol.\ \bibinfo {volume} {577},\ \bibinfo {editor} {edited by\ \bibinfo
  {editor} {\bibfnamefont {V.~E.}\ \bibnamefont {Van~Doren}}, \bibinfo {editor}
  {\bibfnamefont {C.}~\bibnamefont {Van~Alsenoy}},\ and\ \bibinfo {editor}
  {\bibfnamefont {P.}~\bibnamefont {Geerlings}}}\ (\bibinfo  {publisher}
  {American Institute of Physics},\ \bibinfo {year} {2001})\ p.~\bibinfo
  {pages} {1}\BibitemShut {NoStop}%
\bibitem [{\citenamefont {Perdew}\ \emph
  {et~al.}(1996{\natexlab{b}})\citenamefont {Perdew}, \citenamefont {Burke},\
  and\ \citenamefont {Ernzerhof}}]{perdew1996}%
  \BibitemOpen
  \bibfield  {author} {\bibinfo {author} {\bibfnamefont {J.~P.}\ \bibnamefont
  {Perdew}}, \bibinfo {author} {\bibfnamefont {K.}~\bibnamefont {Burke}},\ and\
  \bibinfo {author} {\bibfnamefont {M.}~\bibnamefont {Ernzerhof}},\ }\href
  {https://doi.org/10.1103/PhysRevLett.77.3865} {\bibfield  {journal} {\bibinfo
   {journal} {Phys. Rev. Lett.}\ }\textbf {\bibinfo {volume} {77}},\ \bibinfo
  {pages} {3865} (\bibinfo {year} {1996}{\natexlab{b}})}\BibitemShut {NoStop}%
\bibitem [{\citenamefont {Becke}(1988)}]{becke1988}%
  \BibitemOpen
  \bibfield  {author} {\bibinfo {author} {\bibfnamefont {A.~D.}\ \bibnamefont
  {Becke}},\ }\href {https://doi.org/10.1103/PhysRevA.38.3098} {\bibfield
  {journal} {\bibinfo  {journal} {Phys. Rev. A}\ }\textbf {\bibinfo {volume}
  {38}},\ \bibinfo {pages} {3098} (\bibinfo {year} {1988})}\BibitemShut
  {NoStop}%
\bibitem [{\citenamefont {Lee}\ \emph {et~al.}(1988)\citenamefont {Lee},
  \citenamefont {Yang},\ and\ \citenamefont {Parr}}]{lee1988}%
  \BibitemOpen
  \bibfield  {author} {\bibinfo {author} {\bibfnamefont {C.}~\bibnamefont
  {Lee}}, \bibinfo {author} {\bibfnamefont {W.}~\bibnamefont {Yang}},\ and\
  \bibinfo {author} {\bibfnamefont {R.~G.}\ \bibnamefont {Parr}},\ }\href
  {https://doi.org/10.1103/PhysRevB.37.785} {\bibfield  {journal} {\bibinfo
  {journal} {Phys. Rev. B}\ }\textbf {\bibinfo {volume} {37}},\ \bibinfo
  {pages} {785} (\bibinfo {year} {1988})}\BibitemShut {NoStop}%
\bibitem [{\citenamefont {Miehlich}\ \emph {et~al.}(1989)\citenamefont
  {Miehlich}, \citenamefont {Savin}, \citenamefont {Stoll},\ and\ \citenamefont
  {Preuss}}]{miehlich1989}%
  \BibitemOpen
  \bibfield  {author} {\bibinfo {author} {\bibfnamefont {B.}~\bibnamefont
  {Miehlich}}, \bibinfo {author} {\bibfnamefont {A.}~\bibnamefont {Savin}},
  \bibinfo {author} {\bibfnamefont {H.}~\bibnamefont {Stoll}},\ and\ \bibinfo
  {author} {\bibfnamefont {H.}~\bibnamefont {Preuss}},\ }\href
  {https://doi.org/https://doi.org/10.1016/0009-2614(89)87234-3} {\bibfield
  {journal} {\bibinfo  {journal} {Chem. Phys. Lett.}\ }\textbf {\bibinfo
  {volume} {157}},\ \bibinfo {pages} {200} (\bibinfo {year}
  {1989})}\BibitemShut {NoStop}%
\bibitem [{\citenamefont {Sun}\ \emph {et~al.}(2015)\citenamefont {Sun},
  \citenamefont {Ruzsinszky},\ and\ \citenamefont {Perdew}}]{sun2015}%
  \BibitemOpen
  \bibfield  {author} {\bibinfo {author} {\bibfnamefont {J.}~\bibnamefont
  {Sun}}, \bibinfo {author} {\bibfnamefont {A.}~\bibnamefont {Ruzsinszky}},\
  and\ \bibinfo {author} {\bibfnamefont {J.~P.}\ \bibnamefont {Perdew}},\
  }\href {https://doi.org/10.1103/PhysRevLett.115.036402} {\bibfield  {journal}
  {\bibinfo  {journal} {Phys. Rev. Lett.}\ }\textbf {\bibinfo {volume} {115}},\
  \bibinfo {pages} {036402} (\bibinfo {year} {2015})}\BibitemShut {NoStop}%
\bibitem [{\citenamefont {Furness}\ \emph {et~al.}(2020)\citenamefont
  {Furness}, \citenamefont {Kaplan}, \citenamefont {Ning}, \citenamefont
  {Perdew},\ and\ \citenamefont {Sun}}]{furness2020}%
  \BibitemOpen
  \bibfield  {author} {\bibinfo {author} {\bibfnamefont {J.~W.}\ \bibnamefont
  {Furness}}, \bibinfo {author} {\bibfnamefont {A.~D.}\ \bibnamefont {Kaplan}},
  \bibinfo {author} {\bibfnamefont {J.}~\bibnamefont {Ning}}, \bibinfo {author}
  {\bibfnamefont {J.~P.}\ \bibnamefont {Perdew}},\ and\ \bibinfo {author}
  {\bibfnamefont {J.}~\bibnamefont {Sun}},\ }\href
  {https://doi.org/10.1021/acs.jpclett.0c02405} {\bibfield  {journal} {\bibinfo
   {journal} {J. Phys. Chem. Lett.}\ }\textbf {\bibinfo {volume} {11}},\
  \bibinfo {pages} {8208} (\bibinfo {year} {2020})},\ \bibinfo {note}
  {correction, \textit{ibid.} \textbf{11}, 9248 (2020).}\BibitemShut {Stop}%
\bibitem [{\citenamefont {Zhao}\ and\ \citenamefont
  {Truhlar}(2006)}]{zhao2006}%
  \BibitemOpen
  \bibfield  {author} {\bibinfo {author} {\bibfnamefont {Y.}~\bibnamefont
  {Zhao}}\ and\ \bibinfo {author} {\bibfnamefont {D.~G.}\ \bibnamefont
  {Truhlar}},\ }\href {https://doi.org/10.1063/1.2370993} {\bibfield  {journal}
  {\bibinfo  {journal} {J. Chem. Phys.}\ }\textbf {\bibinfo {volume} {125}},\
  \bibinfo {pages} {194101} (\bibinfo {year} {2006})}\BibitemShut {NoStop}%
\bibitem [{\citenamefont {Yu}\ \emph {et~al.}(2016)\citenamefont {Yu},
  \citenamefont {He},\ and\ \citenamefont {Truhlar}}]{yu2016}%
  \BibitemOpen
  \bibfield  {author} {\bibinfo {author} {\bibfnamefont {H.~S.}\ \bibnamefont
  {Yu}}, \bibinfo {author} {\bibfnamefont {X.}~\bibnamefont {He}},\ and\
  \bibinfo {author} {\bibfnamefont {D.~G.}\ \bibnamefont {Truhlar}},\ }\href
  {https://doi.org/10.1021/acs.jctc.5b01082} {\bibfield  {journal} {\bibinfo
  {journal} {J. Chem. Theory Comput.}\ }\textbf {\bibinfo {volume} {12}},\
  \bibinfo {pages} {1280} (\bibinfo {year} {2016})},\ \bibinfo {note} {pMID:
  26722866}\BibitemShut {NoStop}%
\bibitem [{\citenamefont {Becke}(1993)}]{becke1993}%
  \BibitemOpen
  \bibfield  {author} {\bibinfo {author} {\bibfnamefont {A.~D.}\ \bibnamefont
  {Becke}},\ }\href {https://doi.org/10.1063/1.464913} {\bibfield  {journal}
  {\bibinfo  {journal} {J. Chem. Phys.}\ }\textbf {\bibinfo {volume} {98}},\
  \bibinfo {pages} {5648} (\bibinfo {year} {1993})}\BibitemShut {NoStop}%
\bibitem [{\citenamefont {Stephens}\ \emph {et~al.}(1994)\citenamefont
  {Stephens}, \citenamefont {Devlin}, \citenamefont {Chabalowski},\ and\
  \citenamefont {Frisch}}]{stephens1994}%
  \BibitemOpen
  \bibfield  {author} {\bibinfo {author} {\bibfnamefont {P.~J.}\ \bibnamefont
  {Stephens}}, \bibinfo {author} {\bibfnamefont {F.~J.}\ \bibnamefont
  {Devlin}}, \bibinfo {author} {\bibfnamefont {C.~F.}\ \bibnamefont
  {Chabalowski}},\ and\ \bibinfo {author} {\bibfnamefont {M.~J.}\ \bibnamefont
  {Frisch}},\ }\href {https://doi.org/10.1021/j100096a001} {\bibfield
  {journal} {\bibinfo  {journal} {J. Phys. Chem.}\ }\textbf {\bibinfo {volume}
  {98}},\ \bibinfo {pages} {11623} (\bibinfo {year} {1994})}\BibitemShut
  {NoStop}%
\bibitem [{\citenamefont {Dunning}(1989)}]{dunning1989}%
  \BibitemOpen
  \bibfield  {author} {\bibinfo {author} {\bibfnamefont {T.~H.}\ \bibnamefont
  {Dunning}},\ }\href {https://doi.org/10.1063/1.456153} {\bibfield  {journal}
  {\bibinfo  {journal} {J. Chem. Phys.}\ }\textbf {\bibinfo {volume} {90}},\
  \bibinfo {pages} {1007} (\bibinfo {year} {1989})}\BibitemShut {NoStop}%
\bibitem [{\citenamefont {Weigend}\ \emph {et~al.}(2003)\citenamefont
  {Weigend}, \citenamefont {Furche},\ and\ \citenamefont
  {Ahlrichs}}]{weigend2003}%
  \BibitemOpen
  \bibfield  {author} {\bibinfo {author} {\bibfnamefont {F.}~\bibnamefont
  {Weigend}}, \bibinfo {author} {\bibfnamefont {F.}~\bibnamefont {Furche}},\
  and\ \bibinfo {author} {\bibfnamefont {R.}~\bibnamefont {Ahlrichs}},\ }\href
  {https://doi.org/10.1063/1.1627293} {\bibfield  {journal} {\bibinfo
  {journal} {J. Chem. Phys.}\ }\textbf {\bibinfo {volume} {119}},\ \bibinfo
  {pages} {12753} (\bibinfo {year} {2003})}\BibitemShut {NoStop}%
\bibitem [{\citenamefont {Weigend}\ and\ \citenamefont
  {Ahlrichs}(2005)}]{weigend2005}%
  \BibitemOpen
  \bibfield  {author} {\bibinfo {author} {\bibfnamefont {F.}~\bibnamefont
  {Weigend}}\ and\ \bibinfo {author} {\bibfnamefont {R.}~\bibnamefont
  {Ahlrichs}},\ }\href {https://doi.org/10.1039/B508541A} {\bibfield  {journal}
  {\bibinfo  {journal} {Phys. Chem. Chem. Phys.}\ }\textbf {\bibinfo {volume}
  {7}},\ \bibinfo {pages} {3297} (\bibinfo {year} {2005})}\BibitemShut
  {NoStop}%
\bibitem [{\citenamefont {Pritchard}\ \emph {et~al.}(2019)\citenamefont
  {Pritchard}, \citenamefont {Altarawy}, \citenamefont {Didier}, \citenamefont
  {Gibson},\ and\ \citenamefont {Windus}}]{pritchard2019}%
  \BibitemOpen
  \bibfield  {author} {\bibinfo {author} {\bibfnamefont {B.~P.}\ \bibnamefont
  {Pritchard}}, \bibinfo {author} {\bibfnamefont {D.}~\bibnamefont {Altarawy}},
  \bibinfo {author} {\bibfnamefont {B.}~\bibnamefont {Didier}}, \bibinfo
  {author} {\bibfnamefont {T.~D.}\ \bibnamefont {Gibson}},\ and\ \bibinfo
  {author} {\bibfnamefont {T.~L.}\ \bibnamefont {Windus}},\ }\href
  {https://doi.org/10.1021/acs.jcim.9b00725} {\bibfield  {journal} {\bibinfo
  {journal} {J. Chem. Inf. Model.}\ }\textbf {\bibinfo {volume} {59}},\
  \bibinfo {pages} {4814} (\bibinfo {year} {2019})}\BibitemShut {NoStop}%
\bibitem [{\citenamefont {Perdew}\ \emph
  {et~al.}(1996{\natexlab{c}})\citenamefont {Perdew}, \citenamefont
  {Ernzerhof},\ and\ \citenamefont {Burke}}]{perdew1996b}%
  \BibitemOpen
  \bibfield  {author} {\bibinfo {author} {\bibfnamefont {J.~P.}\ \bibnamefont
  {Perdew}}, \bibinfo {author} {\bibfnamefont {M.}~\bibnamefont {Ernzerhof}},\
  and\ \bibinfo {author} {\bibfnamefont {K.}~\bibnamefont {Burke}},\ }\href
  {https://doi.org/10.1063/1.472933} {\bibfield  {journal} {\bibinfo  {journal}
  {J. Chem. Phys.}\ }\textbf {\bibinfo {volume} {105}},\ \bibinfo {pages}
  {9982} (\bibinfo {year} {1996}{\natexlab{c}})}\BibitemShut {NoStop}%
\bibitem [{\citenamefont {Medvedev}\ \emph {et~al.}(2017)\citenamefont
  {Medvedev}, \citenamefont {Bushmarinov}, \citenamefont {Sun}, \citenamefont
  {Perdew},\ and\ \citenamefont {Lyssenko}}]{medvedev2017}%
  \BibitemOpen
  \bibfield  {author} {\bibinfo {author} {\bibfnamefont {M.~G.}\ \bibnamefont
  {Medvedev}}, \bibinfo {author} {\bibfnamefont {I.~S.}\ \bibnamefont
  {Bushmarinov}}, \bibinfo {author} {\bibfnamefont {J.}~\bibnamefont {Sun}},
  \bibinfo {author} {\bibfnamefont {J.~P.}\ \bibnamefont {Perdew}},\ and\
  \bibinfo {author} {\bibfnamefont {K.~A.}\ \bibnamefont {Lyssenko}},\ }\href
  {https://doi.org/10.1126/science.aah5975} {\bibfield  {journal} {\bibinfo
  {journal} {Science}\ }\textbf {\bibinfo {volume} {355}},\ \bibinfo {pages}
  {49} (\bibinfo {year} {2017})}\BibitemShut {NoStop}%
\bibitem [{\citenamefont {Perdew}(1985)}]{perdew1985a}%
  \BibitemOpen
  \bibfield  {author} {\bibinfo {author} {\bibfnamefont {J.~P.}\ \bibnamefont
  {Perdew}},\ }\bibinfo {title} {What do the {K}ohn--{S}ham orbital energies
  mean? how do atoms dissociate?},\ in\ \href
  {https://doi.org/10.1007/978-1-4757-0818-9} {\emph {\bibinfo {booktitle}
  {Density Functional Methods in Physics}}},\ \bibinfo {editor} {edited by\
  \bibinfo {editor} {\bibfnamefont {R.~M.}\ \bibnamefont {Dreizler}}\ and\
  \bibinfo {editor} {\bibfnamefont {J.}~\bibnamefont {da~Providencia}}}\
  (\bibinfo  {publisher} {Plenum},\ \bibinfo {address} {New York},\ \bibinfo
  {year} {1985})\ pp.\ \bibinfo {pages} {265--308}\BibitemShut {NoStop}%
\bibitem [{\citenamefont {Sham}(1985)}]{sham1985}%
  \BibitemOpen
  \bibfield  {author} {\bibinfo {author} {\bibfnamefont {L.~J.}\ \bibnamefont
  {Sham}},\ }\href {https://doi.org/10.1103/PhysRevB.32.3876} {\bibfield
  {journal} {\bibinfo  {journal} {Phys. Rev. B}\ }\textbf {\bibinfo {volume}
  {32}},\ \bibinfo {pages} {3876} (\bibinfo {year} {1985})}\BibitemShut
  {NoStop}%
\bibitem [{\citenamefont {Almbladh}\ and\ \citenamefont {von
  Barth}(1985)}]{almbladh1985}%
  \BibitemOpen
  \bibfield  {author} {\bibinfo {author} {\bibfnamefont {C.-O.}\ \bibnamefont
  {Almbladh}}\ and\ \bibinfo {author} {\bibfnamefont {U.}~\bibnamefont {von
  Barth}},\ }\href {https://doi.org/10.1103/PhysRevB.31.3231} {\bibfield
  {journal} {\bibinfo  {journal} {Phys. Rev. B}\ }\textbf {\bibinfo {volume}
  {31}},\ \bibinfo {pages} {3231} (\bibinfo {year} {1985})}\BibitemShut
  {NoStop}%
\bibitem [{\citenamefont {Song}\ \emph
  {et~al.}(2022{\natexlab{b}})\citenamefont {Song}, \citenamefont {Vuckovic},
  \citenamefont {Sim},\ and\ \citenamefont {Burke}}]{song2022}%
  \BibitemOpen
  \bibfield  {author} {\bibinfo {author} {\bibfnamefont {S.}~\bibnamefont
  {Song}}, \bibinfo {author} {\bibfnamefont {S.}~\bibnamefont {Vuckovic}},
  \bibinfo {author} {\bibfnamefont {E.}~\bibnamefont {Sim}},\ and\ \bibinfo
  {author} {\bibfnamefont {K.}~\bibnamefont {Burke}},\ }\href
  {https://doi.org/10.1021/acs.jctc.1c01045} {\bibfield  {journal} {\bibinfo
  {journal} {J. Chem. Theory and Comput.}\ }\textbf {\bibinfo {volume} {18}},\
  \bibinfo {pages} {817} (\bibinfo {year} {2022}{\natexlab{b}})}\BibitemShut
  {NoStop}%
\bibitem [{\citenamefont {Shahi}\ \emph {et~al.}(2022)\citenamefont {Shahi},
  \citenamefont {Kanungo}, \citenamefont {Kaplan}, \citenamefont {Perdew},\
  and\ \citenamefont {Gavini}}]{shahi2022}%
  \BibitemOpen
  \bibfield  {author} {\bibinfo {author} {\bibfnamefont {C.}~\bibnamefont
  {Shahi}}, \bibinfo {author} {\bibfnamefont {B.}~\bibnamefont {Kanungo}},
  \bibinfo {author} {\bibfnamefont {A.~D.}\ \bibnamefont {Kaplan}}, \bibinfo
  {author} {\bibfnamefont {J.~P.}\ \bibnamefont {Perdew}},\ and\ \bibinfo
  {author} {\bibfnamefont {V.}~\bibnamefont {Gavini}}} (\bibinfo {year}
  {2022}),\ \bibinfo {note} {work in progress}\BibitemShut {NoStop}%
\end{thebibliography}%

\clearpage

\renewcommand{\thepage}{S\arabic{page}}
\renewcommand{\thesection}{S\arabic{section}}
\renewcommand{\theequation}{S\arabic{equation}}
\renewcommand{\thetable}{S\arabic{table}}
\renewcommand{\thefigure}{S\arabic{figure}}

\setcounter{page}{1}
\setcounter{section}{0}
\setcounter{equation}{0}
\setcounter{table}{0}
\setcounter{figure}{0}

\onecolumngrid
\part*{Supporting Information: Understanding density driven errors via reaction barrier heights}

\twocolumngrid
\tableofcontents

\section{Validation of Methodology}

\subsection{Effect of the self-consistent reference on functional-driven errors \label{sec:exact_v_prox}}

It is reasonable to ask whether using the self-consistent energy of a given proxy to compute the approximate functional-driven error
\begin{equation}
    \Delta E_\text{F} = E_\text{approx}[n_\text{proxy}] - E_\text{proxy}[n_\text{proxy}]
\end{equation}
rather than the accepted, high-level reference energies \cite{goerigk2017} makes meaningful changes to our analysis.
In this section, we use the same definition of approximate density-driven error
\begin{equation}
   \Delta E_\text{D} = E_\text{approx}[n_\text{approx}] - E_\text{approx}[n_\text{proxy}],
\end{equation}
but use the following definition of approximate functional-driven error:
\begin{equation}
    \Delta \widetilde{E}_\text{F} = E_\text{approx}[n_\text{proxy}] - E_\text{exact}[n_\text{exact}]. \label{eq:fde_no_prox_en}
\end{equation}
The exact $E_\text{exact}[n_\text{exact}]$ values are the high-level, correlated wavefunction values \cite{goerigk2017} used to compute errors in the reactions of BH76 and BH76RC.
With the choice of eq \ref{eq:fde_no_prox_en}, the functional and density driven error sum to the total error of a given approach.

\begin{ruledtabular}

\begin{table*}
    \centering
    \begin{tabular}{lrrrrrr}
 & \multicolumn{2}{c}{\lcwpbe{}} & \multicolumn{2}{c}{SX-0.5} & \multicolumn{2}{c}{SCAN-FLOSIC} \\ 
\textit{BH76} & MFE & MDE & MFE & MDE & MFE & MDE \\ \hline 
LSDA & -13.16 & -2.14 & -11.78 & -3.52 & -10.91 & -4.45 \\  
PBE & -7.70 & -1.18 & -6.59 & -2.29 & -5.62 & -3.66 \\  
BLYP & -6.80 & -1.25 & -5.77 & -2.28 &  &  \\  
SCAN & -6.85 & -0.58 & -6.28 & -1.16 & -5.76 & -1.97 \\  
\rrscan{} & -6.35 & -0.57 & -5.64 & -1.27 & -7.24 & 0.00 \\  
M06-L & -2.77 & -0.81 & -1.84 & -1.73 &  &  \\  
MN15-L & -0.32 & -0.54 & 0.53 & -1.40 &  &  \\  
B3LYP & -3.90 & -0.46 & -3.30 & -1.06 &  &  \\  
\lcwpbe{} &  &  & 0.99 & -0.39 &  &  \\  
LSDA@HF & -13.16 & 8.06 & -11.78 & 6.69 & -10.91 & 5.54 \\  
PBE@HF & -7.70 & 6.78 & -6.59 & 5.68 & -5.62 & 4.41 \\  
SCAN@HF & -6.85 & 5.15 & -6.28 & 4.58 & -5.76 & 3.81 \\  
\rrscan{}@HF & -6.35 & 5.25 & -5.64 & 4.54 & -7.24 & 5.88 \\ \hline 
 & \multicolumn{2}{c}{\lcwpbe{}} & \multicolumn{2}{c}{SX-0.5} & \multicolumn{2}{c}{SCAN-FLOSIC} \\ 
\textit{BH76RC} & MFE & MDE & MFE & MDE & MFE & MDE \\ \hline 
LSDA & -0.41 & 0.94 & -0.40 & 0.93 & -0.82 & 0.26 \\  
PBE & 0.97 & 0.08 & 0.94 & 0.10 & -0.02 & 0.73 \\  
BLYP & 0.74 & 0.04 & 0.77 & 0.00 &  &  \\  
SCAN & -0.11 & 0.05 & -0.21 & 0.15 & -0.50 & 0.19 \\  
\rrscan{} & -0.00 & 0.07 & -0.10 & 0.16 & -0.19 & 0.00 \\  
M06-L & 1.54 & 0.04 & 1.45 & 0.14 &  &  \\  
MN15-L & 1.28 & -0.09 & 1.15 & 0.04 &  &  \\  
B3LYP & -0.20 & 0.00 & -0.15 & -0.05 &  &  \\  
\lcwpbe{} &  &  & -0.53 & -0.01 &  &  \\  
LSDA@HF & -0.41 & -0.13 & -0.40 & -0.14 & -0.82 & -0.08 \\  
PBE@HF & 0.97 & -0.26 & 0.94 & -0.24 & -0.02 & 0.44 \\  
SCAN@HF & -0.11 & -0.39 & -0.21 & -0.29 & -0.50 & -0.23 \\  
\rrscan{}@HF & -0.00 & -0.41 & -0.10 & -0.32 & -0.19 & -0.45 \\
    \end{tabular}
    \caption{Mean functional-driven errors (MFEs) and mean density-driven errors (MDEs) for the reaction energy differences in the BH76 and BH76RC sets.
    All calculations used the aug-cc-pVQZ basis set \cite{dunning1989}, except for the SCAN-FLOSIC calculations, which used the NRLMOL basis set \cite{porezag1999} in a Cartesian representation.
    Unlike Table \ref{tab:bh76_fde_dde_dE} of the main text, the functional-driven errors are computed with the ``exact'' values \cite{goerigk2017} used to compute deviations in the reactions of BH76 and BH76RC, as in Eq. \ref{eq:fde_no_prox_en}.
    Thus the sum of MFE and MDE for a fixed proxy would yield the MD for a given set listed in Table \ref{tab:ak_pyscf_aug-cc-pvqz} for the \lcwpbe{} and SX-0.5 columns, and Table \ref{tab:ak_pyscf_nrlmol} for the SCAN-FLOSIC column.
    }
    \label{tab:bh76_fde_dde_sm}
\end{table*}

\end{ruledtabular}

Table \ref{tab:bh76_fde_dde_sm} presents this analysis.
The differences in mean functional-driven error between Table \ref{tab:bh76_fde_dde_dE} (computed with the self-consistent proxy energy) and those in Table \ref{tab:bh76_fde_dde_sm} (computed with the exact values) generally lie within 1 kcal/mol for BH76, and within 2 kcal/mol for BH76RC.
Thus we view our analysis as robust.

\subsection{Removing reactions with large discrepancies between proxies}

It is reasonable to assert that one should only consider reactions where the three proxies agree within some threshold.
Otherwise, one might include reactions where (at least) one proxy produces an unrealistic density, total energy, or both.
Let the uncertainty in the functional- and density-driven errors of reaction $j$ be
\begin{align}
    \delta E_\text{D}^{(j)} = \max_\text{proxies}( \Delta E_\text{D}^{(j)}) - \min_\text{proxies}( \Delta E_\text{D}^{(j)}) \label{eq:dde_ucrt} \\
    \delta \widetilde{E}_\text{F}^{(j)} = \max_\text{proxies}( \Delta \widetilde{E}_\text{F}^{(j)}) - \min_\text{proxies}( \Delta \widetilde{E}_\text{F}^{(j)}) \label{eq:fde_ucrt}
\end{align}
where $\Delta E_\text{D}^{(j)}$ is the density-driven error of reaction $j$, and the maxima and minima are taken over the set of proxies.
In words, for a fixed reaction, we compute the uncertainty as the difference between the largest approximate density- (functional-)driven error and the smallest.
Note that eq \ref{eq:fde_no_prox_en} is used in these calculations.

\begin{table*}
    \centering
    \begin{tabular}{lrrrrrrr}\hline
 & \multicolumn{7}{c}{BH76} \\ 
DFA & $\overline{\Delta E_\mathrm{F}}$ & $\overline{\Delta E_\mathrm{D}}$  & $\delta \widetilde{E}_\mathrm{F}$ & $\delta E_\mathrm{D}$  & $N_\mathrm{cut}^{-1} \sum_{j=1}^\mathrm{N_\mathrm{cut}} \delta \widetilde{E}_\mathrm{F}^{(j)}$ & $N_\mathrm{cut}^{-1} \sum_{j=1}^\mathrm{N_\mathrm{cut}} \delta E_\mathrm{D}^{(j)}$& No. passed \\ \hline 
LSDA & -9.47 & -1.89 & 1.19 & 1.12 & 1.08 & 0.99 & 35 \\  
PBE & -5.14 & -1.41 & 1.07 & 1.01 & 0.96 & 0.87 & 39 \\  
BLYP & -5.87 & -1.53 & 0.96 & 0.96 & 0.77 & 0.77 & 70 \\  
SCAN & -5.48 & -0.81 & 1.01 & 0.83 & 0.85 & 0.68 & 55 \\  
\rrscan{} & -5.80 & -0.41 & 1.12 & 0.96 & 1.00 & 0.89 & 54 \\  
M06-L & -2.14 & -1.21 & 0.92 & 0.92 & 0.79 & 0.79 & 72 \\  
MN15-L & 0.23 & -0.86 & 0.84 & 0.84 & 0.68 & 0.68 & 73 \\  
B3LYP & -3.53 & -0.70 & 0.66 & 0.66 & 0.52 & 0.52 & 74 \\  
\lcwpbe{} & 0.83 & -0.15 & 0.42 & 0.42 & 0.32 & 0.32 & 74 \\  
LSDA@HF & -10.00 & 4.53 & 1.16 & 1.14 & 1.06 & 1.01 & 38 \\  
PBE@HF & -5.28 & 4.12 & 1.10 & 1.08 & 0.98 & 0.94 & 40 \\  
SCAN@HF & -5.51 & 3.62 & 1.02 & 0.89 & 0.86 & 0.74 & 56 \\  
\rrscan{}@HF & -5.80 & 3.43 & 1.12 & 0.95 & 1.00 & 0.88 & 54 \\ \hline 
 & \multicolumn{7}{c}{BH76RC} \\ 
DFA & $\overline{\Delta E_\mathrm{F}}$ & $\overline{\Delta E_\mathrm{D}}$  & $\delta \widetilde{E}_\mathrm{F}$ & $\delta E_\mathrm{D}$  & $N_\mathrm{cut}^{-1} \sum_{j=1}^\mathrm{N_\mathrm{cut}} \delta \widetilde{E}_\mathrm{F}^{(j)}$ & $N_\mathrm{cut}^{-1} \sum_{j=1}^\mathrm{N_\mathrm{cut}} \delta E_\mathrm{D}^{(j)}$& No. passed \\ \hline 
LSDA & -2.01 & 0.03 & 0.99 & 0.73 & 0.81 & 0.64 & 22 \\  
PBE & 0.48 & 0.03 & 0.75 & 0.44 & 0.53 & 0.36 & 25 \\  
BLYP & 0.75 & 0.02 & 0.49 & 0.49 & 0.35 & 0.35 & 30 \\  
SCAN & -0.50 & 0.05 & 0.78 & 0.36 & 0.58 & 0.27 & 29 \\  
\rrscan{} & -0.10 & 0.08 & 0.76 & 0.53 & 0.61 & 0.34 & 30 \\  
M06-L & 1.49 & 0.09 & 0.56 & 0.56 & 0.48 & 0.48 & 30 \\  
MN15-L & 1.22 & -0.03 & 0.40 & 0.40 & 0.33 & 0.33 & 30 \\  
B3LYP & -0.17 & -0.02 & 0.31 & 0.31 & 0.24 & 0.24 & 30 \\  
\lcwpbe{} & -0.53 & -0.01 & 0.20 & 0.20 & 0.15 & 0.15 & 30 \\  
LSDA@HF & -1.00 & -0.30 & 1.02 & 0.80 & 0.85 & 0.70 & 27 \\  
PBE@HF & 0.48 & -0.33 & 0.75 & 0.51 & 0.53 & 0.39 & 25 \\  
SCAN@HF & -0.50 & -0.16 & 0.78 & 0.46 & 0.58 & 0.35 & 29 \\  
\rrscan{}@HF & -0.10 & -0.39 & 0.76 & 0.53 & 0.61 & 0.36 & 30 \\ \hline 
    \end{tabular}
    \caption{Mean functional errors, $\overline{\Delta E_\mathrm{F}}$, and $\overline{\Delta E_\mathrm{D}}$ of Eqs. \ref{eq:fde_avg} and \ref{eq:dde_avg} (second and third columns). 
    These are computed by only considering reactions where the uncertainties of a given reaction, defined in Eqs. \ref{eq:dde_ucrt} and \ref{eq:fde_ucrt}, lie within a cutoff, chosen here to be 2 kcal/mol cutoff.
    The total uncertainties $\delta E_\text{D}$ and $\delta \widetilde E_\text{F}$ are defined in Eqs. \ref{eq:dde_ucrt_tot} and \ref{eq:fde_ucrt_tot}, respectively (fourth and fifth columns).
    Simple averages of the individual uncertainties, as a comparison of the quadrature sum more typical of experimental uncertainty measurement, are given in the sixth and seventh columns.
    ``No. passed'' are the number of reactions that lie within the cutoff out of 76 total for BH76 and 30 for BH76RC.
    Although the density-driven error computed for \lcwpbe{} with \lcwpbe{} as a proxy is identically zero (the functional-driven error computed using Eq. \ref{eq:fde_no_prox_en} is generally nonzero), we include \lcwpbe{} for the possible proxies of \lcwpbe{} to obtain error statistics.
    }
    \label{tab:cut_dc_dft}
\end{table*}

\begin{table*}
    \centering
    \begin{tabular}{lrrrrrrr}\hline
 & \multicolumn{7}{c}{BH76} \\ 
DFA & $\overline{\Delta E_\mathrm{F}}$ & $\overline{\Delta E_\mathrm{D}}$  & $\delta \widetilde{E}_\mathrm{F}$ & $\delta E_\mathrm{D}$  & $N_\mathrm{cut}^{-1} \sum_{j=1}^\mathrm{N_\mathrm{cut}} \delta \widetilde{E}_\mathrm{F}^{(j)}$ & $N_\mathrm{cut}^{-1} \sum_{j=1}^\mathrm{N_\mathrm{cut}} \delta E_\mathrm{D}^{(j)}$& No. passed \\ \hline 
LSDA & -11.29 & -2.29 & 0.93 & 0.93 & 0.78 & 0.78 & 60 \\  
PBE & -6.71 & -1.48 & 0.98 & 0.98 & 0.80 & 0.80 & 69 \\  
BLYP & -5.87 & -1.53 & 0.96 & 0.96 & 0.77 & 0.77 & 70 \\  
SCAN & -6.35 & -0.81 & 0.59 & 0.59 & 0.47 & 0.47 & 72 \\  
\rrscan{} & -5.84 & -0.84 & 0.71 & 0.71 & 0.56 & 0.56 & 72 \\  
M06-L & -2.14 & -1.21 & 0.92 & 0.92 & 0.79 & 0.79 & 72 \\  
MN15-L & 0.23 & -0.86 & 0.84 & 0.84 & 0.68 & 0.68 & 73 \\  
B3LYP & -3.53 & -0.70 & 0.66 & 0.66 & 0.52 & 0.52 & 74 \\  
\lcwpbe{} & 0.83 & -0.15 & 0.42 & 0.42 & 0.32 & 0.32 & 74 \\  
LSDA@HF & -11.29 & 5.35 & 0.93 & 0.93 & 0.78 & 0.78 & 60 \\  
PBE@HF & -6.71 & 5.21 & 0.98 & 0.98 & 0.80 & 0.80 & 69 \\  
SCAN@HF & -6.35 & 4.30 & 0.59 & 0.59 & 0.47 & 0.47 & 72 \\  
\rrscan{}@HF & -5.84 & 4.31 & 0.71 & 0.71 & 0.56 & 0.56 & 72 \\ \hline 
 & \multicolumn{7}{c}{BH76RC} \\ 
DFA & $\overline{\Delta E_\mathrm{F}}$ & $\overline{\Delta E_\mathrm{D}}$  & $\delta \widetilde{E}_\mathrm{F}$ & $\delta E_\mathrm{D}$  & $N_\mathrm{cut}^{-1} \sum_{j=1}^\mathrm{N_\mathrm{cut}} \delta \widetilde{E}_\mathrm{F}^{(j)}$ & $N_\mathrm{cut}^{-1} \sum_{j=1}^\mathrm{N_\mathrm{cut}} \delta E_\mathrm{D}^{(j)}$& No. passed \\ \hline 
LSDA & -0.41 & 0.93 & 0.51 & 0.51 & 0.38 & 0.38 & 30 \\  
PBE & 0.96 & 0.09 & 0.41 & 0.41 & 0.28 & 0.28 & 30 \\  
BLYP & 0.75 & 0.02 & 0.49 & 0.49 & 0.35 & 0.35 & 30 \\  
SCAN & -0.16 & 0.10 & 0.26 & 0.26 & 0.16 & 0.16 & 30 \\  
\rrscan{} & -0.05 & 0.11 & 0.28 & 0.28 & 0.18 & 0.18 & 30 \\  
M06-L & 1.49 & 0.09 & 0.56 & 0.56 & 0.48 & 0.48 & 30 \\  
MN15-L & 1.22 & -0.03 & 0.40 & 0.40 & 0.33 & 0.33 & 30 \\  
B3LYP & -0.17 & -0.02 & 0.31 & 0.31 & 0.24 & 0.24 & 30 \\  
\lcwpbe{} & -0.53 & -0.01 & 0.20 & 0.20 & 0.15 & 0.15 & 30 \\  
LSDA@HF & -0.41 & -0.13 & 0.51 & 0.51 & 0.38 & 0.38 & 30 \\  
PBE@HF & 0.96 & -0.25 & 0.41 & 0.41 & 0.28 & 0.28 & 30 \\  
SCAN@HF & -0.16 & -0.34 & 0.26 & 0.26 & 0.16 & 0.16 & 30 \\  
\rrscan{}@HF & -0.05 & -0.36 & 0.28 & 0.28 & 0.18 & 0.18 & 30 \\ \hline 
    \end{tabular}
    \caption{Same as Table \ref{tab:cut_dc_dft}, but excluding SCAN-FLOSIC as a proxy for LSDA, PBE, SCAN, \rrscan{}, and SCAN@HF.
    A full list of reactions that failed can be found on the data repository \cite{data_repo}.
    We note thatreaction barriers 9, 10, 49, 50, 55, 56, and 65 failed the cutoff test for both LSDA and PBE.
    Reaction barriers 9, 10, 55, and 56 also failed the cutoff test for SCAN, \rrscan{}, and SCAN@HF.
    }
    \label{tab:cut_dc_dft_no_flo}
\end{table*}

The values in Table \ref{tab:cut_dc_dft} are computed for all reactions such that
\begin{equation}
    \delta E_\text{D}^{(j)} < \epsilon \quad \text{and} \quad \delta \widetilde{E}_\text{F}^{(j)} < \epsilon,
\end{equation}
with $\epsilon=2$ kcal/mol here.
For a reaction satisfying this cutoff, we compute the average density- and functional-driven error \textit{for a fixed reaction} $j$
\begin{align}
    \overline{\Delta E_\text{D}}^{(j)} &= \frac{1}{N_\text{proxy}} \sum_\text{proxies} \Delta E_\text{D}^{(j)} \label{eq:dde_avg} \\
    \overline{\Delta E_\text{F}}^{(j)} &= \frac{1}{N_\text{proxy}} \sum_\text{proxies} \Delta \widetilde E_\text{F}^{(j)}. \label{eq:fde_avg}
\end{align}
$N_\text{proxy}$ is the total number of proxies, 3 in the current work.
Suppose there are $N_\text{cut}$ reactions satisfying this cutoff.
We then average these values satisfying the cutoff, and estimate the total uncertainty as the quadrature sum
\begin{align}
    \delta E_\text{D} &=  \left[\frac{1}{N_\text{cut}} \sum_{j=1}^{N_\text{cut}} \left( \delta E_\text{D}^{(j)}\right)^2 \right]^{1/2} \label{eq:dde_ucrt_tot} \\
    \delta \widetilde{E}_\text{F} &=  \left[\frac{1}{N_\text{cut}} \sum_{j=1}^{N_\text{cut}} \left( \delta \widetilde{E}_\text{F}^{(j)}\right)^2 \right]^{1/2}. \label{eq:fde_ucrt_tot}
\end{align}
The mean density-driven and functional driven-errors are, likewise, averages over the values in Eqs. \ref{eq:dde_avg} and \ref{eq:fde_avg}:
\begin{align}
    \overline{\Delta E_\text{D}} &= \frac{1}{N_\text{cut}} \sum_{j=1}^{N_\text{cut}}  \overline{\Delta E_\text{D}}^{(j)} \label{eq:mde_cut} \\
    \overline{\Delta E_\text{F}} &= \frac{1}{N_\text{cut}} \sum_{j=1}^{N_\text{cut}}  \overline{\Delta E_\text{F}}^{(j)}. \label{eq:mfe_cut}
\end{align}
The values in Table \ref{tab:cut_dc_dft} confirm our earlier conclusions, another validation of our methodology.
By excluding SCAN-FLOSIC as a possible proxy in Table \ref{tab:cut_dc_dft_no_flo}, more reactions pass the cutoff test.
For all DFAs in Table \ref{tab:cut_dc_dft_no_flo}, reaction barriers 9 and 10 (F$_2$ + H $\leftrightarrow$ F + HF), and 55 and 56 (F + H$_2$ $\leftrightarrow$ H + HF) failed to reach the uncertainty cutoff.

It should be noted that reactions 9 and 10, the forwards and reverse reactions of F$_2$ + H $\rightarrow$ F + HF, failed to meet the uncertainty cutoff for every DFA in Table \ref{tab:cut_dc_dft}.
The Hartree-Fock density for their common transition state is heavily spin-contaminated, with $\langle S^2 \rangle = 1.213$ ($s \approx 0.710$).
The Hartree-Fock density for atomic F is modestly spin-contaminated, with $\langle S^2 \rangle = 0.754$ ($s \approx 0.502$).
All other systems in that reaction were treated as closed-shell, and spin symmetry was not permitted to break.

\subsection{Replacing the exact energies with proxy values to compute reaction errors}

We also compute the approximate errors (deviations) in the barrier heights and reaction energies of BH76 and BH76RC, respectively, using the energies of the proxies in lieu of the exact values.
This is analogous to Table \ref{tab:bh76_fde_dde_dE}, whereby the self-consistent energies and densities of the exact functional are replaced by those of a proxy.
If the approximate errors, at least on the average, computed with the proxy energies lie close to those computed with the exact values, then a proxy is reasonably faithful to the exact functional.

\begin{ruledtabular}
    \begin{table*}[h]
        \centering
        \begin{tabular}{l *{3}{|rrrR{1.5cm}}}
BH76 & \multicolumn{4}{c}{MD} & \multicolumn{4}{c}{MAD} & \multicolumn{4}{c}{RMSD} \\ 
 & Exact & \lcwpbe{} & SX-0.5 & SCAN-FLOSIC & Exact & \lcwpbe{} & SX-0.5 & SCAN-FLOSIC & Exact & \lcwpbe{} & SX-0.5 & SCAN-FLOSIC \\ \hline 
LSDA & -15.30 & -15.90 & -15.13 & -15.98 & 15.39 & 15.90 & 15.16 & 16.16 & 17.84 & 17.79 & 18.16 & 19.48 \\  
PBE & -8.88 & -9.48 & -8.71 & -9.89 & 8.93 & 9.48 & 8.80 & 10.18 & 10.26 & 10.44 & 10.96 & 12.83 \\  
BLYP & -8.05 & -8.65 & -7.88 &  & 8.06 & 8.65 & 7.99 &  & 9.28 & 9.66 & 9.84 &  \\  
SCAN & -7.44 & -8.04 & -7.27 & -8.35 & 7.50 & 8.04 & 7.27 & 8.46 & 8.22 & 8.61 & 8.47 & 10.41 \\  
\rrscan{} & -6.91 & -7.52 & -6.75 & -7.85 & 6.96 & 7.52 & 6.75 & 8.06 & 7.75 & 8.10 & 8.09 & 10.23 \\  
M06-L & -3.58 & -4.18 & -3.41 &  & 3.84 & 4.29 & 4.88 &  & 4.86 & 5.36 & 6.07 &  \\  
MN15-L & -0.87 & -1.47 & -0.70 &  & 1.80 & 2.34 & 2.90 &  & 2.65 & 3.29 & 4.12 &  \\  
B3LYP & -4.35 & -4.96 & -4.19 &  & 4.41 & 4.97 & 4.41 &  & 5.24 & 5.65 & 5.57 &  \\  
\lcwpbe{} & 0.60 &  & 0.77 &  & 1.87 &  & 2.60 &  & 2.49 &  & 3.41 &  \\  
SCAN@HF & -1.70 & -2.30 & -1.53 & -2.57 & 3.05 & 3.17 & 2.64 & 4.57 & 4.03 & 4.22 & 3.77 & 7.10 \\ \hline 
BH76RC & \multicolumn{4}{c}{MD} & \multicolumn{4}{c}{MAD} & \multicolumn{4}{c}{RMSD} \\ 
 & Exact & \lcwpbe{} & SX-0.5 & SCAN-FLOSIC & Exact & \lcwpbe{} & SX-0.5 & SCAN-FLOSIC & Exact & \lcwpbe{} & SX-0.5 & SCAN-FLOSIC \\ \hline 
LSDA & 0.53 & 1.07 & 2.22 & -0.19 & 8.72 & 7.22 & 10.54 & 10.97 & 11.24 & 9.46 & 14.35 & 16.08 \\  
PBE & 1.04 & 1.58 & 2.74 & 1.09 & 4.09 & 3.00 & 6.81 & 8.12 & 6.00 & 4.51 & 9.91 & 12.08 \\  
BLYP & 0.77 & 1.31 & 2.47 &  & 3.26 & 3.15 & 5.53 &  & 4.35 & 4.05 & 8.07 &  \\  
SCAN & -0.06 & 0.48 & 1.64 & 0.07 & 3.12 & 2.35 & 4.07 & 5.79 & 4.18 & 2.87 & 5.96 & 8.53 \\  
\rrscan{} & 0.06 & 0.60 & 1.76 & 0.19 & 2.98 & 2.32 & 4.38 & 6.27 & 4.03 & 2.80 & 6.42 & 9.16 \\  
M06-L & 1.58 & 2.12 & 3.28 &  & 2.77 & 3.06 & 5.02 &  & 4.16 & 3.81 & 7.50 &  \\  
MN15-L & 1.19 & 1.73 & 2.88 &  & 2.34 & 3.21 & 4.08 &  & 3.14 & 3.93 & 6.39 &  \\  
B3LYP & -0.20 & 0.34 & 1.50 &  & 2.07 & 2.28 & 3.61 &  & 2.66 & 3.01 & 5.21 &  \\  
\lcwpbe{} & -0.54 &  & 1.16 &  & 2.19 &  & 4.37 &  & 2.74 &  & 5.99 &  \\  
SCAN@HF & -0.50 & 0.04 & 1.19 & -0.35 & 2.70 & 2.44 & 3.41 & 5.53 & 3.40 & 2.90 & 4.64 & 7.56 \\  
        \end{tabular}
        \caption{Mean deviations (MDs), mean absolute deviations (MADs), and root-mean-squared deviations (RMSDs) computed using the exact GMTKN values \cite{goerigk2017}, and using each of the proxy funcitonal's computed reaction energies.
        If a functional is to be a good proxy for the exact functional, the deviations computed using either the exact reference values or an approximate reference values should agree within some tolerance.
        The ``Exact'' columns are the values in Table \ref{tab:ak_pyscf_aug-cc-pvqz}.
        }
        \label{tab:sm_devs_summ}
    \end{table*}
\end{ruledtabular}

Table \ref{tab:sm_devs_summ} shows that these average approximate errors generally lie within 1-2 kcal/mol for BH76.
SCAN-FLOSIC appears to be the least reliable proxy from this analysis.
The situation for BH76RC is less clear - many of the MADs and RMSDs differ by 4 kcal/mol or more.

\subsection{Sensitivity to input density}

Recall that the density sensitivity defined in Ref. \cite{sim2018} is
\begin{equation}
    S = |E_\text{approx}[n_\text{LSDA}] - E_\text{approx}[n_\text{HF}] |.
\end{equation}
We define the density \textit{variability} $\mathcal{V}$ as
\begin{equation}
    \mathcal{V} = E_\text{approx}[n_\text{HF}] - E_\text{approx}[n_\text{ref}] , \label{eq:dens_var}
\end{equation}
a metric that shows how localized a reference density is relative to the Hartree-Fock density.
Table \ref{tab:sm_dens_var} tabulates averages of this quantity using the three proxies, and for the non-empirical functionals LSDA, PBE, and SCAN.
For BH76, density variability decreases from LSDA to PBE, and further from PBE to SCAN
For BH76RC, density variability increases from LSDA to PBE

\begin{table}
    \centering
    \begin{tabular}{l|rr|rr|rr} \hline
BH76 & \multicolumn{2}{c}{\lcwpbe{}} & \multicolumn{2}{c}{SX-0.5} & \multicolumn{2}{c}{SCAN-FLOSIC} \\
DFA & Mean & Mean Abs. & Mean & Mean Abs. & Mean & Mean Abs. \\ \hline
LSDA & 8.06 & 8.12 & 6.69 & 6.77 & 5.54 & 5.99 \\
PBE & 6.78 & 6.83 & 5.68 & 5.75 & 4.41 & 5.38 \\
SCAN & 5.15 & 5.18 & 4.58 & 4.63 & 3.81 & 4.12 \\ \hline
BH76RC & \multicolumn{2}{c}{\lcwpbe{}} & \multicolumn{2}{c}{SX-0.5} & \multicolumn{2}{c}{SCAN-FLOSIC} \\
DFA & Mean & Mean Abs. & Mean & Mean Abs. & Mean & Mean Abs. \\ \hline
LSDA & -0.13 & 1.60 & -0.14 & 1.33 & -0.08 & 1.09 \\
PBE & -0.26 & 1.33 & -0.24 & 1.14 & 0.44 & 1.47 \\
SCAN & -0.39 & 0.98 & -0.29 & 0.88 & -0.23 & 0.73 \\ \hline
    \end{tabular}
    \caption{Average signed and absolute density variabilities according to eq \ref{eq:dens_var}.
    The values in the SCAN row and ``Mean'' columns correspond to the SCAN@HF MDE values in Table \ref{tab:bh76_fde_dde_dE}.
    }
    \label{tab:sm_dens_var}
\end{table}

\onecolumngrid

\subsection{Proxy statistical correlation coefficients}

This section demonstrates the consistency of the proxies used for estimating functional and density driven errors using concepts from statistics.

We being by reviewing relevant concepts and terminology.
Define the (non-normalized) covariance matrix of two variables $u$ and $v$ as
\begin{align}
    \sigma_{uv} &= \frac{1}{M} \sum_{i=1}^M \left[u_i - \overline{u} \right] \left[v_i - \overline{v} \right]
    = \frac{1}{M} \sum_{i=1}^M u_i v_i - \overline{u}\, \overline{v},
\end{align}
with means
\begin{equation}
    \overline{u} = \frac{1}{M} \sum_{i=1}^M u_i.
\end{equation}
Then a normalized correlation coefficient can be defined as
\begin{equation}
    \tau_{uv} = \left[\sigma_{uu}\sigma_{vv} \right]^{-1/2} \sigma_{uv},
\end{equation}
with range $-1 \leq \tau_{uv} \leq 1$.
Both $\sigma_{uv}$ and $\tau_{uv}$ are symmetric matrices.
Perfectly correlated variables have $\tau_{uv}=1$ (and $\tau_{uu}=1$ by definition), and perfectly uncorrelated variables have $\tau_{uv}=0$.
Perfectly ``anti-correlated'' variables would have $\tau_{uv}=-1$.

Using the correlation coefficient, we wish to show that the proxies used here, \lcwpbe{}, SX-0.5, and SCAN-FLOSIC, have correlation coefficients that are greater than using another proxy density.
Thus in this section, we will also evaluate functional driven errors using the exact energies, as in Sec. \ref{sec:exact_v_prox}.
This is a basic test that will show if the proxies are \textit{more} consistent with each other than with any other functional considered here.

Because we selected SCAN-FLOSIC as a proxy, we can evaluate correlation coefficients only for a handful of DFAs.
We thus use the functional- and density-driven errors of LSDA, PBE, SCAN, and \rrscan{} to evaluate correlation coefficients.

\begin{figure}
    \centering
    \includegraphics[width=0.8\columnwidth]{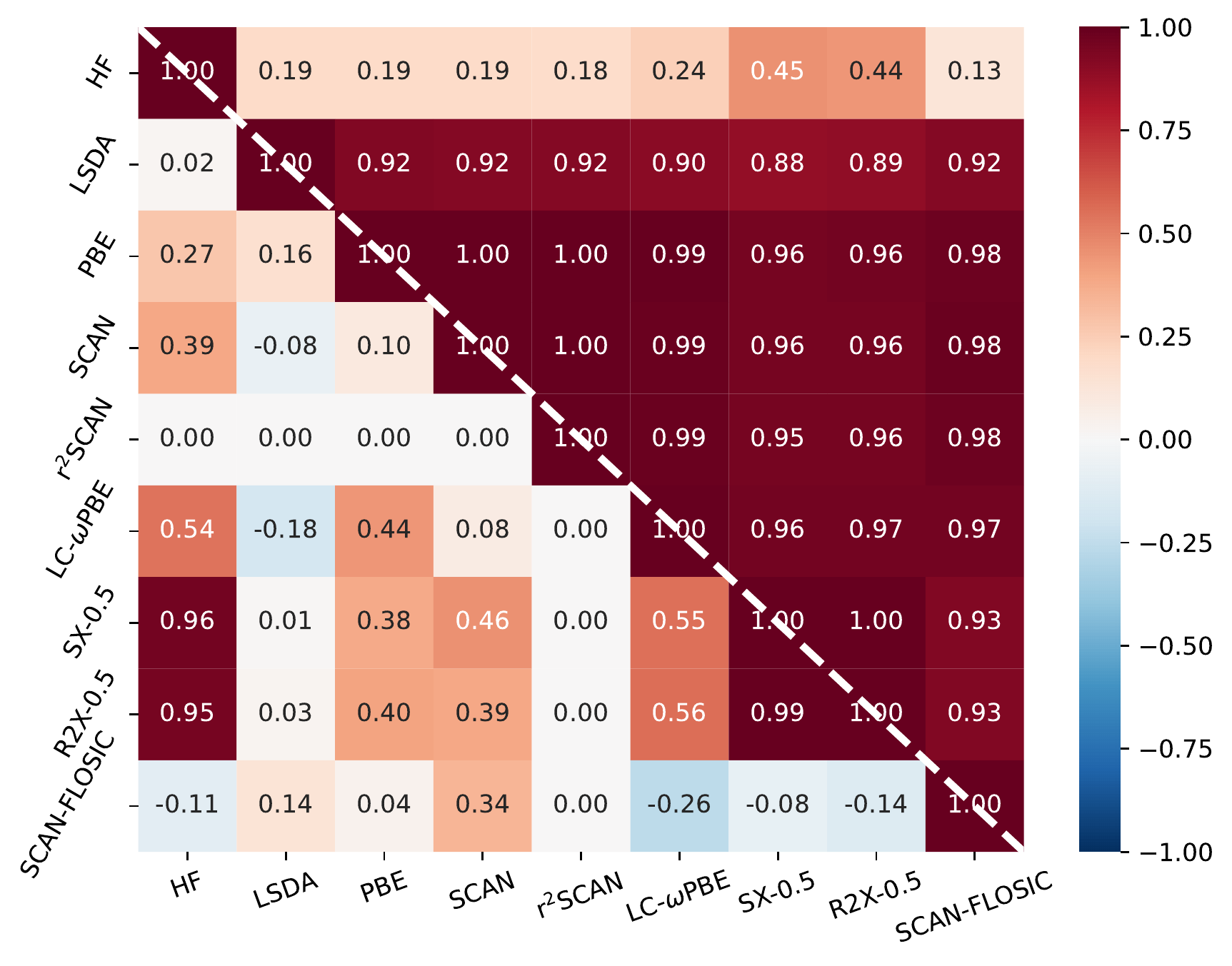}
    \caption{\rrscan{} correlation coefficients in the functional- (upper triangular matrix) and density-driven errors (lower triangular matrix).
    To save space, and using the symmetric property of the correlation coefficient matrix, we present the two sets of error statistics in this form.
    }
    \label{fig:r2SCAN_stat_corr}
\end{figure}

Each column and row of figs \ref{eq:LSDA_stat_corr}--\ref{fig:r2SCAN_stat_corr} labels which functional is used to evaluate functional- (upper triangular half of the matrix) and density-driven errors (lower triangular half).
The scales of each figure are identical, such that darker red tones indicate greater statistical correlation, darker blue tones indicate greater statistical anti-correlation, and white indicates no correlation.
In cases where, e.g., LSDA is used to evaluate density-driven errors of LSDA, the matrix elements are identically zero and thus not statistically meaningful.

\begin{figure}
    \centering
    \includegraphics[width=0.8\columnwidth]{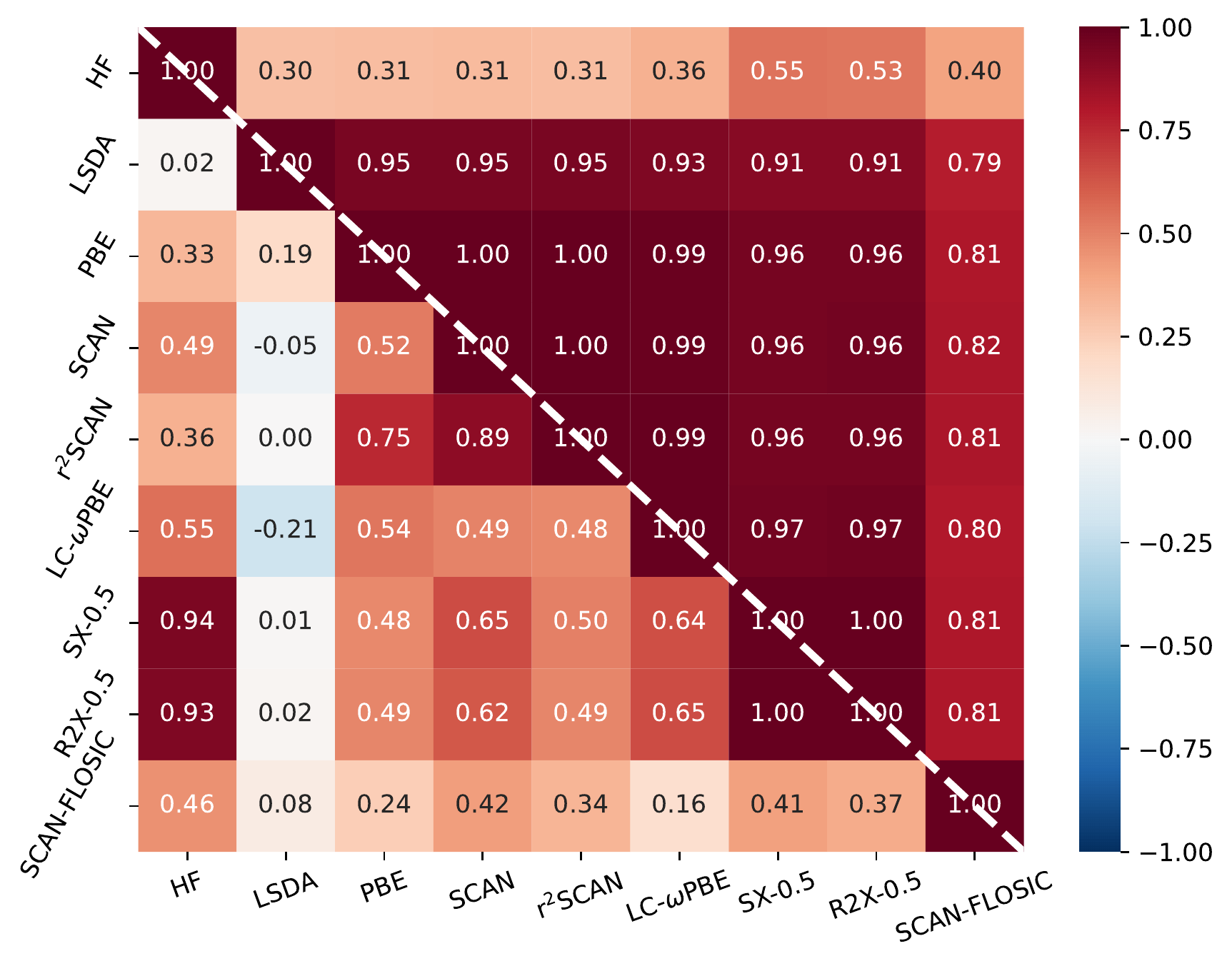}
    \caption{Correlation coefficients averaged over the four ``target'' DFAs: LSDA, PBE, SCAN, and \rrscan{}.
    The functional-driven error correlation coefficients are given in the upper triangular half-matrix, and density-driven errors in the lower triangular half-matrix (to save space).
    Results which are not statistically meaningful, such as the correlation coefficients in density-driven errors using the same functional for the target and proxy, are excluded from the average.
    Each row and column label indicates which functionals are used as proxies for the exact functional.
    Figure \ref{fig:LSDA_stat_corr} shows the statistical correlations between LSDA@DFA$_u$ and LSDA@DFA$_v$ functional- and density-driven errors, where $u$ and $v$ are identified by the row and column labels.
    The same information is presented for PBE, SCAN, and \rrscan{} in Figures \ref{fig:PBE_stat_corr}--\ref{fig:r2SCAN_stat_corr}, and is averaged here over the results in Figures \ref{fig:LSDA_stat_corr}--\ref{fig:r2SCAN_stat_corr}.
    }
    \label{fig:avg_stat_corr}
\end{figure}

Figure \ref{fig:avg_stat_corr}, which averages over the four DFAs (excluding the statistically meaningless quantities) is perhaps the most telling figure.
Essentially all densities used here, except Hartree-Fock (HF) and SCAN-FLOSIC, give consistent functional-driven errors.
For evaluating density-driven errors, \lcwpbe{} is most correlated with SX-0.5 or its \rrscan{} counterpart, R2X-0.5.
Both 50\% global hybrids are most correlated with the HF density, as might be expected from the high fraction of exact exchange admixture.
For evaluating density-driven errors, none of the proxies are statistically correlated with the SCAN-FLOSIC density.
Thus \lcwpbe{}, SX-0.5, and R2X-0.5 are likely the most reliable proxies in this set.

\begin{figure}
    \centering
    \includegraphics[width=0.8\columnwidth]{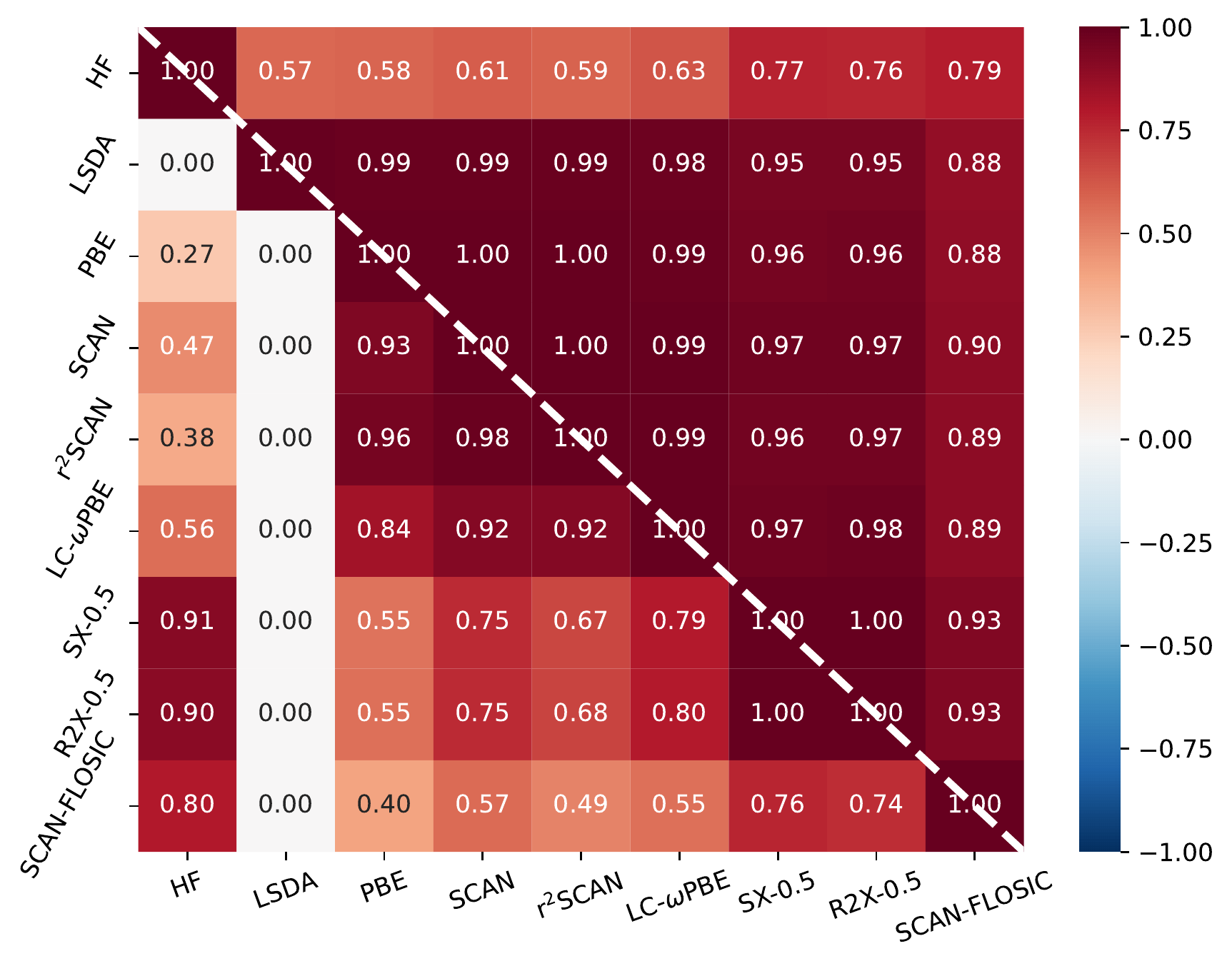}
    \caption{LSDA correlation coefficients in the functional- (upper triangular matrix) and density-driven errors (lower triangular matrix).
    To save space, and using the symmetric property of the correlation coefficient matrix, we present the two sets of error statistics in this form.
    }
    \label{fig:LSDA_stat_corr}
\end{figure}

\begin{figure}
    \centering
    \includegraphics[width=0.8\columnwidth]{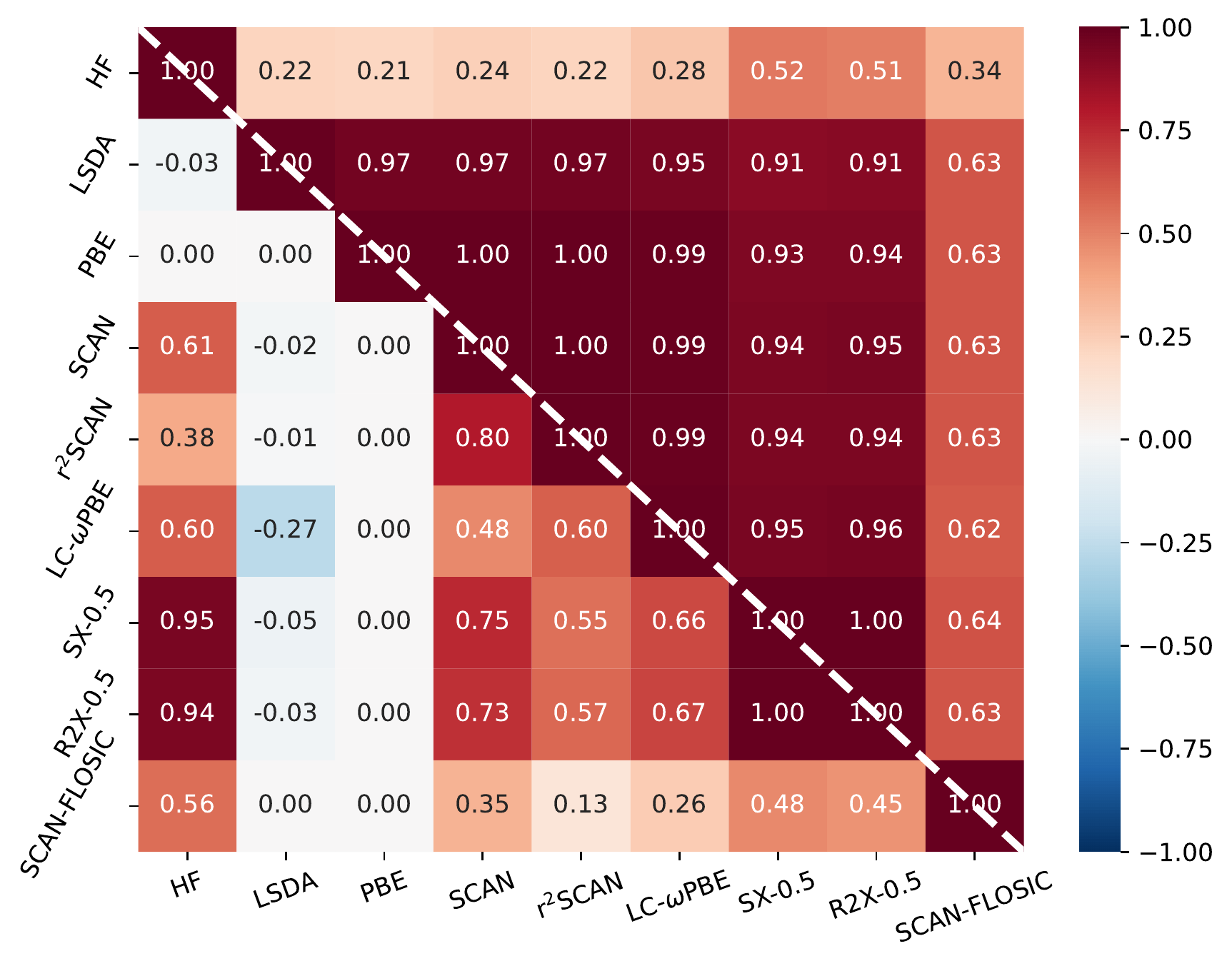}
    \caption{PBE correlation coefficients in the functional- (upper triangular matrix) and density-driven errors (lower triangular matrix).
    To save space, and using the symmetric property of the correlation coefficient matrix, we present the two sets of error statistics in this form.
    }
    \label{fig:PBE_stat_corr}
\end{figure}

\begin{figure}
    \centering
    \includegraphics[width=0.8\columnwidth]{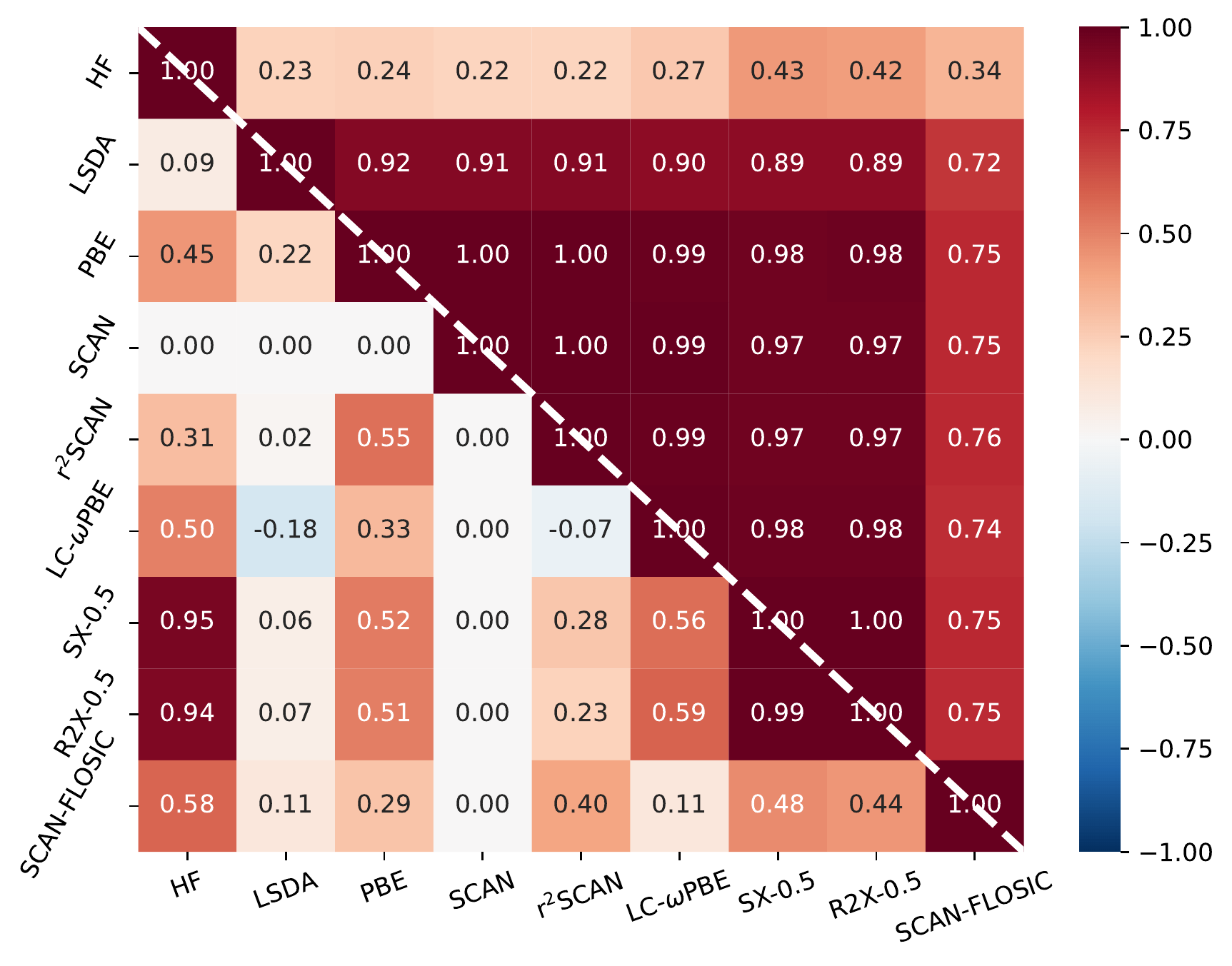}
    \caption{SCAN correlation coefficients in the functional- (upper triangular matrix) and density-driven errors (lower triangular matrix).
    To save space, and using the symmetric property of the correlation coefficient matrix, we present the two sets of error statistics in this form.
    }
    \label{fig:SCAN_stat_corr}
\end{figure}

\section{Charge transfer error analysis \label{sec:ctan}}

Errors in absolute and relative total energies from variational methods (e.g., DFT and Hartree-Fock but not coupled cluster) are typically second order in the error of the density.
Suppose that $n_\text{exact}^{(0)}$ ($n_\text{approx}^{(0)}$) is the ground state density found from the exact (an approximate) functional.
Then their total energies have Taylor series near their respective ground-state densities \cite{vuckovic2019}
\begin{align}
    & E_i[n_i^{(0)} + \Delta n] = E_i[n_i^{(0)}] + \int \frac{\delta E_i}{\delta n(\br)}\bigg|_{n_i^{(0)}} \Delta n(\br) d^3 r
    + \frac{1}{2} \int \frac{\delta^2 E_i}{\delta n(\br) \delta n(\br')} \bigg|_{n_i^{(0)}} \Delta n(\br) \Delta n(\br') d^3 r' d^3 r + \mathcal{O}(\Delta n)^3,
\end{align}
where $i=$ ``exact'' or ``approx''.
By the variational principle, the first functional derivative evaluated at the ground-state density is zero, and thus
\begin{align}
    & E_i[n_i^{(0)} + \Delta n] =
    E_i[n_i^{(0)}] \label{eq:fnl_taylor}
    + \frac{1}{2} \int K_i([n_i^{(0)}];\br,\br') \Delta n(\br) \Delta n(\br') d^3 r' d^3 r + \mathcal{O}(\Delta n)^3,
\end{align}
with $K$ the kernel
\begin{equation}
    K_i([\widetilde{n}];\br,\br') \equiv \frac{\delta^2 E_i}{\delta n(\br) \delta n(\br')}\bigg|_{\widetilde{n}}.
\end{equation}
Letting $\Delta n(\br) = n_\text{exact}^{(0)} - n_\text{approx}^{(0)}$,
the functional-driven error is, by eq \ref{eq:fnl_taylor},
\begin{align}
    & \Delta E_\text{F} = E_\text{approx}[n_\text{approx}^{(0)}] - E_\text{exact}[n_\text{exact}^{(0)}]
    \label{eq:fde_taylor} \\
    & + \frac{1}{2} \int K_\text{approx}([n_\text{approx}^{(0)}];\br,\br')
    \left[
    n_\text{exact}^{(0)}(\br) - n_\text{approx}^{(0)}(\br)
    \right]
    \left[
    n_\text{exact}^{(0)}(\br') - n_\text{approx}^{(0)}(\br')
    \right] d^3 r' d^3 r
    + \mathcal{O}\left[
    n_\text{exact}^{(0)} - n_\text{approx}^{(0)}
    \right]^3,
    \nonumber
\end{align}
a constant plus a second-order term.
Analogously, the density-driven error is second-order to lowest order,
\begin{align}
    & \Delta E_\text{D} =
    - \frac{1}{2} \int K_\text{approx}([n_\text{approx}^{(0)}];\br,\br')
    \left[
    n_\text{exact}^{(0)}(\br) - n_\text{approx}^{(0)}(\br)
    \right]
    \left[
    n_\text{exact}^{(0)}(\br') - n_\text{approx}^{(0)}(\br')
    \right] d^3 r' d^3 r  \\
    &+ \mathcal{O}\left[
    n_\text{exact}^{(0)} - n_\text{approx}^{(0)}
    \right]^3. \nonumber
\end{align}
Clearly if the ground-state density corresponds to a saddle point, these errors will be third-order to a lowest approximation.

Consider two infinitely separated fragments.
The Perdew-Parr-Levy Balduz (PPLB) theorem \cite{perdew1982} asserts that energy errors arising from a spurious transfer of charge between these infinitely-separated fragments are first order in errors of the density.
Let $0^+$ be a positive infinitesimal and $-1 < \eta < 1$, then the PPLB theorem can be written as
\begin{equation}
    E_\text{exact}(N + \eta) = E_\text{exact}(N)
    + \eta \frac{\partial E_\text{exact}}{\partial N}\bigg|_{N +\text{sign}(\eta) \, 0^+}, \label{eq:pplb_taylor}
\end{equation}
with no higher-order terms.
If the fragments are instead found at large, finite separations, eq \ref{eq:pplb_taylor} will also have a second-order term, effectively rounding the piecewise linear behavior of eq \ref{eq:pplb_taylor}.
Figure \ref{fig:scan_pplb} suggests that the energy errors due to spurious charge transfers in the transition states of chemical reactions are primarily quadratic.
Thus the stretched bonds typical of transition states are not sufficiently stretched to yield linear, PPLB-like behavior of the energy.

Of course, the stretched bonds of transition states are far from the infinitely-stretched bonds for which eq \ref{eq:pplb_taylor} is valid.
The relevance of the PPLB theorem is that a proxy functional that yields small or zero electron charge-transfer errors even in the infinitely-stretched limit is  likely to yield very small charge-transfer errors in transition states.

\begin{figure}
    \centering
    \includegraphics[width=0.8\columnwidth]{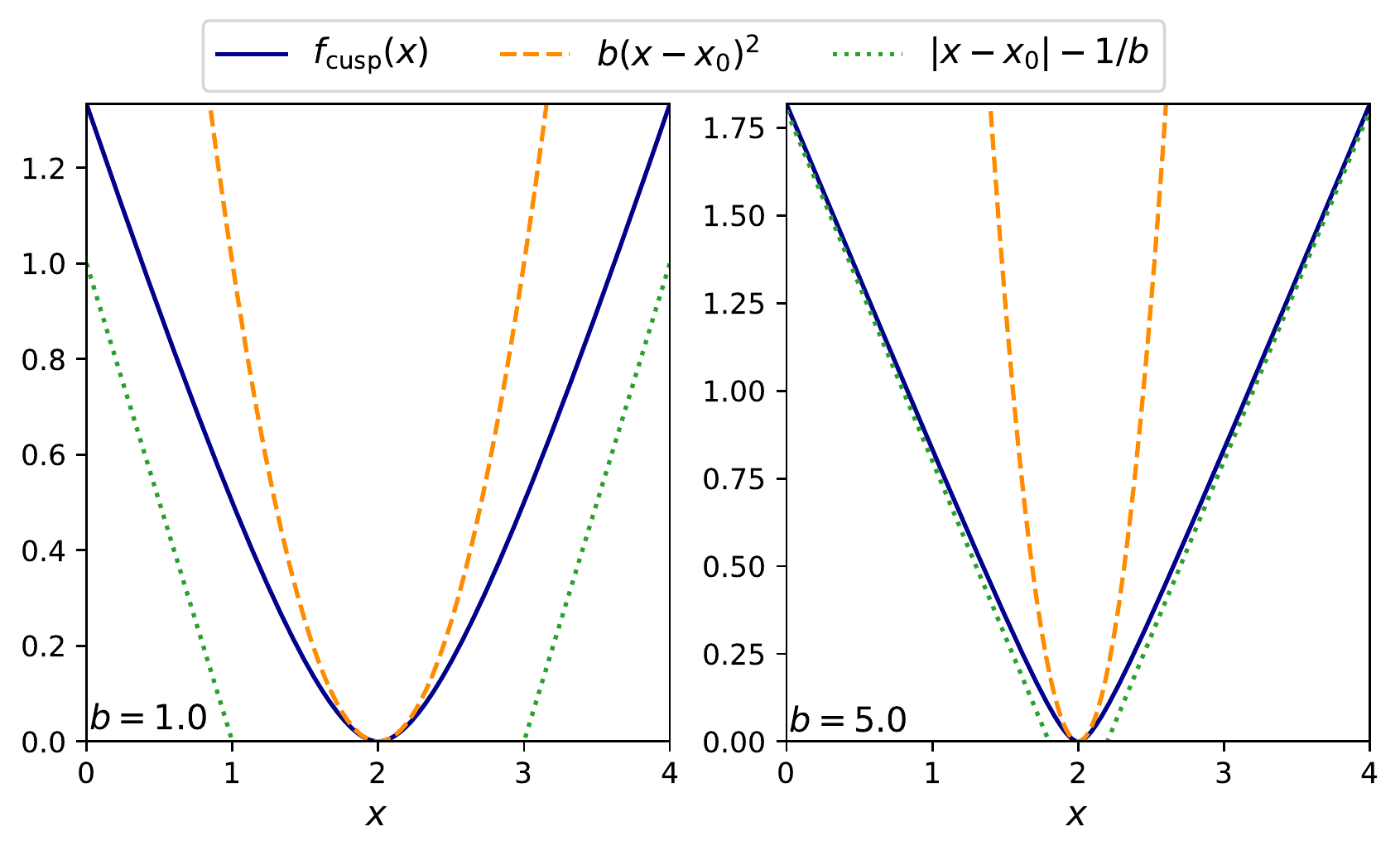}
    \caption{Visual representation of the relative order of energy errors arising from errors in the density, using the model function of eq \ref{eq:cuspy}.
    In both panels, $x_0=2$.
    In the left panel, $b$ is relatively small, and the function appears to be dominantly quadratic near the minimum $x_0$.
    This typifies the second-order nature of energy errors arising from spurious transfers of charge, even in the stretched-bonds of transition states.
    In the right panel, $b$ is relatively large, and the function reduces to $|x - x_0| - 1/b$ away from the minimum.
    This would typify the energy errors associated with charge transfers of highly-separated fragments.
    }
    \label{fig:cuspy}
\end{figure}

A simple mathematical model of the relative order of energy errors is given by
\begin{equation}
    f(x;b,x_0) = \frac{b(x - x_0)^2}{1 + b |x - x_0|}.
    \label{eq:cuspy}
\end{equation}
$b > 0$ is a control parameter that models the relative strength of the first- and second-order energy errors.
For finite $b$, the behavior of this function is dominantly quadratic, $b (x - x_0)^2$, near the minimum $x_0$.
As $b \to \infty$, $f(x;b,x_0) \to |x - x_0| - 1/b$ for $|x- x_0| > 0$.
These behaviors can be seen in fig \ref{fig:cuspy}.

\clearpage
\onecolumngrid

\section{Additional data and figures}

This Appendix presents additional data and figures that support the conclusions of the main text.

Tables \ref{tab:ak_pyscf_def2-qzvp} and \ref{tab:ak_pyscf_nrlmol} analyze the BH76 and BH76RC error statistics for the non-empirical DFAs (LSDA, PBE, SCAN, and \rrscan{}) and \lcwpbe{} considered in the main text, but using the def2-QZVP and NRLMOL basis sets, respectively.
These tables are analogous to Table \ref{tab:ak_pyscf_aug-cc-pvqz}.

\begin{figure}[h]
    \centering
    \includegraphics[width=0.6\columnwidth]{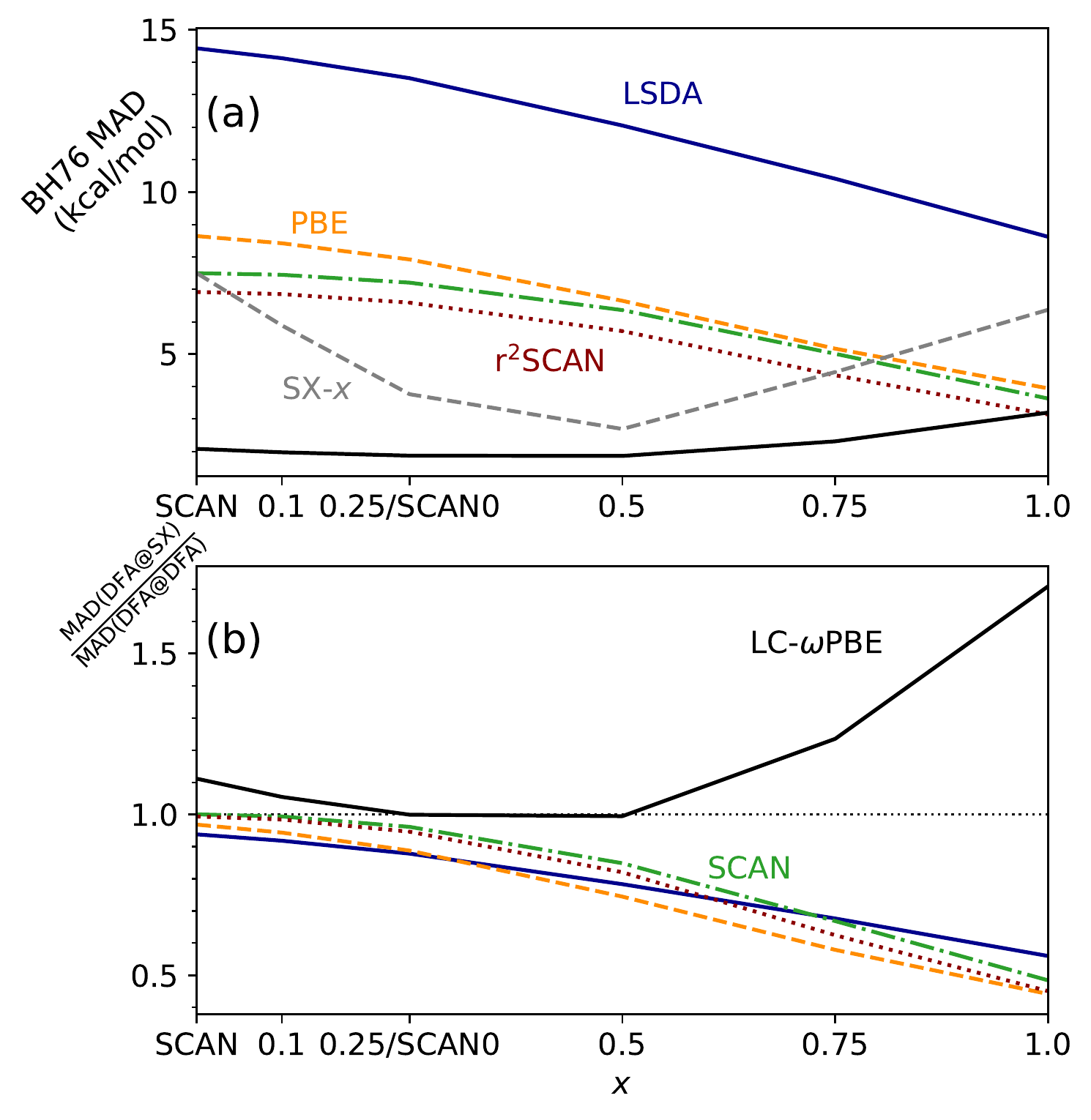}
    \caption{(a) Plot of the non-selfconsistent BH76 MAD for a few common DFAs using the global hybrid SX-$x$ density.
    (b) The same MADs (DFA@SX) rescaled by the self-consistent MADs (DFA@DFA) of Table \ref{tab:ak_pyscf_aug-cc-pvqz}.
    SX-0 is equivalent to SCAN; SX-0.25 is commonly called SCAN0; SX-1 uses the Kohn-Sham exact exchange only approximation (EXOA) with SCAN correlation.
    As we cannot easily compute the EXOA, which is not equivalent to the HF approximation, we use HF densities as a stand-in.
    For the numeric values presented here, see Table \ref{tab:sx_hyb_mads}.
    }
    \label{fig:sx_hyb_mads}
\end{figure}

\begin{figure}[h]
    \centering
    \includegraphics[width=0.6\columnwidth]{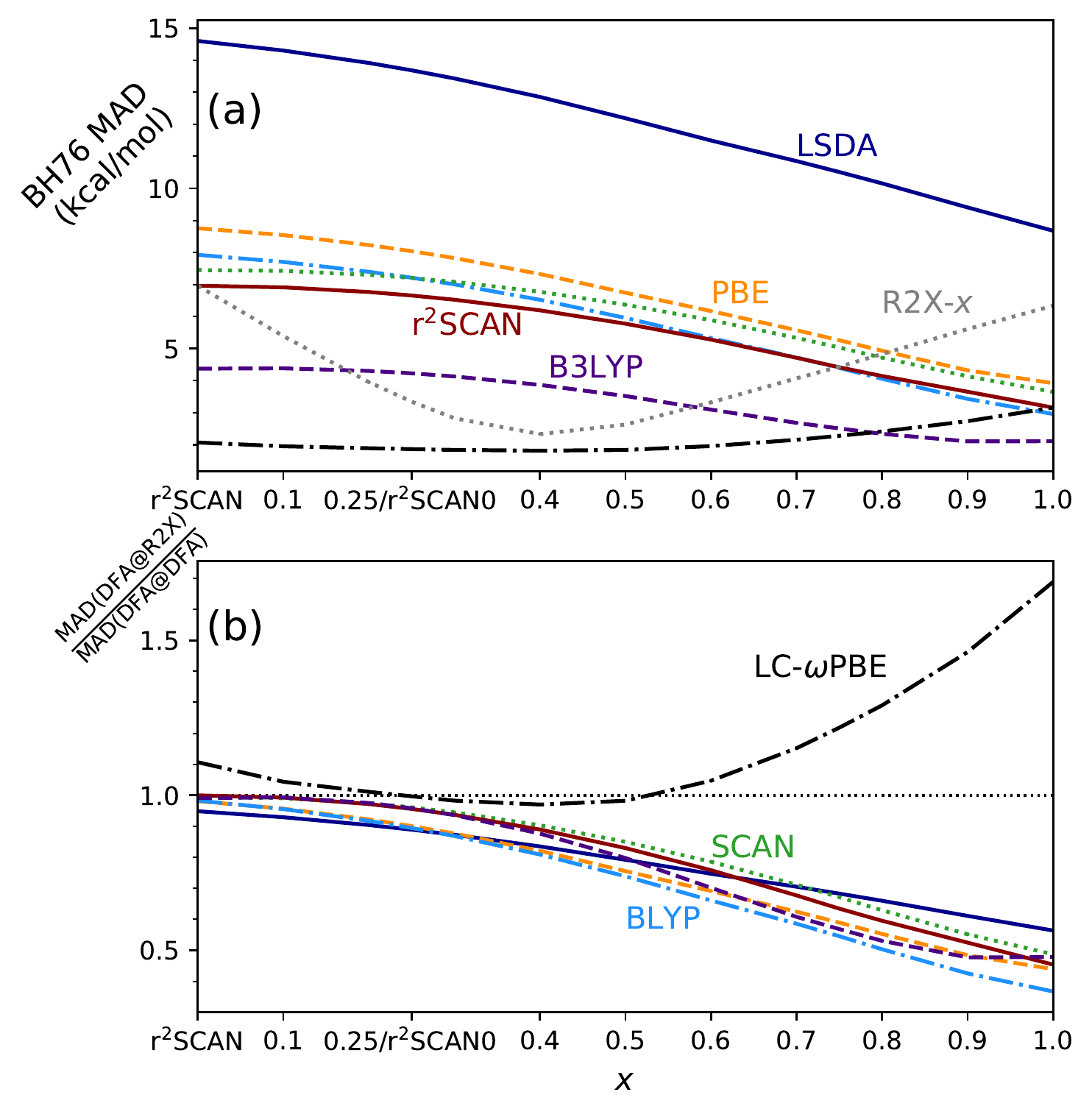}
    \caption{Same as Figure \ref{fig:sx_hyb_mads}, but using an \rrscan{} global hybrid, called R2X-$x$.
    \lcwpbe{}, when evaluated on the R2X-0.25 (commonly called \rrscan{}0) or R2X-0.5 (\rrscan{}50) densities, has essentially the same MAD.
    However, R2X-0.4 appears to be the best \rrscan{} global hybrid for this set.
    }
    \label{fig:R2X_hyb_mads}
\end{figure}

Figure \ref{fig:sx_hyb_mads} is a visual representation of the data presented in Table \ref{tab:sx_hyb_mads}.

Table \ref{tab:bh76_fde_dde_SP} computes the approximate functional- and density-driven errors for the calculated \textit{single-point} or total energies of the BH76 set, in the same way that table \ref{tab:bh76_fde_dde_dE} computes these metrics for energy differences.
Table \ref{tab:dE_HF_DFA} presents the average increases in the total energy for the transition states and, separately, for the reactant and product states, found by evaluating a DFA on the HF density, rather than its self-consistent density.
This table also shows that the magnitudes of the density-driven errors are generally larger for the transition states than for the reactant and product states.

\subsection{SCAN and \rrscan{} global hybrids}

\begin{ruledtabular}

\begin{table*}
    \centering
    \begin{tabular}{rrrrrrrrr}
$x$ & LSDA & PBE & BLYP & SCAN & r$^2$SCAN & B3LYP & \lcwpbe{} & R2X-$x$ \\ 
SCF & 15.39 & 8.93 & 8.06 & 7.50 & 6.96 & 4.41 & 1.87 & \\ \hline 
0.00 & 14.60 & 8.76 & 7.93 & 7.45 & 6.96 & 4.38 & 2.07 & 6.96 \\  
0.10 & 14.30 & 8.54 & 7.71 & 7.43 & 6.91 & 4.39 & 1.95 & 5.39 \\  
0.20 & 13.91 & 8.23 & 7.40 & 7.30 & 6.77 & 4.31 & 1.89 & 3.96 \\  
0.25 & 13.68 & 8.04 & 7.21 & 7.21 & 6.66 & 4.23 & 1.87 & 3.35 \\  
0.30 & 13.43 & 7.83 & 7.01 & 7.09 & 6.53 & 4.13 & 1.84 & 2.83 \\  
0.40 & 12.85 & 7.33 & 6.52 & 6.77 & 6.20 & 3.87 & 1.82 & 2.34 \\  
0.50 & 12.19 & 6.74 & 5.96 & 6.37 & 5.78 & 3.52 & 1.84 & 2.63 \\  
0.60 & 11.49 & 6.17 & 5.33 & 5.89 & 5.28 & 3.10 & 1.96 & 3.33 \\  
0.70 & 10.85 & 5.58 & 4.72 & 5.34 & 4.71 & 2.68 & 2.16 & 4.08 \\  
0.75 & 10.51 & 5.26 & 4.40 & 5.03 & 4.42 & 2.51 & 2.28 & 4.44 \\  
0.80 & 10.16 & 4.94 & 4.06 & 4.72 & 4.15 & 2.34 & 2.42 & 4.83 \\  
0.90 & 9.40 & 4.32 & 3.43 & 4.14 & 3.66 & 2.11 & 2.74 & 5.61 \\  
1.00 & 8.68 & 3.92 & 2.96 & 3.65 & 3.16 & 2.11 & 3.16 & 6.34 \\ \hline 
@HF & 7.82 & 3.85 & 2.84 & 3.05 & 2.84 & 2.71 & 4.18 & \\ 
    \end{tabular}
    \caption{Same as Table \ref{tab:sx_hyb_mads}, but using an \rrscan{} global hybrid instead of SCAN.
    Appendix Figure \ref{fig:R2X_hyb_mads} plots the non-self-consistent MADs as a function of $x$.
    }
    \label{tab:R2X_hyb_mads}
\end{table*}

\end{ruledtabular}

\clearpage
\subsection{BH76 results with different basis sets}

\begin{ruledtabular} 
    \begin{table*}[h]
        \centering
        \begin{tabular}{l|rrr|rrr|rrr}
BH76 & \multicolumn{3}{c|}{MD} & \multicolumn{3}{c|}{MAD} & \multicolumn{3}{c}{RMSD}\\ 
DFA & @DFA & @HF & @LC-$\omega$PBE & @DFA & @HF & @LC-$\omega$PBE & @DFA & @HF & @LC-$\omega$PBE \\ \hline 
HF & 10.54 &  &  & 11.20 &  &  & 13.17 &  & \\  
LSDA & -15.64 & -5.33 & -13.48 & 15.73 & 7.87 & 13.71 & 18.01 & 9.92 & 15.75\\  
PBE & -9.26 & -1.18 & -8.05 & 9.30 & 3.56 & 8.11 & 10.51 & 5.16 & 9.28\\  
SCAN & -7.71 & -1.90 & -7.12 & 7.76 & 3.06 & 7.20 & 8.45 & 3.97 & 7.86\\  
r$^2$SCAN & -7.19 & -1.30 & -6.61 & 7.23 & 2.78 & 6.68 & 7.95 & 3.98 & 7.37\\  
LC-$\omega$PBE & 0.36 & 3.90 &  & 1.61 & 3.97 &  & 2.22 & 5.46 & \\ \hline 
BH76RC & \multicolumn{3}{c|}{MD} & \multicolumn{3}{c|}{MAD} & \multicolumn{3}{c}{RMSD}\\ 
DFA & @DFA & @HF & @LC-$\omega$PBE & @DFA & @HF & @LC-$\omega$PBE & @DFA & @HF & @LC-$\omega$PBE \\ \hline 
HF & -0.31 &  &  & 8.57 &  &  & 11.37 &  & \\  
LSDA & 0.33 & -0.70 & -0.61 & 8.88 & 6.83 & 7.92 & 11.33 & 8.73 & 10.42\\  
PBE & 0.83 & 0.53 & 0.77 & 4.09 & 3.03 & 3.81 & 6.05 & 4.21 & 5.71\\  
SCAN & -0.23 & -0.65 & -0.28 & 3.33 & 2.93 & 3.26 & 4.36 & 3.59 & 4.19\\  
r$^2$SCAN & -0.11 & -0.56 & -0.18 & 3.15 & 2.79 & 3.12 & 4.21 & 3.38 & 4.05\\  
LC-$\omega$PBE & -0.65 & -0.75 &  & 2.29 & 1.96 &  & 2.81 & 2.48 & \\  
        \end{tabular}
        \caption{BH76 error statistics (in kcal/mol) using PySCF and the def2-QZVP \cite{weigend2003,weigend2005} basis set.
        The same notation as in Table \ref{tab:ak_pyscf_aug-cc-pvqz} is used here.
        }
        \label{tab:ak_pyscf_def2-qzvp}
    \end{table*}
\end{ruledtabular}

\begin{ruledtabular} 
    \begin{table*}[h]
        \centering
        \begin{tabular}{l|rrr|rrr|rrr}
BH76 & \multicolumn{3}{c|}{MD} & \multicolumn{3}{c|}{MAD} & \multicolumn{3}{c}{RMSD}\\ 
DFA & @DFA & @HF & @LC-$\omega$PBE & @DFA & @HF & @LC-$\omega$PBE & @DFA & @HF & @LC-$\omega$PBE \\ \hline 
HF & 10.39 &  &  & 11.03 &  &  & 13.08 &  & \\  
LSDA & -15.50 & -5.38 & -13.50 & 15.59 & 7.77 & 13.72 & 17.78 & 9.91 & 15.74\\  
PBE & -9.28 & -1.21 & -8.07 & 9.33 & 3.66 & 8.13 & 10.52 & 5.23 & 9.31\\  
SCAN & -7.76 & -1.96 & -7.19 & 7.83 & 3.06 & 7.28 & 8.50 & 4.07 & 7.93\\  
r$^2$SCAN & -7.26 & -1.36 & -6.69 & 7.32 & 2.77 & 6.77 & 8.03 & 4.08 & 7.45\\  
LC-$\omega$PBE & 0.35 & 3.89 &  & 1.63 & 3.99 &  & 2.20 & 5.49 & \\ \hline 
BH76RC & \multicolumn{3}{c|}{MD} & \multicolumn{3}{c|}{MAD} & \multicolumn{3}{c}{RMSD}\\ 
DFA & @DFA & @HF & @LC-$\omega$PBE & @DFA & @HF & @LC-$\omega$PBE & @DFA & @HF & @LC-$\omega$PBE \\ \hline 
HF & -0.50 &  &  & 8.61 &  &  & 11.56 &  & \\  
LSDA & -0.23 & -0.90 & -0.79 & 8.30 & 6.47 & 7.66 & 10.70 & 8.36 & 10.06\\  
PBE & 0.71 & 0.42 & 0.65 & 4.03 & 2.89 & 3.75 & 5.72 & 3.92 & 5.38\\  
SCAN & -0.31 & -0.73 & -0.36 & 3.08 & 2.66 & 3.00 & 4.09 & 3.41 & 3.93\\  
r$^2$SCAN & -0.20 & -0.64 & -0.26 & 3.00 & 2.60 & 2.93 & 3.94 & 3.19 & 3.79\\  
LC-$\omega$PBE & -0.84 & -0.94 &  & 2.18 & 2.00 &  & 2.74 & 2.55 & \\  
        \end{tabular}
        \caption{BH76 error statistics (in kcal/mol) using PySCF and the default NRLMOL density-functional optimized (DFO) \cite{porezag1999} basis set in a Cartesian representation.
        The same notation as in Table \ref{tab:ak_pyscf_aug-cc-pvqz} is used here.
        }
        \label{tab:ak_pyscf_nrlmol}
    \end{table*}
\end{ruledtabular}

\clearpage
\subsection{Auxiliary density-corrected DFT tables}

\begin{ruledtabular}

\begin{table*}[h]
    \centering
    \begin{tabular}{lrrrrrr}
 & \multicolumn{6}{c}{Proxy reference or proxy exact} \\
 & \multicolumn{2}{c}{\lcwpbe{}} & \multicolumn{2}{c}{SX-0.5} & \multicolumn{2}{c}{SCAN-FLOSIC} \\ 
\textit{Transition states} & MFE & MDE & MFE & MDE & MFE & MDE \\ \hline 
LSDA & 794.09 & -6.52 & 816.50 & -14.28 & 677.44 & -19.43 \\  
PBE & 50.34 & -2.79 & 69.87 & -7.67 & -68.14 & -12.76 \\  
BLYP & -50.36 & -4.14 & -29.46 & -10.39 &  &  \\  
SCAN & -49.72 & -2.41 & -34.43 & -3.05 & -173.13 & -6.47 \\  
\rrscan{} & -25.50 & -2.45 & -10.10 & -3.21 & -155.27 & 0.00 \\  
M06-L & -48.37 & -8.14 & -35.48 & -6.37 &  &  \\  
MN15-L & 26.34 & -13.57 & 37.85 & -10.43 &  &  \\  
B3LYP & -76.33 & -1.77 & -58.43 & -5.02 &  &  \\  
\lcwpbe{} &  &  & 18.63 & -3.97 &  &  \\  
SCAN@HF & -49.72 & 10.99 & -34.43 & 10.35 & -173.13 & 6.78 \\ \hline 
\textit{Forwards} & \multicolumn{2}{c}{\lcwpbe{}} & \multicolumn{2}{c}{SX-0.5} & \multicolumn{2}{c}{SCAN-FLOSIC} \\ 
\textit{Reactants and products} & MFE & MDE & MFE & MDE & MFE & MDE \\ \hline 
LSDA & 807.77 & -4.75 & 827.54 & -11.13 & 689.10 & -15.09 \\  
PBE & 58.00 & -1.64 & 75.17 & -5.42 & -62.11 & -9.39 \\  
BLYP & -43.50 & -2.90 & -24.91 & -8.11 &  &  \\  
SCAN & -42.46 & -1.85 & -28.96 & -1.96 & -166.75 & -4.57 \\  
\rrscan{} & -18.79 & -1.91 & -5.33 & -1.99 & -147.54 & -0.00 \\  
M06-L & -45.86 & -7.36 & -35.12 & -4.72 &  &  \\  
MN15-L & 26.59 & -12.99 & 36.03 & -9.05 &  &  \\  
B3LYP & -71.98 & -1.31 & -55.96 & -3.94 &  &  \\  
\lcwpbe{} &  &  & 16.97 & -3.58 &  &  \\  
SCAN@HF & -42.46 & 6.00 & -28.96 & 5.89 & -166.75 & 3.06 \\ \hline 
\textit{Reverse} & \multicolumn{2}{c}{\lcwpbe{}} & \multicolumn{2}{c}{SX-0.5} & \multicolumn{2}{c}{SCAN-FLOSIC} \\ 
\textit{Reactants and products} & MFE & MDE & MFE & MDE & MFE & MDE \\ \hline 
LSDA & 807.91 & -4.01 & 828.68 & -10.40 & 688.83 & -14.88 \\  
PBE & 59.28 & -1.58 & 77.42 & -5.35 & -61.70 & -8.82 \\  
BLYP & -42.41 & -2.88 & -22.80 & -8.11 &  &  \\  
SCAN & -42.08 & -1.80 & -27.68 & -1.84 & -166.76 & -4.42 \\  
\rrscan{} & -18.32 & -1.86 & -3.92 & -1.87 & -147.29 & -0.00 \\  
M06-L & -44.15 & -7.30 & -32.50 & -4.57 &  &  \\  
MN15-L & 27.94 & -13.06 & 38.28 & -9.03 &  &  \\  
B3LYP & -71.68 & -1.31 & -54.63 & -3.98 &  &  \\  
\lcwpbe{} &  &  & 17.96 & -3.58 &  &  \\  
SCAN@HF & -42.08 & 5.68 & -27.68 & 5.65 & -166.76 & 2.87 \\ 
    \end{tabular}
    \caption{Mean functional-driven errors (MFEs) and mean density-driven errors (MDEs) in the calculated single-point or total energies of the BH76 set.
    In each subset, the average is weighted by the number of times a system appears in the subset.
    This table can be used to reconstruct Table \ref{tab:bh76_fde_dde_dE}.
    The vertical columns indicate which DFA is used as a proxy for the exact functional and density in eqs \ref{eq:fde} and \ref{eq:dde}.
    Both LC-$\omega$PBE and SX-0.5 are reliable estimators of density-driven errors for SCAN and \rrscan{}.
    All calculations used the aug-cc-pVQZ basis set \cite{dunning1989} in PySCF; SCAN-FLOSIC calculations used the NRLMOL FLOSIC code, as well as calculations with the NRLMOL basis set in PySCF as described in the main text.
    }
    \label{tab:bh76_fde_dde_SP}
\end{table*}

\end{ruledtabular}

\begin{ruledtabular}
\begin{table}[h]
    \centering
\begin{tabular}{rrrr}
DFA & TS & RP & Difference \\ \hline 
LSDA & $29.56 \pm 12.93$ & $17.01 \pm 10.26$ & 12.55 \\ 
PBE & $20.48 \pm 9.80$ & $11.34 \pm 7.00$ & 9.14 \\ 
BLYP & $23.66 \pm 10.68$ & $13.88 \pm 8.14$ & 9.78 \\ 
SCAN & $12.78 \pm 6.48$ & $6.66 \pm 4.40$ & 6.12 \\ 
r$^2$SCAN & $13.07 \pm 6.64$ & $6.79 \pm 4.56$ & 6.28 \\ 
M06-L & $15.59 \pm 6.90$ & $8.22 \pm 5.62$ & 7.37 \\ 
MN15-L & $16.90 \pm 7.60$ & $9.51 \pm 5.42$ & 7.40 \\ 
B3LYP & $15.45 \pm 7.08$ & $8.80 \pm 5.13$ & 6.65 \\ 
SX-0.25 & $7.71 \pm 3.87$ & $3.95 \pm 2.59$ & 3.76 \\ 
LC-$\omega$PBE & $12.87 \pm 6.33$ & $8.23 \pm 4.77$ & 4.64 \\ 
    \end{tabular}
    \caption{Average energy shifts in the total energies of the BH76 transition states (TS) and reactant or product states (RP) and their standard deviations.
    The positive shift, in kcal/mol, is defined as the difference in the total energies found from non-self-consistent evaluation on the HF density and from evaluation on the self-consistent density.
    }
    \label{tab:dE_HF_DFA}
\end{table}

\end{ruledtabular}

\clearpage
\section{Complete density- and functional-driven errors data \label{sec:fde_dde_full}}

\subsection{LSDA}

\begin{longtable}{ll|rr|rr|rr}
  \caption{
Functional (FE) and density error (DE) metrics for LSDA on the BH76 set, all in kcal/mol.
Each set of two columns uses a different proxy for the exact functional as in Table V of the main text.
Error statistics are reported in the bottom three rows.
The reaction index used by the GMTKN and the symbolic index are given in the first two columns.}
  \label{tab:dc_dft_BH76_LSDA}
\\ \hline
 &  & \multicolumn{2}{c}{\lcwpbe{}} & \multicolumn{2}{c}{SX-0.5} & \multicolumn{2}{c}{SCAN-FLOSIC} \\ 
Index & Reaction & FE & DE & FE & DE & FE & DE \\ \hline 
  \endfirsthead
\\ \hline
 &  & \multicolumn{2}{c}{\lcwpbe{}} & \multicolumn{2}{c}{SX-0.5} & \multicolumn{2}{c}{SCAN-FLOSIC} \\ 
Index & Reaction & FE & DE & FE & DE & FE & DE \\ \hline 
  \endhead
1 & H + N$_2$O $\rightarrow$ N$_2$ + OH & -13.76 & -1.60 & -8.84 & -2.72 & 4.69 & -13.10 \\  
2 & 1 reverse & -42.44 & -3.26 & -50.06 & -5.11 & -46.98 & -19.50 \\  
3 & H + HF $\rightarrow$ HF + H & -21.36 & -0.81 & -18.33 & -1.04 & -22.34 & -8.33 \\  
4 & 3 reverse & -21.36 & -0.81 & -18.33 & -1.04 & -22.34 & -8.33 \\  
5 & H + HCl $\rightarrow$ HCl + H & -14.59 & -1.64 & -8.52 & -2.53 & -11.91 & -3.95 \\  
6 & 5 reverse & -14.59 & -1.64 & -8.52 & -2.53 & -11.91 & -3.95 \\  
7 & CH$_3$F + H $\rightarrow$ CH$_3$ + HF & -17.51 & -1.54 & -14.38 & -2.12 & -16.49 & -8.44 \\  
8 & 7 reverse & -23.40 & -1.90 & -24.38 & -2.97 & -19.26 & -9.06 \\
\hline 
\end{longtable}

\begin{longtable}{ll|rr|rr|rr}
  \caption{
Functional (FE) and density error (DE) metrics for LSDA on the BH76RC set, all in kcal/mol.
Each set of two columns uses a different proxy for the exact functional as in Table V of the main text.
Error statistics are reported in the bottom three rows.
The reaction index used by the GMTKN and the symbolic index are given in the first two columns.}
  \label{tab:dc_dft_BH76RC_LSDA}
\\ \hline
 &  & \multicolumn{2}{c}{\lcwpbe{}} & \multicolumn{2}{c}{SX-0.5} & \multicolumn{2}{c}{SCAN-FLOSIC} \\ 
Index & Reaction & FE & DE & FE & DE & FE & DE \\ \hline 
  \endfirsthead
\\ \hline
 &  & \multicolumn{2}{c}{\lcwpbe{}} & \multicolumn{2}{c}{SX-0.5} & \multicolumn{2}{c}{SCAN-FLOSIC} \\ 
Index & Reaction & FE & DE & FE & DE & FE & DE \\ \hline 
  \endhead
1 & H + N$_2$O $\rightarrow$ OH + N$_2$ & 28.68 & 1.66 & 41.22 & 2.38 & 51.67 & 6.40 \\  
2 & H + CH$_3$F $\rightarrow$ HF + CH$_3$ & 5.88 & 0.36 & 10.00 & 0.85 & 2.77 & 0.61 \\  
3 & H + F$_2$ $\rightarrow$ HF + F & 16.70 & 0.34 & 35.26 & 1.04 & 23.29 & 0.88 \\  
4 & CH$_3$ + ClF $\rightarrow$ CH$_3$F + Cl & 4.06 & 5.87 & 6.93 & 5.74 & 3.42 & 1.21 \\  
5 & CH$_3$Cl + F$^-$ $\rightarrow$ CH$_3$F + Cl$^-$ & -0.22 & -0.07 & 6.03 & 0.80 & 8.84 & 0.82 \\  
6 & F$^-$ $\hdots$ CH$_3$Cl $\rightarrow$ CH$_3$F $\hdots$ Cl$^-$ & 2.74 & 1.02 & 6.66 & 1.94 & 8.85 & 2.26 \\  
7 & CH$_3$F + OH$^-$ $\rightarrow$ HOCH$_3$ + F$^-$ & 2.58 & 0.40 & 1.55 & 0.03 & -0.84 & -0.01 \\  
8 & CH$_3$F $\hdots$ OH$^-$ $\rightarrow$ HOCH$_3$ $\hdots$ F$^-$ & -3.75 & 0.47 & -3.40 & 0.27 & -5.80 & 0.18 \\
\hline 
\end{longtable}

\clearpage
\subsection{PBE}

\begin{longtable}{ll|rr|rr|rr}
  \caption{
Functional (FE) and density error (DE) metrics for PBE on the BH76 set, all in kcal/mol.
Each set of two columns uses a different proxy for the exact functional as in Table V of the main text.
Error statistics are reported in the bottom three rows.
The reaction index used by the GMTKN and the symbolic index are given in the first two columns.}
  \label{tab:dc_dft_BH76_PBE}
\\ \hline
 &  & \multicolumn{2}{c}{\lcwpbe{}} & \multicolumn{2}{c}{SX-0.5} & \multicolumn{2}{c}{SCAN-FLOSIC} \\ 
Index & Reaction & FE & DE & FE & DE & FE & DE \\ \hline 
  \endfirsthead
\\ \hline
 &  & \multicolumn{2}{c}{\lcwpbe{}} & \multicolumn{2}{c}{SX-0.5} & \multicolumn{2}{c}{SCAN-FLOSIC} \\ 
Index & Reaction & FE & DE & FE & DE & FE & DE \\ \hline 
  \endhead
1 & H + N$_2$O $\rightarrow$ N$_2$ + OH & -7.26 & -0.84 & -2.58 & -1.73 & 9.62 & -10.76 \\  
2 & 1 reverse & -22.98 & -2.32 & -30.87 & -3.90 & -29.63 & -16.82 \\  
3 & H + HF $\rightarrow$ HF + H & -13.19 & -0.35 & -10.24 & -0.49 & -15.84 & -6.24 \\  
4 & 3 reverse & -13.19 & -0.35 & -10.24 & -0.49 & -15.84 & -6.24 \\  
5 & H + HCl $\rightarrow$ HCl + H & -7.93 & -0.95 & -2.14 & -1.57 & -5.65 & -2.93 \\  
6 & 5 reverse & -7.93 & -0.95 & -2.14 & -1.57 & -5.65 & -2.93 \\  
7 & CH$_3$F + H $\rightarrow$ CH$_3$ + HF & -13.10 & -0.82 & -10.16 & -1.20 & -13.29 & -6.46 \\  
8 & 7 reverse & -14.90 & -1.05 & -16.28 & -1.72 & -12.59 & -6.49 \\
\hline 
\end{longtable}

\begin{longtable}{ll|rr|rr|rr}
  \caption{
Functional (FE) and density error (DE) metrics for PBE on the BH76RC set, all in kcal/mol.
Each set of two columns uses a different proxy for the exact functional as in Table V of the main text.
Error statistics are reported in the bottom three rows.
The reaction index used by the GMTKN and the symbolic index are given in the first two columns.}
  \label{tab:dc_dft_BH76RC_PBE}
\\ \hline
 &  & \multicolumn{2}{c}{\lcwpbe{}} & \multicolumn{2}{c}{SX-0.5} & \multicolumn{2}{c}{SCAN-FLOSIC} \\ 
Index & Reaction & FE & DE & FE & DE & FE & DE \\ \hline 
  \endfirsthead
\\ \hline
 &  & \multicolumn{2}{c}{\lcwpbe{}} & \multicolumn{2}{c}{SX-0.5} & \multicolumn{2}{c}{SCAN-FLOSIC} \\ 
Index & Reaction & FE & DE & FE & DE & FE & DE \\ \hline 
  \endhead
1 & H + N$_2$O $\rightarrow$ OH + N$_2$ & 15.72 & 1.48 & 28.28 & 2.16 & 39.25 & 6.06 \\  
2 & H + CH$_3$F $\rightarrow$ HF + CH$_3$ & 1.80 & 0.23 & 6.12 & 0.51 & -0.70 & 0.03 \\  
3 & H + F$_2$ $\rightarrow$ HF + F & 11.30 & 0.24 & 29.98 & 0.83 & 18.57 & 0.69 \\  
4 & CH$_3$ + ClF $\rightarrow$ CH$_3$F + Cl & 5.14 & 0.39 & 7.87 & 0.40 & 4.55 & 0.41 \\  
5 & CH$_3$Cl + F$^-$ $\rightarrow$ CH$_3$F + Cl$^-$ & 1.56 & 0.04 & 7.65 & 1.07 & 10.22 & 1.20 \\  
6 & F$^-$ $\hdots$ CH$_3$Cl $\rightarrow$ CH$_3$F $\hdots$ Cl$^-$ & 3.16 & 1.16 & 7.10 & 2.07 & 9.21 & 2.40 \\  
7 & CH$_3$F + OH$^-$ $\rightarrow$ HOCH$_3$ + F$^-$ & 2.84 & 0.36 & 1.92 & -0.12 & -0.48 & -0.14 \\  
8 & CH$_3$F $\hdots$ OH$^-$ $\rightarrow$ HOCH$_3$ $\hdots$ F$^-$ & 1.15 & 0.42 & 1.56 & 0.16 & -0.96 & -0.06 \\
\hline 
\end{longtable}

\clearpage
\subsection{BLYP}

\begin{longtable}{ll|rr|rr}
  \caption{
Functional (FE) and density error (DE) metrics for BLYP on the BH76 set, all in kcal/mol.
Each set of two columns uses a different proxy for the exact functional as in Table V of the main text.
Error statistics are reported in the bottom three rows.
The reaction index used by the GMTKN and the symbolic index are given in the first two columns.}
  \label{tab:dc_dft_BH76_BLYP}
\\ \hline
 &  & \multicolumn{2}{c}{\lcwpbe{}} & \multicolumn{2}{c}{SX-0.5} \\ 
Index & Reaction & FE & DE & FE & DE \\ \hline 
  \endfirsthead
\\ \hline
 &  & \multicolumn{2}{c}{\lcwpbe{}} & \multicolumn{2}{c}{SX-0.5} \\ 
Index & Reaction & FE & DE & FE & DE \\ \hline 
  \endhead
1 & H + N$_2$O $\rightarrow$ N$_2$ + OH & -8.43 & -1.09 & -3.78 & -1.95 \\  
2 & 1 reverse & -13.57 & -2.57 & -21.34 & -4.27 \\  
3 & H + HF $\rightarrow$ HF + H & -14.09 & -0.45 & -11.17 & -0.56 \\  
4 & 3 reverse & -14.09 & -0.45 & -11.17 & -0.56 \\  
5 & H + HCl $\rightarrow$ HCl + H & -7.54 & -1.16 & -1.71 & -1.81 \\  
6 & 5 reverse & -7.54 & -1.16 & -1.71 & -1.81 \\  
7 & CH$_3$F + H $\rightarrow$ CH$_3$ + HF & -15.21 & -1.06 & -12.14 & -1.57 \\  
8 & 7 reverse & -13.62 & -1.10 & -15.17 & -1.61 \\
\hline 
\end{longtable}

\begin{longtable}{ll|rr|rr}
  \caption{
Functional (FE) and density error (DE) metrics for BLYP on the BH76RC set, all in kcal/mol.
Each set of two columns uses a different proxy for the exact functional as in Table V of the main text.
Error statistics are reported in the bottom three rows.
The reaction index used by the GMTKN and the symbolic index are given in the first two columns.}
  \label{tab:dc_dft_BH76RC_BLYP}
\\ \hline
 &  & \multicolumn{2}{c}{\lcwpbe{}} & \multicolumn{2}{c}{SX-0.5} \\ 
Index & Reaction & FE & DE & FE & DE \\ \hline 
  \endfirsthead
\\ \hline
 &  & \multicolumn{2}{c}{\lcwpbe{}} & \multicolumn{2}{c}{SX-0.5} \\ 
Index & Reaction & FE & DE & FE & DE \\ \hline 
  \endhead
1 & H + N$_2$O $\rightarrow$ OH + N$_2$ & 5.15 & 1.48 & 17.56 & 2.32 \\  
2 & H + CH$_3$F $\rightarrow$ HF + CH$_3$ & -1.59 & 0.04 & 3.02 & 0.04 \\  
3 & H + F$_2$ $\rightarrow$ HF + F & 7.96 & 0.27 & 26.89 & 0.60 \\  
4 & CH$_3$ + ClF $\rightarrow$ CH$_3$F + Cl & 4.31 & 0.41 & 7.09 & 0.38 \\  
5 & CH$_3$Cl + F$^-$ $\rightarrow$ CH$_3$F + Cl$^-$ & 1.12 & -0.01 & 6.90 & 1.32 \\  
6 & F$^-$ $\hdots$ CH$_3$Cl $\rightarrow$ CH$_3$F $\hdots$ Cl$^-$ & 2.80 & 1.01 & 6.57 & 2.10 \\  
7 & CH$_3$F + OH$^-$ $\rightarrow$ HOCH$_3$ + F$^-$ & 3.80 & 0.36 & 3.02 & -0.27 \\  
8 & CH$_3$F $\hdots$ OH$^-$ $\rightarrow$ HOCH$_3$ $\hdots$ F$^-$ & 4.12 & 0.39 & 4.65 & 0.02 \\
\hline 
\end{longtable}

\clearpage
\subsection{SCAN}

\begin{longtable}{ll|rr|rr|rr}
  \caption{
Functional (FE) and density error (DE) metrics for SCAN on the BH76 set, all in kcal/mol.
Each set of two columns uses a different proxy for the exact functional as in Table V of the main text.
Error statistics are reported in the bottom three rows.
The reaction index used by the GMTKN and the symbolic index are given in the first two columns.}
  \label{tab:dc_dft_BH76_SCAN}
\\ \hline
 &  & \multicolumn{2}{c}{\lcwpbe{}} & \multicolumn{2}{c}{SX-0.5} & \multicolumn{2}{c}{SCAN-FLOSIC} \\ 
Index & Reaction & FE & DE & FE & DE & FE & DE \\ \hline 
  \endfirsthead
\\ \hline
 &  & \multicolumn{2}{c}{\lcwpbe{}} & \multicolumn{2}{c}{SX-0.5} & \multicolumn{2}{c}{SCAN-FLOSIC} \\ 
Index & Reaction & FE & DE & FE & DE & FE & DE \\ \hline 
  \endhead
1 & H + N$_2$O $\rightarrow$ N$_2$ + OH & -8.29 & -0.43 & -4.21 & -0.72 & 5.83 & -7.53 \\  
2 & 1 reverse & -11.64 & -1.43 & -20.13 & -2.40 & -21.91 & -12.64 \\  
3 & H + HF $\rightarrow$ HF + H & -11.62 & -0.26 & -8.61 & -0.46 & -15.67 & -4.80 \\  
4 & 3 reverse & -11.62 & -0.26 & -8.61 & -0.46 & -15.67 & -4.80 \\  
5 & H + HCl $\rightarrow$ HCl + H & -8.95 & -0.40 & -3.64 & -0.54 & -7.50 & -1.52 \\  
6 & 5 reverse & -8.95 & -0.40 & -3.64 & -0.54 & -7.50 & -1.52 \\  
7 & CH$_3$F + H $\rightarrow$ CH$_3$ + HF & -11.74 & -0.57 & -9.02 & -0.74 & -13.52 & -4.65 \\  
8 & 7 reverse & -9.91 & -0.69 & -11.46 & -1.19 & -9.21 & -4.69 \\
\hline 
\end{longtable}

\begin{longtable}{ll|rr|rr|rr}
  \caption{
Functional (FE) and density error (DE) metrics for SCAN on the BH76RC set, all in kcal/mol.
Each set of two columns uses a different proxy for the exact functional as in Table V of the main text.
Error statistics are reported in the bottom three rows.
The reaction index used by the GMTKN and the symbolic index are given in the first two columns.}
  \label{tab:dc_dft_BH76RC_SCAN}
\\ \hline
 &  & \multicolumn{2}{c}{\lcwpbe{}} & \multicolumn{2}{c}{SX-0.5} & \multicolumn{2}{c}{SCAN-FLOSIC} \\ 
Index & Reaction & FE & DE & FE & DE & FE & DE \\ \hline 
  \endfirsthead
\\ \hline
 &  & \multicolumn{2}{c}{\lcwpbe{}} & \multicolumn{2}{c}{SX-0.5} & \multicolumn{2}{c}{SCAN-FLOSIC} \\ 
Index & Reaction & FE & DE & FE & DE & FE & DE \\ \hline 
  \endhead
1 & H + N$_2$O $\rightarrow$ OH + N$_2$ & 3.35 & 0.99 & 15.92 & 1.68 & 27.74 & 5.11 \\  
2 & H + CH$_3$F $\rightarrow$ HF + CH$_3$ & -1.83 & 0.12 & 2.44 & 0.45 & -4.31 & 0.04 \\  
3 & H + F$_2$ $\rightarrow$ HF + F & 0.67 & 0.31 & 19.07 & 1.17 & 8.16 & 1.07 \\  
4 & CH$_3$ + ClF $\rightarrow$ CH$_3$F + Cl & 3.11 & 0.41 & 5.75 & 0.51 & 2.88 & 0.50 \\  
5 & CH$_3$Cl + F$^-$ $\rightarrow$ CH$_3$F + Cl$^-$ & -4.01 & 0.05 & 2.61 & 0.55 & 5.28 & 0.79 \\  
6 & F$^-$ $\hdots$ CH$_3$Cl $\rightarrow$ CH$_3$F $\hdots$ Cl$^-$ & -0.78 & 0.68 & 3.61 & 1.13 & 5.75 & 1.38 \\  
7 & CH$_3$F + OH$^-$ $\rightarrow$ HOCH$_3$ + F$^-$ & 3.45 & 0.26 & 2.20 & 0.11 & 0.03 & 0.15 \\  
8 & CH$_3$F $\hdots$ OH$^-$ $\rightarrow$ HOCH$_3$ $\hdots$ F$^-$ & 1.61 & 0.23 & 1.92 & 0.07 & -0.16 & 0.09 \\
\hline 
\end{longtable}

\clearpage
\subsection{\rrscan{}}

\begin{longtable}{ll|rr|rr|rr}
  \caption{
Functional (FE) and density error (DE) metrics for \rrscan{} on the BH76 set, all in kcal/mol.
Each set of two columns uses a different proxy for the exact functional as in Table V of the main text.
Error statistics are reported in the bottom three rows.
The reaction index used by the GMTKN and the symbolic index are given in the first two columns.}
  \label{tab:dc_dft_BH76_r2SCAN}
\\ \hline
 &  & \multicolumn{2}{c}{\lcwpbe{}} & \multicolumn{2}{c}{SX-0.5} & \multicolumn{2}{c}{SCAN-FLOSIC} \\ 
Index & Reaction & FE & DE & FE & DE & FE & DE \\ \hline 
  \endfirsthead
\\ \hline
 &  & \multicolumn{2}{c}{\lcwpbe{}} & \multicolumn{2}{c}{SX-0.5} & \multicolumn{2}{c}{SCAN-FLOSIC} \\ 
Index & Reaction & FE & DE & FE & DE & FE & DE \\ \hline 
  \endhead
1 & H + N$_2$O $\rightarrow$ N$_2$ + OH & -7.26 & -0.44 & -2.90 & -1.01 & -0.78 & 0.00 \\  
2 & 1 reverse & -13.02 & -1.51 & -21.20 & -2.79 & -36.02 & 0.00 \\  
3 & H + HF $\rightarrow$ HF + H & -11.27 & -0.24 & -8.26 & -0.44 & -20.10 & -0.00 \\  
4 & 3 reverse & -11.27 & -0.24 & -8.26 & -0.44 & -20.10 & -0.00 \\  
5 & H + HCl $\rightarrow$ HCl + H & -8.63 & -0.51 & -3.11 & -0.85 & -8.79 & 0.00 \\  
6 & 5 reverse & -8.63 & -0.51 & -3.11 & -0.85 & -8.79 & 0.00 \\  
7 & CH$_3$F + H $\rightarrow$ CH$_3$ + HF & -12.59 & -0.42 & -9.85 & -0.61 & -18.83 & 0.00 \\  
8 & 7 reverse & -9.35 & -0.70 & -11.00 & -1.10 & -13.31 & 0.00 \\
\hline 
\end{longtable}

\begin{longtable}{ll|rr|rr|rr}
  \caption{
Functional (FE) and density error (DE) metrics for \rrscan{} on the BH76RC set, all in kcal/mol.
Each set of two columns uses a different proxy for the exact functional as in Table V of the main text.
Error statistics are reported in the bottom three rows.
The reaction index used by the GMTKN and the symbolic index are given in the first two columns.}
  \label{tab:dc_dft_BH76RC_r2SCAN}
\\ \hline
 &  & \multicolumn{2}{c}{\lcwpbe{}} & \multicolumn{2}{c}{SX-0.5} & \multicolumn{2}{c}{SCAN-FLOSIC} \\ 
Index & Reaction & FE & DE & FE & DE & FE & DE \\ \hline 
  \endfirsthead
\\ \hline
 &  & \multicolumn{2}{c}{\lcwpbe{}} & \multicolumn{2}{c}{SX-0.5} & \multicolumn{2}{c}{SCAN-FLOSIC} \\ 
Index & Reaction & FE & DE & FE & DE & FE & DE \\ \hline 
  \endhead
1 & H + N$_2$O $\rightarrow$ OH + N$_2$ & 5.76 & 1.06 & 18.30 & 1.78 & 35.25 & -0.00 \\  
2 & H + CH$_3$F $\rightarrow$ HF + CH$_3$ & -3.24 & 0.28 & 1.15 & 0.49 & -5.52 & -0.00 \\  
3 & H + F$_2$ $\rightarrow$ HF + F & 1.05 & 0.25 & 19.54 & 1.02 & 9.75 & -0.00 \\  
4 & CH$_3$ + ClF $\rightarrow$ CH$_3$F + Cl & 4.08 & 0.28 & 6.53 & 0.58 & 4.33 & 0.00 \\  
5 & CH$_3$Cl + F$^-$ $\rightarrow$ CH$_3$F + Cl$^-$ & -2.73 & 0.02 & 3.81 & 0.60 & 7.26 & 0.00 \\  
6 & F$^-$ $\hdots$ CH$_3$Cl $\rightarrow$ CH$_3$F $\hdots$ Cl$^-$ & 0.03 & 0.75 & 4.36 & 1.26 & 8.01 & 0.00 \\  
7 & CH$_3$F + OH$^-$ $\rightarrow$ HOCH$_3$ + F$^-$ & 3.17 & 0.21 & 1.98 & -0.00 & -0.10 & 0.00 \\  
8 & CH$_3$F $\hdots$ OH$^-$ $\rightarrow$ HOCH$_3$ $\hdots$ F$^-$ & 1.41 & 0.14 & 1.81 & -0.11 & -0.36 & -0.00 \\
\hline 
\end{longtable}

\clearpage
\subsection{M06-L}

\begin{longtable}{ll|rr|rr}
  \caption{
Functional (FE) and density error (DE) metrics for M06-L on the BH76 set, all in kcal/mol.
Each set of two columns uses a different proxy for the exact functional as in Table V of the main text.
Error statistics are reported in the bottom three rows.
The reaction index used by the GMTKN and the symbolic index are given in the first two columns.}
  \label{tab:dc_dft_BH76_M06-L}
\\ \hline
 &  & \multicolumn{2}{c}{\lcwpbe{}} & \multicolumn{2}{c}{SX-0.5} \\ 
Index & Reaction & FE & DE & FE & DE \\ \hline 
  \endfirsthead
\\ \hline
 &  & \multicolumn{2}{c}{\lcwpbe{}} & \multicolumn{2}{c}{SX-0.5} \\ 
Index & Reaction & FE & DE & FE & DE \\ \hline 
  \endhead
1 & H + N$_2$O $\rightarrow$ N$_2$ + OH & 2.24 & -0.99 & 7.35 & -2.31 \\  
2 & 1 reverse & -8.16 & -3.27 & -16.58 & -4.30 \\  
3 & H + HF $\rightarrow$ HF + H & 0.83 & -0.77 & 4.58 & -1.71 \\  
4 & 3 reverse & 0.83 & -0.77 & 4.58 & -1.71 \\  
5 & H + HCl $\rightarrow$ HCl + H & 0.22 & -0.97 & 6.18 & -1.75 \\  
6 & 5 reverse & 0.22 & -0.97 & 6.18 & -1.75 \\  
7 & CH$_3$F + H $\rightarrow$ CH$_3$ + HF & -3.11 & -0.86 & 0.16 & -1.58 \\  
8 & 7 reverse & -5.02 & -1.10 & -6.10 & -2.07 \\
\hline 
\end{longtable}

\begin{longtable}{ll|rr|rr}
  \caption{
Functional (FE) and density error (DE) metrics for M06-L on the BH76RC set, all in kcal/mol.
Each set of two columns uses a different proxy for the exact functional as in Table V of the main text.
Error statistics are reported in the bottom three rows.
The reaction index used by the GMTKN and the symbolic index are given in the first two columns.}
  \label{tab:dc_dft_BH76RC_M06-L}
\\ \hline
 &  & \multicolumn{2}{c}{\lcwpbe{}} & \multicolumn{2}{c}{SX-0.5} \\ 
Index & Reaction & FE & DE & FE & DE \\ \hline 
  \endfirsthead
\\ \hline
 &  & \multicolumn{2}{c}{\lcwpbe{}} & \multicolumn{2}{c}{SX-0.5} \\ 
Index & Reaction & FE & DE & FE & DE \\ \hline 
  \endhead
1 & H + N$_2$O $\rightarrow$ OH + N$_2$ & 10.39 & 2.28 & 23.93 & 1.99 \\  
2 & H + CH$_3$F $\rightarrow$ HF + CH$_3$ & 1.91 & 0.23 & 6.26 & 0.49 \\  
3 & H + F$_2$ $\rightarrow$ HF + F & 2.90 & -0.23 & 21.05 & 0.88 \\  
4 & CH$_3$ + ClF $\rightarrow$ CH$_3$F + Cl & 1.13 & 0.34 & 3.78 & 0.44 \\  
5 & CH$_3$Cl + F$^-$ $\rightarrow$ CH$_3$F + Cl$^-$ & -3.95 & -0.11 & 2.54 & 0.52 \\  
6 & F$^-$ $\hdots$ CH$_3$Cl $\rightarrow$ CH$_3$F $\hdots$ Cl$^-$ & -1.22 & 0.69 & 2.97 & 1.35 \\  
7 & CH$_3$F + OH$^-$ $\rightarrow$ HOCH$_3$ + F$^-$ & 3.26 & 0.46 & 1.88 & 0.43 \\  
8 & CH$_3$F $\hdots$ OH$^-$ $\rightarrow$ HOCH$_3$ $\hdots$ F$^-$ & 2.92 & 0.29 & 2.90 & 0.46 \\
\hline 
\end{longtable}

\clearpage
\subsection{MN15-L}

\begin{longtable}{ll|rr|rr}
  \caption{
Functional (FE) and density error (DE) metrics for MN15-L on the BH76 set, all in kcal/mol.
Each set of two columns uses a different proxy for the exact functional as in Table V of the main text.
Error statistics are reported in the bottom three rows.
The reaction index used by the GMTKN and the symbolic index are given in the first two columns.}
  \label{tab:dc_dft_BH76_MN15-L}
\\ \hline
 &  & \multicolumn{2}{c}{\lcwpbe{}} & \multicolumn{2}{c}{SX-0.5} \\ 
Index & Reaction & FE & DE & FE & DE \\ \hline 
  \endfirsthead
\\ \hline
 &  & \multicolumn{2}{c}{\lcwpbe{}} & \multicolumn{2}{c}{SX-0.5} \\ 
Index & Reaction & FE & DE & FE & DE \\ \hline 
  \endhead
1 & H + N$_2$O $\rightarrow$ N$_2$ + OH & 0.74 & -0.98 & 5.53 & -1.99 \\  
2 & 1 reverse & -3.57 & -2.42 & -12.45 & -3.00 \\  
3 & H + HF $\rightarrow$ HF + H & -2.05 & -0.31 & 0.90 & -0.44 \\  
4 & 3 reverse & -2.05 & -0.31 & 0.90 & -0.44 \\  
5 & H + HCl $\rightarrow$ HCl + H & -0.08 & -0.97 & 5.87 & -1.74 \\  
6 & 5 reverse & -0.08 & -0.97 & 5.87 & -1.74 \\  
7 & CH$_3$F + H $\rightarrow$ CH$_3$ + HF & -2.49 & -0.42 & 0.67 & -1.03 \\  
8 & 7 reverse & -2.62 & -0.71 & -4.32 & -1.07 \\
\hline 
\end{longtable}

\begin{longtable}{ll|rr|rr}
  \caption{
Functional (FE) and density error (DE) metrics for MN15-L on the BH76RC set, all in kcal/mol.
Each set of two columns uses a different proxy for the exact functional as in Table V of the main text.
Error statistics are reported in the bottom three rows.
The reaction index used by the GMTKN and the symbolic index are given in the first two columns.}
  \label{tab:dc_dft_BH76RC_MN15-L}
\\ \hline
 &  & \multicolumn{2}{c}{\lcwpbe{}} & \multicolumn{2}{c}{SX-0.5} \\ 
Index & Reaction & FE & DE & FE & DE \\ \hline 
  \endfirsthead
\\ \hline
 &  & \multicolumn{2}{c}{\lcwpbe{}} & \multicolumn{2}{c}{SX-0.5} \\ 
Index & Reaction & FE & DE & FE & DE \\ \hline 
  \endhead
1 & H + N$_2$O $\rightarrow$ OH + N$_2$ & 4.31 & 1.43 & 17.98 & 1.02 \\  
2 & H + CH$_3$F $\rightarrow$ HF + CH$_3$ & 0.13 & 0.29 & 4.99 & 0.03 \\  
3 & H + F$_2$ $\rightarrow$ HF + F & 0.75 & 0.54 & 20.25 & 0.30 \\  
4 & CH$_3$ + ClF $\rightarrow$ CH$_3$F + Cl & 2.84 & -0.01 & 5.18 & 0.39 \\  
5 & CH$_3$Cl + F$^-$ $\rightarrow$ CH$_3$F + Cl$^-$ & -6.40 & -0.31 & -0.07 & 0.48 \\  
6 & F$^-$ $\hdots$ CH$_3$Cl $\rightarrow$ CH$_3$F $\hdots$ Cl$^-$ & -4.42 & -0.02 & -0.25 & 0.65 \\  
7 & CH$_3$F + OH$^-$ $\rightarrow$ HOCH$_3$ + F$^-$ & 4.35 & 0.89 & 3.28 & 0.56 \\  
8 & CH$_3$F $\hdots$ OH$^-$ $\rightarrow$ HOCH$_3$ $\hdots$ F$^-$ & 6.00 & 0.52 & 6.23 & 0.44 \\
\hline 
\end{longtable}

\clearpage
\subsection{B3LYP}

\begin{longtable}{ll|rr|rr}
  \caption{
Functional (FE) and density error (DE) metrics for B3LYP on the BH76 set, all in kcal/mol.
Each set of two columns uses a different proxy for the exact functional as in Table V of the main text.
Error statistics are reported in the bottom three rows.
The reaction index used by the GMTKN and the symbolic index are given in the first two columns.}
  \label{tab:dc_dft_BH76_B3LYP}
\\ \hline
 &  & \multicolumn{2}{c}{\lcwpbe{}} & \multicolumn{2}{c}{SX-0.5} \\ 
Index & Reaction & FE & DE & FE & DE \\ \hline 
  \endfirsthead
\\ \hline
 &  & \multicolumn{2}{c}{\lcwpbe{}} & \multicolumn{2}{c}{SX-0.5} \\ 
Index & Reaction & FE & DE & FE & DE \\ \hline 
  \endhead
1 & H + N$_2$O $\rightarrow$ N$_2$ + OH & -6.10 & -0.57 & -1.75 & -1.13 \\  
2 & 1 reverse & -4.00 & -0.99 & -12.45 & -2.01 \\  
3 & H + HF $\rightarrow$ HF + H & -9.28 & -0.28 & -6.42 & -0.34 \\  
4 & 3 reverse & -9.28 & -0.28 & -6.42 & -0.34 \\  
5 & H + HCl $\rightarrow$ HCl + H & -5.57 & -0.53 & 0.13 & -1.05 \\  
6 & 5 reverse & -5.57 & -0.53 & 0.13 & -1.05 \\  
7 & CH$_3$F + H $\rightarrow$ CH$_3$ + HF & -10.03 & -0.57 & -7.11 & -0.94 \\  
8 & 7 reverse & -7.77 & -0.57 & -9.62 & -0.77 \\
\hline 
\end{longtable}

\begin{longtable}{ll|rr|rr}
  \caption{
Functional (FE) and density error (DE) metrics for B3LYP on the BH76RC set, all in kcal/mol.
Each set of two columns uses a different proxy for the exact functional as in Table V of the main text.
Error statistics are reported in the bottom three rows.
The reaction index used by the GMTKN and the symbolic index are given in the first two columns.}
  \label{tab:dc_dft_BH76RC_B3LYP}
\\ \hline
 &  & \multicolumn{2}{c}{\lcwpbe{}} & \multicolumn{2}{c}{SX-0.5} \\ 
Index & Reaction & FE & DE & FE & DE \\ \hline 
  \endfirsthead
\\ \hline
 &  & \multicolumn{2}{c}{\lcwpbe{}} & \multicolumn{2}{c}{SX-0.5} \\ 
Index & Reaction & FE & DE & FE & DE \\ \hline 
  \endhead
1 & H + N$_2$O $\rightarrow$ OH + N$_2$ & -2.10 & 0.42 & 10.70 & 0.88 \\  
2 & H + CH$_3$F $\rightarrow$ HF + CH$_3$ & -2.26 & 0.00 & 2.52 & -0.17 \\  
3 & H + F$_2$ $\rightarrow$ HF + F & -1.62 & 0.19 & 17.72 & 0.11 \\  
4 & CH$_3$ + ClF $\rightarrow$ CH$_3$F + Cl & 0.96 & 0.16 & 3.72 & 0.14 \\  
5 & CH$_3$Cl + F$^-$ $\rightarrow$ CH$_3$F + Cl$^-$ & -1.90 & -0.17 & 4.49 & 0.57 \\  
6 & F$^-$ $\hdots$ CH$_3$Cl $\rightarrow$ CH$_3$F $\hdots$ Cl$^-$ & -0.09 & 0.15 & 4.15 & 0.75 \\  
7 & CH$_3$F + OH$^-$ $\rightarrow$ HOCH$_3$ + F$^-$ & 2.92 & 0.18 & 1.86 & -0.16 \\  
8 & CH$_3$F $\hdots$ OH$^-$ $\rightarrow$ HOCH$_3$ $\hdots$ F$^-$ & 2.90 & 0.16 & 3.27 & -0.06 \\
\hline 
\end{longtable}

\clearpage
\subsection{\lcwpbe{}}

\begin{longtable}{ll|rr}
  \caption{
Functional (FE) and density error (DE) metrics for \lcwpbe{} on the BH76 set, all in kcal/mol.
Each set of two columns uses a different proxy for the exact functional as in Table V of the main text.
Error statistics are reported in the bottom three rows.
The reaction index used by the GMTKN and the symbolic index are given in the first two columns.}
  \label{tab:dc_dft_BH76_LCwPBE}
\\ \hline
 &  & \multicolumn{2}{c}{SX-0.5} \\ 
Index & Reaction & FE & DE \\ \hline 
  \endfirsthead
\\ \hline
 &  & \multicolumn{2}{c}{SX-0.5} \\ 
Index & Reaction & FE & DE \\ \hline 
  \endhead
1 & H + N$_2$O $\rightarrow$ N$_2$ + OH & 4.18 & -0.39 \\  
2 & 1 reverse & -8.90 & -0.57 \\  
3 & H + HF $\rightarrow$ HF + H & 2.90 & -0.10 \\  
4 & 3 reverse & 2.90 & -0.10 \\  
5 & H + HCl $\rightarrow$ HCl + H & 5.35 & -0.17 \\  
6 & 5 reverse & 5.35 & -0.17 \\  
7 & CH$_3$F + H $\rightarrow$ CH$_3$ + HF & 2.85 & -0.29 \\  
8 & 7 reverse & -1.65 & -0.41 \\
\hline 
\end{longtable}

\begin{longtable}{ll|rr}
  \caption{
Functional (FE) and density error (DE) metrics for \lcwpbe{} on the BH76RC set, all in kcal/mol.
Each set of two columns uses a different proxy for the exact functional as in Table V of the main text.
Error statistics are reported in the bottom three rows.
The reaction index used by the GMTKN and the symbolic index are given in the first two columns.}
  \label{tab:dc_dft_BH76RC_LCwPBE}
\\ \hline
 &  & \multicolumn{2}{c}{SX-0.5} \\ 
Index & Reaction & FE & DE \\ \hline 
  \endfirsthead
\\ \hline
 &  & \multicolumn{2}{c}{SX-0.5} \\ 
Index & Reaction & FE & DE \\ \hline 
  \endhead
1 & H + N$_2$O $\rightarrow$ OH + N$_2$ & 13.07 & 0.18 \\  
2 & H + CH$_3$F $\rightarrow$ HF + CH$_3$ & 4.49 & 0.11 \\  
3 & H + F$_2$ $\rightarrow$ HF + F & 18.80 & 0.46 \\  
4 & CH$_3$ + ClF $\rightarrow$ CH$_3$F + Cl & 2.79 & -0.05 \\  
5 & CH$_3$Cl + F$^-$ $\rightarrow$ CH$_3$F + Cl$^-$ & 6.75 & 0.36 \\  
6 & F$^-$ $\hdots$ CH$_3$Cl $\rightarrow$ CH$_3$F $\hdots$ Cl$^-$ & 4.60 & 0.25 \\  
7 & CH$_3$F + OH$^-$ $\rightarrow$ HOCH$_3$ + F$^-$ & -1.16 & -0.24 \\  
8 & CH$_3$F $\hdots$ OH$^-$ $\rightarrow$ HOCH$_3$ $\hdots$ F$^-$ & 0.25 & -0.10 \\
\hline 
\end{longtable}

\clearpage
\subsection{LSDA@HF}

\begin{longtable}{ll|rr|rr|rr}
  \caption{
Functional (FE) and density error (DE) metrics for LSDA@HF on the BH76 set, all in kcal/mol.
Each set of two columns uses a different proxy for the exact functional as in Table V of the main text.
Error statistics are reported in the bottom three rows.
The reaction index used by the GMTKN and the symbolic index are given in the first two columns.}
  \label{tab:dc_dft_BH76_LSDA@HF}
\\ \hline
 &  & \multicolumn{2}{c}{\lcwpbe{}} & \multicolumn{2}{c}{SX-0.5} & \multicolumn{2}{c}{SCAN-FLOSIC} \\ 
Index & Reaction & FE & DE & FE & DE & FE & DE \\ \hline 
  \endfirsthead
\\ \hline
 &  & \multicolumn{2}{c}{\lcwpbe{}} & \multicolumn{2}{c}{SX-0.5} & \multicolumn{2}{c}{SCAN-FLOSIC} \\ 
Index & Reaction & FE & DE & FE & DE & FE & DE \\ \hline 
  \endhead
1 & H + N$_2$O $\rightarrow$ N$_2$ + OH & -13.76 & 13.32 & -8.84 & 12.19 & 4.69 & 1.97 \\  
2 & 1 reverse & -42.44 & 20.33 & -50.06 & 18.49 & -46.98 & 4.31 \\  
3 & H + HF $\rightarrow$ HF + H & -21.36 & 3.86 & -18.33 & 3.63 & -22.34 & -3.70 \\  
4 & 3 reverse & -21.36 & 3.86 & -18.33 & 3.63 & -22.34 & -3.70 \\  
5 & H + HCl $\rightarrow$ HCl + H & -14.59 & 4.41 & -8.52 & 3.53 & -11.91 & 1.99 \\  
6 & 5 reverse & -14.59 & 4.41 & -8.52 & 3.53 & -11.91 & 1.99 \\  
7 & CH$_3$F + H $\rightarrow$ CH$_3$ + HF & -17.51 & 6.12 & -14.38 & 5.54 & -16.49 & -0.83 \\  
8 & 7 reverse & -23.40 & 9.79 & -24.38 & 8.72 & -19.26 & 2.57 \\
\hline 
\end{longtable}

\begin{longtable}{ll|rr|rr|rr}
  \caption{
Functional (FE) and density error (DE) metrics for LSDA@HF on the BH76RC set, all in kcal/mol.
Each set of two columns uses a different proxy for the exact functional as in Table V of the main text.
Error statistics are reported in the bottom three rows.
The reaction index used by the GMTKN and the symbolic index are given in the first two columns.}
  \label{tab:dc_dft_BH76RC_LSDA@HF}
\\ \hline
 &  & \multicolumn{2}{c}{\lcwpbe{}} & \multicolumn{2}{c}{SX-0.5} & \multicolumn{2}{c}{SCAN-FLOSIC} \\ 
Index & Reaction & FE & DE & FE & DE & FE & DE \\ \hline 
  \endfirsthead
\\ \hline
 &  & \multicolumn{2}{c}{\lcwpbe{}} & \multicolumn{2}{c}{SX-0.5} & \multicolumn{2}{c}{SCAN-FLOSIC} \\ 
Index & Reaction & FE & DE & FE & DE & FE & DE \\ \hline 
  \endhead
1 & H + N$_2$O $\rightarrow$ OH + N$_2$ & 28.68 & -7.02 & 41.22 & -6.30 & 51.67 & -2.34 \\  
2 & H + CH$_3$F $\rightarrow$ HF + CH$_3$ & 5.88 & -3.67 & 10.00 & -3.18 & 2.77 & -3.39 \\  
3 & H + F$_2$ $\rightarrow$ HF + F & 16.70 & -4.47 & 35.26 & -3.77 & 23.29 & -4.08 \\  
4 & CH$_3$ + ClF $\rightarrow$ CH$_3$F + Cl & 4.06 & -0.54 & 6.93 & -0.67 & 3.42 & -1.16 \\  
5 & CH$_3$Cl + F$^-$ $\rightarrow$ CH$_3$F + Cl$^-$ & -0.22 & -1.46 & 6.03 & -0.59 & 8.84 & 0.21 \\  
6 & F$^-$ $\hdots$ CH$_3$Cl $\rightarrow$ CH$_3$F $\hdots$ Cl$^-$ & 2.74 & -2.89 & 6.66 & -1.97 & 8.85 & -1.36 \\  
7 & CH$_3$F + OH$^-$ $\rightarrow$ HOCH$_3$ + F$^-$ & 2.58 & -0.22 & 1.55 & -0.60 & -0.84 & -0.60 \\  
8 & CH$_3$F $\hdots$ OH$^-$ $\rightarrow$ HOCH$_3$ $\hdots$ F$^-$ & -3.75 & 0.20 & -3.40 & 0.00 & -5.80 & 0.43 \\
\hline 
\end{longtable}

\clearpage
\subsection{PBE@HF}

\begin{longtable}{ll|rr|rr|rr}
  \caption{
Functional (FE) and density error (DE) metrics for PBE@HF on the BH76 set, all in kcal/mol.
Each set of two columns uses a different proxy for the exact functional as in Table V of the main text.
Error statistics are reported in the bottom three rows.
The reaction index used by the GMTKN and the symbolic index are given in the first two columns.}
  \label{tab:dc_dft_BH76_PBE@HF}
\\ \hline
 &  & \multicolumn{2}{c}{\lcwpbe{}} & \multicolumn{2}{c}{SX-0.5} & \multicolumn{2}{c}{SCAN-FLOSIC} \\ 
Index & Reaction & FE & DE & FE & DE & FE & DE \\ \hline 
  \endfirsthead
\\ \hline
 &  & \multicolumn{2}{c}{\lcwpbe{}} & \multicolumn{2}{c}{SX-0.5} & \multicolumn{2}{c}{SCAN-FLOSIC} \\ 
Index & Reaction & FE & DE & FE & DE & FE & DE \\ \hline 
  \endhead
1 & H + N$_2$O $\rightarrow$ N$_2$ + OH & -7.26 & 10.79 & -2.58 & 9.90 & 9.62 & 1.05 \\  
2 & 1 reverse & -22.98 & 17.64 & -30.87 & 16.06 & -29.63 & 3.30 \\  
3 & H + HF $\rightarrow$ HF + H & -13.19 & 3.12 & -10.24 & 2.98 & -15.84 & -2.78 \\  
4 & 3 reverse & -13.19 & 3.12 & -10.24 & 2.98 & -15.84 & -2.78 \\  
5 & H + HCl $\rightarrow$ HCl + H & -7.93 & 3.27 & -2.14 & 2.65 & -5.65 & 1.24 \\  
6 & 5 reverse & -7.93 & 3.27 & -2.14 & 2.65 & -5.65 & 1.24 \\  
7 & CH$_3$F + H $\rightarrow$ CH$_3$ + HF & -13.10 & 4.99 & -10.16 & 4.60 & -13.29 & -0.68 \\  
8 & 7 reverse & -14.90 & 8.13 & -16.28 & 7.46 & -12.59 & 2.64 \\
\hline 
\end{longtable}

\begin{longtable}{ll|rr|rr|rr}
  \caption{
Functional (FE) and density error (DE) metrics for PBE@HF on the BH76RC set, all in kcal/mol.
Each set of two columns uses a different proxy for the exact functional as in Table V of the main text.
Error statistics are reported in the bottom three rows.
The reaction index used by the GMTKN and the symbolic index are given in the first two columns.}
  \label{tab:dc_dft_BH76RC_PBE@HF}
\\ \hline
 &  & \multicolumn{2}{c}{\lcwpbe{}} & \multicolumn{2}{c}{SX-0.5} & \multicolumn{2}{c}{SCAN-FLOSIC} \\ 
Index & Reaction & FE & DE & FE & DE & FE & DE \\ \hline 
  \endfirsthead
\\ \hline
 &  & \multicolumn{2}{c}{\lcwpbe{}} & \multicolumn{2}{c}{SX-0.5} & \multicolumn{2}{c}{SCAN-FLOSIC} \\ 
Index & Reaction & FE & DE & FE & DE & FE & DE \\ \hline 
  \endhead
1 & H + N$_2$O $\rightarrow$ OH + N$_2$ & 15.72 & -6.85 & 28.28 & -6.16 & 39.25 & -2.24 \\  
2 & H + CH$_3$F $\rightarrow$ HF + CH$_3$ & 1.80 & -3.14 & 6.12 & -2.85 & -0.70 & -3.32 \\  
3 & H + F$_2$ $\rightarrow$ HF + F & 11.30 & -4.23 & 29.98 & -3.64 & 18.57 & -3.85 \\  
4 & CH$_3$ + ClF $\rightarrow$ CH$_3$F + Cl & 5.14 & -0.98 & 7.87 & -0.97 & 4.55 & -1.07 \\  
5 & CH$_3$Cl + F$^-$ $\rightarrow$ CH$_3$F + Cl$^-$ & 1.56 & -1.72 & 7.65 & -0.69 & 10.22 & 0.37 \\  
6 & F$^-$ $\hdots$ CH$_3$Cl $\rightarrow$ CH$_3$F $\hdots$ Cl$^-$ & 3.16 & -2.83 & 7.10 & -1.92 & 9.21 & -1.23 \\  
7 & CH$_3$F + OH$^-$ $\rightarrow$ HOCH$_3$ + F$^-$ & 2.84 & -0.07 & 1.92 & -0.55 & -0.48 & -0.58 \\  
8 & CH$_3$F $\hdots$ OH$^-$ $\rightarrow$ HOCH$_3$ $\hdots$ F$^-$ & 1.15 & 0.36 & 1.56 & 0.11 & -0.96 & 0.57 \\
\hline 
\end{longtable}

\clearpage
\subsection{SCAN@HF}

\begin{longtable}{ll|rr|rr|rr}
  \caption{
Functional (FE) and density error (DE) metrics for SCAN@HF on the BH76 set, all in kcal/mol.
Each set of two columns uses a different proxy for the exact functional as in Table V of the main text.
Error statistics are reported in the bottom three rows.
The reaction index used by the GMTKN and the symbolic index are given in the first two columns.}
  \label{tab:dc_dft_BH76_SCAN@HF}
\\ \hline
 &  & \multicolumn{2}{c}{\lcwpbe{}} & \multicolumn{2}{c}{SX-0.5} & \multicolumn{2}{c}{SCAN-FLOSIC} \\ 
Index & Reaction & FE & DE & FE & DE & FE & DE \\ \hline 
  \endfirsthead
\\ \hline
 &  & \multicolumn{2}{c}{\lcwpbe{}} & \multicolumn{2}{c}{SX-0.5} & \multicolumn{2}{c}{SCAN-FLOSIC} \\ 
Index & Reaction & FE & DE & FE & DE & FE & DE \\ \hline 
  \endhead
1 & H + N$_2$O $\rightarrow$ N$_2$ + OH & -8.29 & 7.29 & -4.21 & 7.00 & 5.83 & 0.23 \\  
2 & 1 reverse & -11.64 & 13.31 & -20.13 & 12.33 & -21.91 & 2.20 \\  
3 & H + HF $\rightarrow$ HF + H & -11.62 & 2.78 & -8.61 & 2.58 & -15.67 & -1.83 \\  
4 & 3 reverse & -11.62 & 2.78 & -8.61 & 2.58 & -15.67 & -1.83 \\  
5 & H + HCl $\rightarrow$ HCl + H & -8.95 & 1.69 & -3.64 & 1.55 & -7.50 & 0.54 \\  
6 & 5 reverse & -8.95 & 1.69 & -3.64 & 1.55 & -7.50 & 0.54 \\  
7 & CH$_3$F + H $\rightarrow$ CH$_3$ + HF & -11.74 & 4.44 & -9.02 & 4.28 & -13.52 & 0.32 \\  
8 & 7 reverse & -9.91 & 6.95 & -11.46 & 6.45 & -9.21 & 2.98 \\
\hline 
\end{longtable}

\begin{longtable}{ll|rr|rr|rr}
  \caption{
Functional (FE) and density error (DE) metrics for SCAN@HF on the BH76RC set, all in kcal/mol.
Each set of two columns uses a different proxy for the exact functional as in Table V of the main text.
Error statistics are reported in the bottom three rows.
The reaction index used by the GMTKN and the symbolic index are given in the first two columns.}
  \label{tab:dc_dft_BH76RC_SCAN@HF}
\\ \hline
 &  & \multicolumn{2}{c}{\lcwpbe{}} & \multicolumn{2}{c}{SX-0.5} & \multicolumn{2}{c}{SCAN-FLOSIC} \\ 
Index & Reaction & FE & DE & FE & DE & FE & DE \\ \hline 
  \endfirsthead
\\ \hline
 &  & \multicolumn{2}{c}{\lcwpbe{}} & \multicolumn{2}{c}{SX-0.5} & \multicolumn{2}{c}{SCAN-FLOSIC} \\ 
Index & Reaction & FE & DE & FE & DE & FE & DE \\ \hline 
  \endhead
1 & H + N$_2$O $\rightarrow$ OH + N$_2$ & 3.35 & -6.02 & 15.92 & -5.33 & 27.74 & -1.97 \\  
2 & H + CH$_3$F $\rightarrow$ HF + CH$_3$ & -1.83 & -2.51 & 2.44 & -2.17 & -4.31 & -2.67 \\  
3 & H + F$_2$ $\rightarrow$ HF + F & 0.67 & -4.42 & 19.07 & -3.56 & 8.16 & -3.55 \\  
4 & CH$_3$ + ClF $\rightarrow$ CH$_3$F + Cl & 3.11 & -1.21 & 5.75 & -1.11 & 2.88 & -1.05 \\  
5 & CH$_3$Cl + F$^-$ $\rightarrow$ CH$_3$F + Cl$^-$ & -4.01 & -0.79 & 2.61 & -0.29 & 5.28 & 0.51 \\  
6 & F$^-$ $\hdots$ CH$_3$Cl $\rightarrow$ CH$_3$F $\hdots$ Cl$^-$ & -0.78 & -1.79 & 3.61 & -1.34 & 5.75 & -0.86 \\  
7 & CH$_3$F + OH$^-$ $\rightarrow$ HOCH$_3$ + F$^-$ & 3.45 & -0.51 & 2.20 & -0.67 & 0.03 & -0.50 \\  
8 & CH$_3$F $\hdots$ OH$^-$ $\rightarrow$ HOCH$_3$ $\hdots$ F$^-$ & 1.61 & 0.13 & 1.92 & -0.03 & -0.16 & 0.43 \\
\hline 
\end{longtable}

\clearpage
\subsection{\rrscan{}@HF}

\begin{longtable}{ll|rr|rr|rr}
  \caption{
Functional (FE) and density error (DE) metrics for \rrscan{}@HF on the BH76 set, all in kcal/mol.
Each set of two columns uses a different proxy for the exact functional as in Table V of the main text.
Error statistics are reported in the bottom three rows.
The reaction index used by the GMTKN and the symbolic index are given in the first two columns.}
  \label{tab:dc_dft_BH76_r2SCAN@HF}
\\ \hline
 &  & \multicolumn{2}{c}{\lcwpbe{}} & \multicolumn{2}{c}{SX-0.5} & \multicolumn{2}{c}{SCAN-FLOSIC} \\ 
Index & Reaction & FE & DE & FE & DE & FE & DE \\ \hline 
  \endfirsthead
\\ \hline
 &  & \multicolumn{2}{c}{\lcwpbe{}} & \multicolumn{2}{c}{SX-0.5} & \multicolumn{2}{c}{SCAN-FLOSIC} \\ 
Index & Reaction & FE & DE & FE & DE & FE & DE \\ \hline 
  \endhead
1 & H + N$_2$O $\rightarrow$ N$_2$ + OH & -7.26 & 8.22 & -2.90 & 7.65 & -0.78 & 8.79 \\  
2 & 1 reverse & -13.02 & 14.55 & -21.20 & 13.26 & -36.02 & 16.29 \\  
3 & H + HF $\rightarrow$ HF + H & -11.27 & 2.68 & -8.26 & 2.48 & -20.10 & 2.85 \\  
4 & 3 reverse & -11.27 & 2.68 & -8.26 & 2.48 & -20.10 & 2.85 \\  
5 & H + HCl $\rightarrow$ HCl + H & -8.63 & 2.32 & -3.11 & 1.98 & -8.79 & 2.76 \\  
6 & 5 reverse & -8.63 & 2.32 & -3.11 & 1.98 & -8.79 & 2.76 \\  
7 & CH$_3$F + H $\rightarrow$ CH$_3$ + HF & -12.59 & 4.28 & -9.85 & 4.09 & -18.83 & 4.60 \\  
8 & 7 reverse & -9.35 & 6.72 & -11.00 & 6.32 & -13.31 & 7.42 \\
\hline 
\end{longtable}

\begin{longtable}{ll|rr|rr|rr}
  \caption{
Functional (FE) and density error (DE) metrics for \rrscan{}@HF on the BH76RC set, all in kcal/mol.
Each set of two columns uses a different proxy for the exact functional as in Table V of the main text.
Error statistics are reported in the bottom three rows.
The reaction index used by the GMTKN and the symbolic index are given in the first two columns.}
  \label{tab:dc_dft_BH76RC_r2SCAN@HF}
\\ \hline
 &  & \multicolumn{2}{c}{\lcwpbe{}} & \multicolumn{2}{c}{SX-0.5} & \multicolumn{2}{c}{SCAN-FLOSIC} \\ 
Index & Reaction & FE & DE & FE & DE & FE & DE \\ \hline 
  \endfirsthead
\\ \hline
 &  & \multicolumn{2}{c}{\lcwpbe{}} & \multicolumn{2}{c}{SX-0.5} & \multicolumn{2}{c}{SCAN-FLOSIC} \\ 
Index & Reaction & FE & DE & FE & DE & FE & DE \\ \hline 
  \endhead
1 & H + N$_2$O $\rightarrow$ OH + N$_2$ & 5.76 & -6.33 & 18.30 & -5.61 & 35.25 & -7.50 \\  
2 & H + CH$_3$F $\rightarrow$ HF + CH$_3$ & -3.24 & -2.44 & 1.15 & -2.23 & -5.52 & -2.82 \\  
3 & H + F$_2$ $\rightarrow$ HF + F & 1.05 & -4.22 & 19.54 & -3.45 & 9.75 & -4.44 \\  
4 & CH$_3$ + ClF $\rightarrow$ CH$_3$F + Cl & 4.08 & -1.65 & 6.53 & -1.36 & 4.33 & -1.86 \\  
5 & CH$_3$Cl + F$^-$ $\rightarrow$ CH$_3$F + Cl$^-$ & -2.73 & -1.17 & 3.81 & -0.59 & 7.26 & -0.58 \\  
6 & F$^-$ $\hdots$ CH$_3$Cl $\rightarrow$ CH$_3$F $\hdots$ Cl$^-$ & 0.03 & -2.17 & 4.36 & -1.65 & 8.01 & -2.66 \\  
7 & CH$_3$F + OH$^-$ $\rightarrow$ HOCH$_3$ + F$^-$ & 3.17 & -0.24 & 1.98 & -0.45 & -0.10 & -0.41 \\  
8 & CH$_3$F $\hdots$ OH$^-$ $\rightarrow$ HOCH$_3$ $\hdots$ F$^-$ & 1.41 & 0.58 & 1.81 & 0.33 & -0.36 & 0.82 \\
\hline 
\end{longtable}

\section{Deviations computed with high-level reference values and approximate functionals}

\subsection{LSDA}

\begin{longtable}{llrrrrr}
  \caption{
Deviations (in kcal/mol) from the GMTKN reference values (Exact) and various functionals for the BH76 set using LSDA.
The deviations are approximate errors; the error statistics in the ``Exact'' column correspond to the values in Table \ref{tab:ak_pyscf_aug-cc-pvqz} of the main text.
The other columns replace the exact energies with those of a proxy.
For an approximate functional to be a good proxy for the exact one (in evaluating density- and functional-driven errors), the deviations computed with the high-level reference values and with the approximate functional should be comparable.
Mean deviations (MDs), mean absolute deviations (MADs), and variances (VARs) are reported in the bottom three rows.
The reaction index used by the GMTKN and the symbolic index are given in the first two columns.
The sum of ``Barrier Height'' and ``Exact'' columns yields the corresponding LSDA barrier height.}
  \label{tab:devs_BH76_LSDA}
\\ \hline
Index & Reaction & Barrier Height & Exact & \lcwpbe{} & SX-0.5 & SCAN-FLOSIC \\ \hline 
  \endfirsthead
\\ \hline
Index & Reaction & Barrier Height & Exact & \lcwpbe{} & SX-0.5 & SCAN-FLOSIC \\ \hline 
  \endhead
1 & H + N$_2$O $\rightarrow$ N$_2$ + OH & 17.70 & -15.35 & -15.35 & -11.56 & -8.41 \\  
2 & 1 reverse & 82.60 & -48.42 & -45.70 & -55.17 & -66.48 \\  
3 & H + HF $\rightarrow$ HF + H & 42.10 & -23.66 & -22.17 & -19.37 & -30.67 \\  
4 & 3 reverse & 42.10 & -23.66 & -22.17 & -19.37 & -30.67 \\  
5 & H + HCl $\rightarrow$ HCl + H & 17.80 & -14.80 & -16.23 & -11.05 & -15.86 \\  
6 & 5 reverse & 17.80 & -14.80 & -16.23 & -11.05 & -15.86 \\  
7 & CH$_3$F + H $\rightarrow$ CH$_3$ + HF & 30.50 & -17.71 & -19.05 & -16.50 & -24.94 \\  
8 & 7 reverse & 56.90 & -25.00 & -25.30 & -27.35 & -28.32 \\
\hline 
\end{longtable}

\begin{longtable}{llrrrrr}
  \caption{
Deviations (in kcal/mol) from the GMTKN reference values (Exact) and various functionals for the BH76RC set using LSDA.
The deviations are approximate errors; the error statistics in the ``Exact'' column correspond to the values in Table \ref{tab:ak_pyscf_aug-cc-pvqz} of the main text.
The other columns replace the exact energies with those of a proxy.
For an approximate functional to be a good proxy for the exact one (in evaluating density- and functional-driven errors), the deviations computed with the high-level reference values and with the approximate functional should be comparable.
Mean deviations (MDs), mean absolute deviations (MADs), and variances (VARs) are reported in the bottom three rows.
The reaction index used by the GMTKN and the symbolic index are given in the first two columns.
The sum of ``Reaction Energy'' and ``Exact'' columns yields the corresponding LSDA reaction energy.}
  \label{tab:devs_BH76RC_LSDA}
\\ \hline
Index & Reaction & Reaction Energy & Exact & \lcwpbe{} & SX-0.5 & SCAN-FLOSIC \\ \hline 
  \endfirsthead
\\ \hline
Index & Reaction & Reaction Energy & Exact & \lcwpbe{} & SX-0.5 & SCAN-FLOSIC \\ \hline 
  \endhead
1 & H + N$_2$O $\rightarrow$ OH + N$_2$ & -64.91 & 33.08 & 30.35 & 43.60 & 58.07 \\  
2 & H + CH$_3$F $\rightarrow$ HF + CH$_3$ & -26.32 & 7.20 & 6.25 & 10.85 & 3.38 \\  
3 & H + F$_2$ $\rightarrow$ HF + F & -103.28 & 18.42 & 17.05 & 36.31 & 24.17 \\  
4 & CH$_3$ + ClF $\rightarrow$ CH$_3$F + Cl & -52.64 & 2.93 & 9.93 & 12.67 & 4.63 \\  
5 & CH$_3$Cl + F$^-$ $\rightarrow$ CH$_3$F + Cl$^-$ & -32.16 & -0.31 & -0.29 & 6.83 & 9.66 \\  
6 & F$^-$ $\hdots$ CH$_3$Cl $\rightarrow$ CH$_3$F $\hdots$ Cl$^-$ & -26.12 & 3.67 & 3.76 & 8.61 & 11.12 \\  
7 & CH$_3$F + OH$^-$ $\rightarrow$ HOCH$_3$ + F$^-$ & -20.32 & -1.73 & 2.98 & 1.58 & -0.85 \\  
8 & CH$_3$F $\hdots$ OH$^-$ $\rightarrow$ HOCH$_3$ $\hdots$ F$^-$ & -36.71 & -8.81 & -3.28 & -3.13 & -5.62 \\
\hline 
\end{longtable}

\clearpage
\subsection{PBE}

\begin{longtable}{llrrrrr}
  \caption{
Deviations (in kcal/mol) from the GMTKN reference values (Exact) and various functionals for the BH76 set using PBE.
The deviations are approximate errors; the error statistics in the ``Exact'' column correspond to the values in Table \ref{tab:ak_pyscf_aug-cc-pvqz} of the main text.
The other columns replace the exact energies with those of a proxy.
For an approximate functional to be a good proxy for the exact one (in evaluating density- and functional-driven errors), the deviations computed with the high-level reference values and with the approximate functional should be comparable.
Mean deviations (MDs), mean absolute deviations (MADs), and variances (VARs) are reported in the bottom three rows.
The reaction index used by the GMTKN and the symbolic index are given in the first two columns.
The sum of ``Barrier Height'' and ``Exact'' columns yields the corresponding PBE barrier height.}
  \label{tab:devs_BH76_PBE}
\\ \hline
Index & Reaction & Barrier Height & Exact & \lcwpbe{} & SX-0.5 & SCAN-FLOSIC \\ \hline 
  \endfirsthead
\\ \hline
Index & Reaction & Barrier Height & Exact & \lcwpbe{} & SX-0.5 & SCAN-FLOSIC \\ \hline 
  \endhead
1 & H + N$_2$O $\rightarrow$ N$_2$ + OH & 17.70 & -8.08 & -8.11 & -4.32 & -1.14 \\  
2 & 1 reverse & 82.60 & -28.38 & -25.30 & -34.77 & -46.44 \\  
3 & H + HF $\rightarrow$ HF + H & 42.10 & -15.07 & -13.53 & -10.73 & -22.08 \\  
4 & 3 reverse & 42.10 & -15.07 & -13.53 & -10.73 & -22.08 \\  
5 & H + HCl $\rightarrow$ HCl + H & 17.80 & -7.53 & -8.88 & -3.71 & -8.59 \\  
6 & 5 reverse & 17.80 & -7.53 & -8.88 & -3.71 & -8.59 \\  
7 & CH$_3$F + H $\rightarrow$ CH$_3$ + HF & 30.50 & -12.53 & -13.92 & -11.36 & -19.75 \\  
8 & 7 reverse & 56.90 & -15.76 & -15.94 & -18.00 & -19.08 \\
\hline 
\end{longtable}

\begin{longtable}{llrrrrr}
  \caption{
Deviations (in kcal/mol) from the GMTKN reference values (Exact) and various functionals for the BH76RC set using PBE.
The deviations are approximate errors; the error statistics in the ``Exact'' column correspond to the values in Table \ref{tab:ak_pyscf_aug-cc-pvqz} of the main text.
The other columns replace the exact energies with those of a proxy.
For an approximate functional to be a good proxy for the exact one (in evaluating density- and functional-driven errors), the deviations computed with the high-level reference values and with the approximate functional should be comparable.
Mean deviations (MDs), mean absolute deviations (MADs), and variances (VARs) are reported in the bottom three rows.
The reaction index used by the GMTKN and the symbolic index are given in the first two columns.
The sum of ``Reaction Energy'' and ``Exact'' columns yields the corresponding PBE reaction energy.}
  \label{tab:devs_BH76RC_PBE}
\\ \hline
Index & Reaction & Reaction Energy & Exact & \lcwpbe{} & SX-0.5 & SCAN-FLOSIC \\ \hline 
  \endfirsthead
\\ \hline
Index & Reaction & Reaction Energy & Exact & \lcwpbe{} & SX-0.5 & SCAN-FLOSIC \\ \hline 
  \endhead
1 & H + N$_2$O $\rightarrow$ OH + N$_2$ & -64.91 & 20.32 & 17.19 & 30.45 & 45.30 \\  
2 & H + CH$_3$F $\rightarrow$ HF + CH$_3$ & -26.32 & 3.15 & 2.03 & 6.64 & -0.67 \\  
3 & H + F$_2$ $\rightarrow$ HF + F & -103.28 & 13.51 & 11.55 & 30.81 & 19.26 \\  
4 & CH$_3$ + ClF $\rightarrow$ CH$_3$F + Cl & -52.64 & 3.26 & 5.52 & 8.27 & 4.96 \\  
5 & CH$_3$Cl + F$^-$ $\rightarrow$ CH$_3$F + Cl$^-$ & -32.16 & 1.45 & 1.60 & 8.72 & 11.42 \\  
6 & F$^-$ $\hdots$ CH$_3$Cl $\rightarrow$ CH$_3$F $\hdots$ Cl$^-$ & -26.12 & 4.16 & 4.32 & 9.17 & 11.61 \\  
7 & CH$_3$F + OH$^-$ $\rightarrow$ HOCH$_3$ + F$^-$ & -20.32 & -1.49 & 3.20 & 1.80 & -0.61 \\  
8 & CH$_3$F $\hdots$ OH$^-$ $\rightarrow$ HOCH$_3$ $\hdots$ F$^-$ & -36.71 & -4.21 & 1.57 & 1.72 & -1.02 \\
\hline 
\end{longtable}

\clearpage
\subsection{BLYP}

\begin{longtable}{llrrrr}
  \caption{
Deviations (in kcal/mol) from the GMTKN reference values (Exact) and various functionals for the BH76 set using BLYP.
The deviations are approximate errors; the error statistics in the ``Exact'' column correspond to the values in Table \ref{tab:ak_pyscf_aug-cc-pvqz} of the main text.
The other columns replace the exact energies with those of a proxy.
For an approximate functional to be a good proxy for the exact one (in evaluating density- and functional-driven errors), the deviations computed with the high-level reference values and with the approximate functional should be comparable.
Mean deviations (MDs), mean absolute deviations (MADs), and variances (VARs) are reported in the bottom three rows.
The reaction index used by the GMTKN and the symbolic index are given in the first two columns.
The sum of ``Barrier Height'' and ``Exact'' columns yields the corresponding BLYP barrier height.}
  \label{tab:devs_BH76_BLYP}
\\ \hline
Index & Reaction & Barrier Height & Exact & \lcwpbe{} & SX-0.5 \\ \hline 
  \endfirsthead
\\ \hline
Index & Reaction & Barrier Height & Exact & \lcwpbe{} & SX-0.5 \\ \hline 
  \endhead
1 & H + N$_2$O $\rightarrow$ N$_2$ + OH & 17.70 & -9.16 & -9.52 & -5.73 \\  
2 & 1 reverse & 82.60 & -20.64 & -16.15 & -25.61 \\  
3 & H + HF $\rightarrow$ HF + H & 42.10 & -16.16 & -14.54 & -11.74 \\  
4 & 3 reverse & 42.10 & -16.16 & -14.54 & -11.74 \\  
5 & H + HCl $\rightarrow$ HCl + H & 17.80 & -7.65 & -8.70 & -3.52 \\  
6 & 5 reverse & 17.80 & -7.65 & -8.70 & -3.52 \\  
7 & CH$_3$F + H $\rightarrow$ CH$_3$ + HF & 30.50 & -14.59 & -16.27 & -13.72 \\  
8 & 7 reverse & 56.90 & -14.67 & -14.72 & -16.78 \\
\hline 
\end{longtable}

\begin{longtable}{llrrrr}
  \caption{
Deviations (in kcal/mol) from the GMTKN reference values (Exact) and various functionals for the BH76RC set using BLYP.
The deviations are approximate errors; the error statistics in the ``Exact'' column correspond to the values in Table \ref{tab:ak_pyscf_aug-cc-pvqz} of the main text.
The other columns replace the exact energies with those of a proxy.
For an approximate functional to be a good proxy for the exact one (in evaluating density- and functional-driven errors), the deviations computed with the high-level reference values and with the approximate functional should be comparable.
Mean deviations (MDs), mean absolute deviations (MADs), and variances (VARs) are reported in the bottom three rows.
The reaction index used by the GMTKN and the symbolic index are given in the first two columns.
The sum of ``Reaction Energy'' and ``Exact'' columns yields the corresponding BLYP reaction energy.}
  \label{tab:devs_BH76RC_BLYP}
\\ \hline
Index & Reaction & Reaction Energy & Exact & \lcwpbe{} & SX-0.5 \\ \hline 
  \endfirsthead
\\ \hline
Index & Reaction & Reaction Energy & Exact & \lcwpbe{} & SX-0.5 \\ \hline 
  \endhead
1 & H + N$_2$O $\rightarrow$ OH + N$_2$ & -64.91 & 11.49 & 6.63 & 19.88 \\  
2 & H + CH$_3$F $\rightarrow$ HF + CH$_3$ & -26.32 & 0.00 & -1.55 & 3.06 \\  
3 & H + F$_2$ $\rightarrow$ HF + F & -103.28 & 10.76 & 8.23 & 27.49 \\  
4 & CH$_3$ + ClF $\rightarrow$ CH$_3$F + Cl & -52.64 & 4.13 & 4.72 & 7.46 \\  
5 & CH$_3$Cl + F$^-$ $\rightarrow$ CH$_3$F + Cl$^-$ & -32.16 & 2.17 & 1.10 & 8.22 \\  
6 & F$^-$ $\hdots$ CH$_3$Cl $\rightarrow$ CH$_3$F $\hdots$ Cl$^-$ & -26.12 & 3.94 & 3.82 & 8.66 \\  
7 & CH$_3$F + OH$^-$ $\rightarrow$ HOCH$_3$ + F$^-$ & -20.32 & 1.45 & 4.16 & 2.75 \\  
8 & CH$_3$F $\hdots$ OH$^-$ $\rightarrow$ HOCH$_3$ $\hdots$ F$^-$ & -36.71 & 1.25 & 4.51 & 4.66 \\
\hline 
\end{longtable}

\clearpage
\subsection{SCAN}

%

\clearpage
\subsection{\rrscan{}}

%

\clearpage
\subsection{M06-L}

%

\clearpage
\subsection{MN15-L}

%

\clearpage
\subsection{B3LYP}

%

\clearpage
\subsection{\lcwpbe{}}

%

\clearpage
\subsection{SCAN@HF}

%

\section{Individual reaction errors}

\subsection{LSDA}

%

\clearpage
\subsection{PBE}

%

\clearpage
\subsection{BLYP}

%

\clearpage
\subsection{SCAN}

%

\clearpage
\subsection{\rrscan{}}

%

\clearpage
\subsection{M06-L}

%

\clearpage
\subsection{MN15-L}

%

\clearpage
\subsection{\lcwpbe{}}

%

\clearpage
\subsection{B3LYP}

%

\clearpage
\subsection{SX-0.5}

\begin{longtable}{llrr}
  \caption{
Individual BH76 errors for SX-0.5 including its self-consistent (SCF) and non-selfconsistent values (indicated with an ``@ symbol) used throughout.
All values are in kcal/mol.
The ``Exact'' column lists the reference values, such that the sum of the ``Exact'' column and any other column would yield the barrier height computed with SX-0.5 at that density.
The reaction index used by the GMTKN and the symbolic name are given in the first two columns.}
  \label{tab:BH76_errors_S50X}
\\ \hline 
Index & Reaction & Exact & SCF \\ \hline 
  \endfirsthead
\\ \hline 
Index & Reaction & Exact & SCF \\ \hline 
  \endhead
1 & H + N$_2$O $\rightarrow$ N$_2$ + OH & 17.70 & -3.43 \\  
2 & 1 reverse & 82.60 & 4.97 \\  
3 & H + HF $\rightarrow$ HF + H & 42.10 & -4.43 \\  
4 & 3 reverse & 42.10 & -4.43 \\  
5 & H + HCl $\rightarrow$ HCl + H & 17.80 & -4.13 \\  
6 & 5 reverse & 17.80 & -4.13 \\  
7 & CH$_3$F + H $\rightarrow$ CH$_3$ + HF & 30.50 & -0.87 \\  
8 & 7 reverse & 56.90 & 2.11 \\
\hline 
\end{longtable}%

\clearpage
\subsection{LSDA-FLOSIC}

%

\clearpage
\subsection{PBE-FLOSIC}

%

\clearpage
\subsection{SCAN-FLOSIC}

%

\clearpage
\subsection{\rrscan{}-FLOSIC}

%

\end{document}